\documentclass[showpacs,amsmath,amssymb,twocolumn,superscriptaddress,notitlepage,preprintnumbers,pra]{revtex4-1}

\usepackage[dvips]{graphicx}

\usepackage{amsmath,amssymb,amsthm,mathrsfs,amsfonts,dsfont}

\usepackage{dcolumn}% Align table columns on decimal point
\usepackage{bm}% bold math
\usepackage{amsmath}
\usepackage{amssymb}
\usepackage{braket}
\usepackage{physics}
\usepackage{amsthm}
\usepackage{natbib}
\usepackage{url}
\usepackage{mathtools}
\usepackage{diagbox}

\usepackage[utf8]{inputenc}
\usepackage[english]{babel}
\usepackage{scalerel}[2014/03/10]
\usepackage{stackengine}
\usepackage{algorithm}
\usepackage[noend]{algpseudocode} 
\usepackage{subfigure}

\usepackage{tikz}
\usetikzlibrary{quantikz}
\usetikzlibrary{intersections,through}
\usepackage{hyperref}
\hypersetup{colorlinks=true, linkcolor=blue, citecolor=blue, urlcolor=black }

\renewcommand{\tr}{\textup{Tr}}

\algrenewcommand\algorithmicrequire{\textbf{Input:}}
\algrenewcommand\algorithmicensure{\textbf{Output:}}

\makeatletter
\newcommand{\algmargin}{\the\ALG@thistlm}
\makeatother
\newlength{\whilewidth}
\algdef{SE}[parWHILE]{parWhile}{EndparWhile}[1]
  {\parbox[t]{\dimexpr\linewidth-\algmargin}{%
     \hangindent\whilewidth\strut\algorithmicwhile\ #1\ \algorithmicdo\strut}}{\algorithmicend\ \algorithmicwhile}%
\algnewcommand{\parState}[1]{\State%
  \parbox[t]{\dimexpr\linewidth-\algmargin}{\strut #1\strut}}

\theoremstyle{definition}

\newcommand{\thu}{Department of Mathematics, Tsinghua University,  Beijing 100084, China}
\newcommand{\YMSC}{Yau Mathematical Sciences Center, Tsinghua University,  Beijing 100084, China}
\newcommand{\bimsa}{Yanqi Lake Beijing Institute of Mathematical Sciences and Applications, Beijing 100407, China }
\newcommand{\nju}{National Laboratory of Solid State Microstructures and College of Engineering and Applied Sciences, Nanjing University, Nanjing 210093, China}%c

\begin{document}

\title{Variational Graphical Quantum Error Correction Codes}

\author{Yuguo Shao}
\thanks{These authors contributed equally to this work.}
\affiliation{\YMSC}
\affiliation{\thu}

\author{Yong-Chang Li}
\thanks{These authors contributed equally to this work.}
\affiliation{\nju}

\author{Fuchuan Wei}
\affiliation{\YMSC}
\affiliation{\thu}

\author{Hao Zhan}
\affiliation{\nju}

\author{Ben Wang}%
\affiliation{\nju}

\author{Zhaohui Wei}
\thanks{weizhaohui@gmail.com}
\affiliation{\YMSC}
\affiliation{\bimsa}

\author{Lijian Zhang}%
\thanks{lijian.zhang@nju.edu.cn}
\affiliation{\nju}

\author{Zhengwei Liu}
\thanks{liuzhengwei@mail.tsinghua.edu.cn}
\affiliation{\YMSC}
\affiliation{\thu}
\affiliation{\bimsa}

\begin{abstract}
    Quantum error correction is essential for achieving fault-tolerant quantum computation. 
    However, most typical quantum error-correcting codes are designed for generic noise models, which may fail to accurately capture the intricate noise characteristics of real quantum devices, limiting their practical performance.
    This work introduces a learning-based framework for the constructing of quantum error-correcting codes, termed Variational Graphical Quantum Error Correction (VGQEC) codes, which adapts to specific noise profiles of different quantum devices, enabling the design of noise-tailored codes. 
    Specifically, inspired by Quon, a graphical language for quantum information, VGQEC codes incorporate tunable parameters embedded within their Quon graphs, allowing dynamic reconfigurations of the graph structures through parameter adjustments.
    As the first application of this approach, we show that this flexibility in code designs facilitates seamless transitions between various code families, exemplified by the establishment of a bridge between the five-qubit repetition code and the [[5,1,3]] code, thereby combining their respective advantages.
    Additionally, a VGQEC code derived from the three-qubit repetition code is fine-tuned for the amplitude damping noise, showcasing the approach’s ability for noise-specific code design.
    Moreover, we experimentally demonstrate the effectiveness of the three-qubit VGQEC code in the low-to-medium noise regime with a photonic system, highlighting its potential for real-world applications.
\end{abstract}

\maketitle

\section{Introduction}
In recent years, quantum computation has undergone substantial growth and demonstrated its potential in effectively solving certain hard computational problems~\cite{nielsen2002quantum,shor1999polynomial,grover1996fast,dunjko2018computational,campbell2019applying,jones2012faster,harrow2009quantum}.
However, physical implementations of quantum information processing tasks are unavoidably subject to noise, which could even eradicate useful quantum information.
To address this challenge, it is essential to implement active quantum error correction~(QEC) to protect information against errors that occur dynamically during the storage and the processing of quantum information~\cite{terhal2015quantum,shor1995scheme,lidar2013quantum,girvin2021introduction}.
A common approach to designing QEC codes involves utilizing the Pauli framework to construct and analyze codes for generic noise acting on a small but unknown subset of qubits~\cite{kitaev1997quantum, fowler2012surface, dennis2002topological, fowler2009high, bravyi1998quantum, calderbank1996good, steane1996multiple, gottesman1997stabilizer, cross2008codeword, chuang2009codeword, breuckmann2021quantum, panteleev2022asymptotically}.
However, in realistic devices, quantum information is subject to hardware-specific noise processes, which differ significantly across various physical platforms~\cite{dawson2006noise,wilen2021correlated,guo2021testing}.
Furthermore, the error model may change over time, and some qubits in a real-world device might be more prone to errors than others.
Therefore, general-purpose quantum error correction codes mentioned above may not be optimal choices for a given quantum device with specific noise characteristics~\cite{leung1997approximate}.

Because of this issue, many efforts have been devoted into the development of noise-tailored QEC codes~\cite{grassl2018quantum,mao2024optimized}. For instance, a 4-qubit code was tailored for amplitude damping noise~\cite{leung1997approximate}, while cat-codes and binomial codes were devised to safeguard information stored in boson modes~\cite{cochrane1999macroscopically,li2017cat,michael2016new}.
However, tailoring QEC codes to specific noise models remains a labor-intensive task, requiring extensive manual analysis for each individual case. The lack of general and systematic approaches for this task limits not only the efficiency of code designs but also the widespread applications of QEC in practical quantum systems.

Adaptive methodologies for searching QEC codes have been explored as a means to address this challenge. 
One common strategy involves reformulating the search for effective error correction schemes as an optimization problem~\cite{fletcher2008channel,kosut2008robust,taghavi2010channel,berta2022semidefinite,kosut2009quantum,fletcher2008structured,reimpell2005iterative}.
Nevertheless, these methods typically require explicit characterizations of noise models, and the computational cost of simulating a quantum process often scales exponentially, making them impractical for large-scale quantum systems.
An alternative approach employs the variational quantum algorithm framework to construct noise-tailored QEC codes through experimental interactions with quantum devices ~\cite{nautrup2019optimizing,fosel2018reinforcement,locher2023quantum,bausch2020quantum,johnson2017qvector,cao2022quantum,zoratti2023improving}.
Yet, these methods attempt to develop target codes from scratch, without leveraging prior knowledge of known general-purpose QEC codes.
This makes it challenging to identify high-quality noise-tailored codes because of the limited efficiency.

To address this limitation, in this work we propose a novel learning-based strategy that can develop various noise-tailored codes based on established general-purpose codes.
This approach capitalizes on the strengths of established codes, potentially reducing the time and resources required for code searching, and improving the adaptability of obtained noise-tailored codes.
The key to our approach is a mathematical picture language~\cite{jaffe2018mathematical,liu2017quon}, known as the \textit{Quon 3D language}.
This language represents quantum information objects using graphs, termed Quon graphs.
Building on this foundation, we introduce \textit{Variational Graphical Quantum Error Correction (VGQEC) codes}, a general framework that defines a family of QEC codes through Quon graphs with multiple adjustable parameters.
By tuning these parameters, the structure of the Quon graph changes accordingly, enabling the VGQEC code to smoothly transition between different error correction codes and adapt to various noise models.

As a demonstration of the potential of VGQEC codes, we first design such a code based on the repetition code, which is optimized for the amplitude damping noise. In particular, we identify a specific three-qubit code that effectively mitigates the noise, which, to our knowledge, is the most compact code known for this type of noise.
Furthermore, we design another VGQEC code that interpolates between the repetition code and the $[[5,1,3]]$ code. 
This code possesses the ability to seamlessly transition between the two codes, enabling it to adapt to different noises by switching between them and leveraging their respective advantages.

To physically implement VGQEC codes on Noisy Intermediate-Scale Quantum (NISQ) devices~\cite{preskill2018quantum}, we propose a quantum-classical hybrid approach using the parameterized quantum circuit framework, which employs parameterized quantum circuits to realize both encoding and recovery maps, simplifying the constructions of VGQEC codes from established QEC codes.
Our numerical simulations for the amplitude damping noise and the thermal relaxation processes demonstrate the effectiveness of this approach.
For example, we confirm that the proposed scheme successfully implements the three-qubit code for the amplitude damping noise mentioned above. Similarly, in scenarios where qubits exhibit varying coherence times, our approach optimizes a specific VGQEC code based on the $[[5,1,3]]$ code, which significantly improves the performance compared with the original code.

To showcase the practical feasibility of our approach, we experimentally demonstrate the above three-qubit VGQEC code for the amplitude damping noise with a photonic system. 
Through the experimentally measured fidelity of the composite channel composed of the encoding map, the damping channel, and the optimal recovery map, we confirm that the obtained three-qubit VGQEC code exhibits superior performance in the low-to-medium damping noise regime, in comparison to the original three-qubit repetition code and the [[5,1,3]] code.
This serves as a proof-of-concept demonstration of how VGQEC codes mitigate the amplitude damping noise in real-world quantum devices.

\section{Variational Graphical Codes} \label{sec:VMQECC}

\begin{figure*}[htbp]
    \centering
    \includegraphics[width=\linewidth]{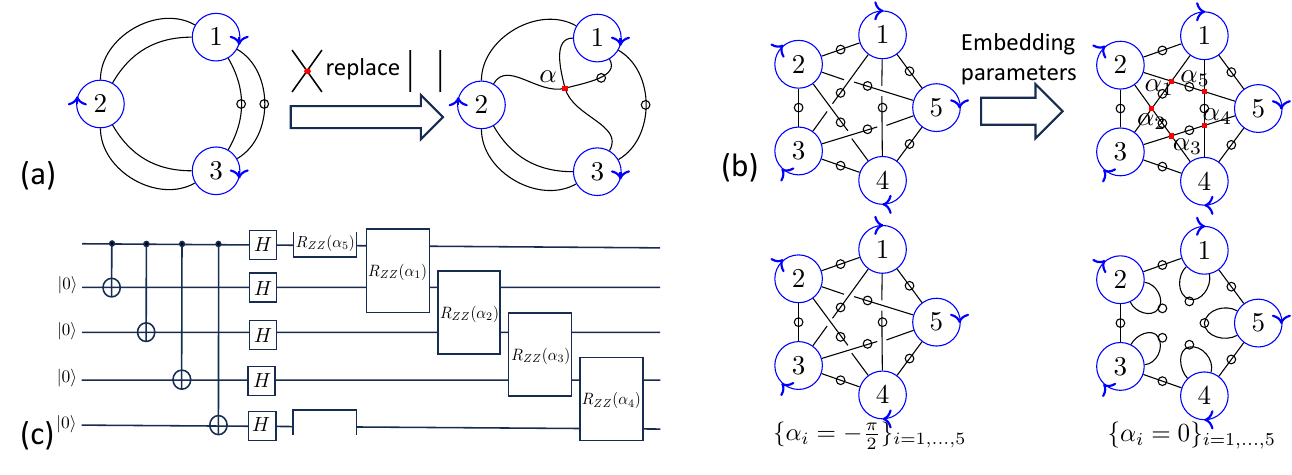}
    \caption{
        Illustration of VGQEC codes.
        \textbf{(a)} The Quon graph of the three-qubit repetition code is derived from a circular chain of three vertices. The symbol $\circ$ indicates the positions where charge pairs are placed. In this representation, the Quon graph corresponds to the logical state $\ket{+}_L=(\ket{000}+\ket{111})/\sqrt{2}$ when no charge is added, and to $\ket{-}_L=(\ket{000}-\ket{111})/\sqrt{2}$ when all charges are added.
        A heuristic VGQEC code is constructed by replacing a parallel edge with a variable crossing, and the parameter $\alpha$ is embedded in the crossing.
        \textbf{(b)} The Quon graphs of the logical states of $[[5,1,3]]$ code are constructed from the complete graph $K_5$, where the logical state $\ket{0}_L$ corresponds to the absence of charges, and $\ket{1}_L$ corresponds to the presence of all charges.
 Embedding five parameters $\{\alpha_i\}_{i=1,\dots,5}$ into the crossings generates a five-qubit VGQEC code.
 When all parameters $\alpha_i$ are set to $-\frac{\pi}{2}$, the discs are pairwise connected, resulting in the $[[5,1,3]]$ code.
Conversely, if $\{\alpha_i=0\}_{i=1,\dots, 5}$, the five crossings are removed, transforming the code into a five-qubit repetition code in the $X$-basis.
\textbf{(c)} The encoding circuit for the VGQEC code in Fig.~\ref{fig:513}(b). The $R_{ZZ}(\theta)$ gates represent the two-qubit Pauli rotation $e^{-i\frac{\theta}{2}Z\otimes Z}$, and a half-open $R_{ZZ}(\alpha_5)$ gate is applied to the first and fifth qubits.
}\label{fig:513}
\end{figure*}
    
\begin{figure}[tbp]
    \centering
    \includegraphics[scale=0.45]{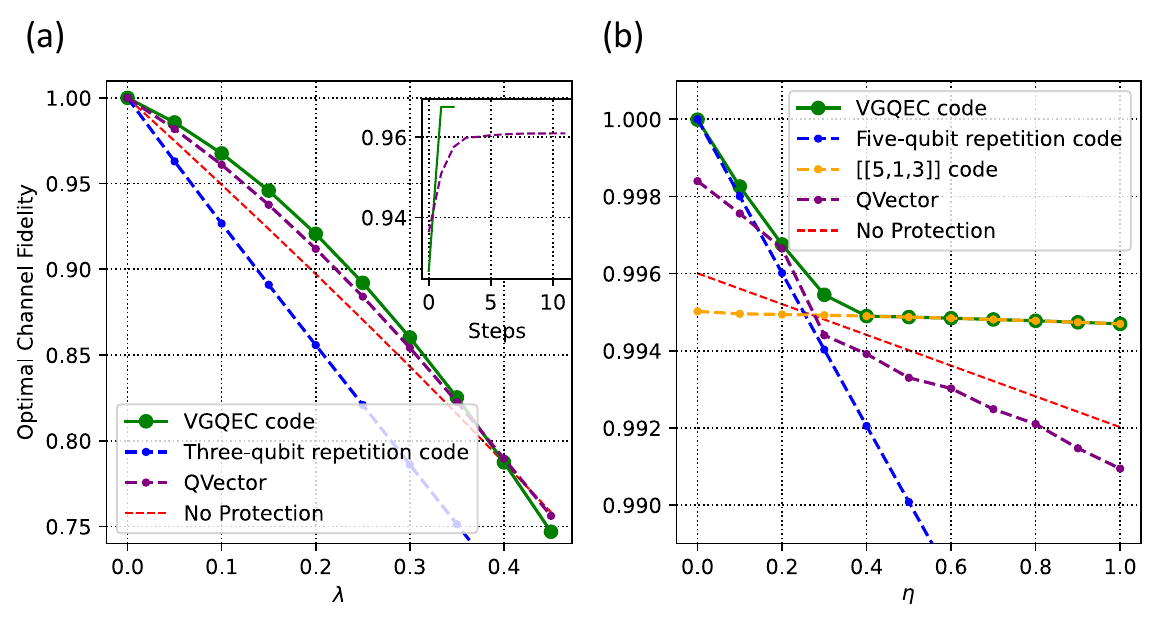}
    \caption{
    Performance of the VGQEC codes outlined in Fig.~\ref{fig:513}.
        \textbf{(a)} The three-qubit VGQEC code (Fig.~\ref{fig:513}(a)) is optimized and evaluated under the amplitude damping noise.
    The horizontal axis denotes the amplitude damping strength $\lambda$, while the vertical axis shows the best achievable channel fidelity after recovery~(see Methods).
    The “QVector” data is derived from the numerical optimization method described in Ref.~\cite{johnson2017qvector}.
    The subfigure demonstrates the convergence behavior of the VGQEC code and the QVector method for $\lambda = 0.1$.
    \textbf{(b)} The five-qubit VGQEC code (Fig.~\ref{fig:513}(b)) is optimized for a specific time-varying noise described in Supplementary Section D.
    The horizontal axis, labeled $\eta$, indicates the progression of noise evolution.
  }\label{fig:sdp_opt}
  \end{figure}

Quon is a 3D topological language designed for quantum information~\cite{liu2017quon}.
In the context of quantum error correction, the Quon language provides a general tool for studying codes and introduces an innovative pictorial method to construct them~\cite{liu2019quantized}. 
The basic rules of Quon are reviewed in Supplementary Section A.
In this paper, we enhance the power of error correction codes by integrating parameters into their Quon graphs.
These parameters enable dynamic reconfigurations of graph structures, allowing the obtained codes to adapt to a variety of noise models by optimizing parameters.
We refer to this new class of codes as Variational Graphical Quantum Error Correction (VGQEC) codes.

To illustrate the basic idea of VGQEC, we consider the 3-qubit repetition code, whose Quon graph is shaped as a chain of three vertices arranged in a circle, as depicted in Fig.~\ref{fig:513}(a). In the graph, the vertexes correspond to physical qubits, and the edges between them represent the correlations among the qubits, which can be interpreted as the code's stabilizers. Further details are elaborated in Supplementary Section B.

To enhance the error correction capability of the code, we replace the edges with variable crossings, as depicted in Fig.~\ref{fig:513}(a), and introduce a parameter $\alpha$ to the graph\footnote{This is a simple heuristic modification, where parallel edges are replaced with variable crossings to make the graphical transformation appear most dramatic.}. The variable crossing represents an extension of the braiding crossing, which can be graphically expressed as \begin{tikzpicture}
    \begin{scope}[shift={(0,0)}]
        \draw[fill=white,draw=none,opacity=0] (0,0) rectangle (0.01,0.01);
    \end{scope}
    \begin{scope}
        \draw[-] (0.5,-0.25) -- (0,0.25);
        \node at(0.1,0) {$\alpha$};
        %\draw[fill=white,draw=none] (0.4,-0.1) rectangle +(0.2,0.2);
        \draw[-] (0,-0.25) -- (0.5,0.25);
    \end{scope}
    \node at(0.8,0) {$=$};
    \node at(1.5,0) {$\frac{1+e^{i\alpha}}{2}$};
    \begin{scope}[shift={(2.1,0)}]
        \draw[-] (0,-0.25) -- (0,0.25);
        %\draw[fill=white,draw=none] (0.4,-0.1) rectangle +(0.2,0.2);
        \draw[-] (0.5,-0.25) -- (0.5,0.25);
        %\fill (0.25,0.25) circle (1.5pt);
    \end{scope}
    \node at(3.3,0) {$+\frac{1-e^{i\alpha}}{2}$};
    \begin{scope}[shift={(4,0)}]
        \draw[-] (0,-0.25) -- (0,0.25);
        %\draw[fill=white,draw=none] (0.4,-0.1) rectangle +(0.2,0.2);
        \draw[-] (0.5,-0.25) -- (0.5,0.25);
        \fill (0,0) circle (1.5pt);
        \fill (0.5,0) circle (1.5pt);
    \end{scope}
\end{tikzpicture}.
Moreover, the specific cases of positive braiding crossings \begin{tikzpicture}
    \begin{scope}[shift={(0,0)}]
        \draw[fill=white,draw=none,opacity=0] (0,0) rectangle (0.01,0.01);
    \end{scope}
    \begin{scope}[shift={(0.1,0)}]
        \draw[-] (0.5,-0.25) -- (0,0.25);
        %\node at(0,0) {$\alpha$};
        \draw[fill=white,draw=none] (0.15,-0.1) rectangle +(0.2,0.2);
        \draw[-] (0,-0.25) -- (0.5,0.25);
    \end{scope}
\end{tikzpicture} and negative braiding crossings \begin{tikzpicture}
    \begin{scope}[shift={(0,0)}]
        \draw[fill=white,draw=none,opacity=0] (0,0) rectangle (0.01,0.01);
    \end{scope}
\begin{scope}[shift={(0.1,0)}]
    \draw[-] (0,-0.25) -- (0.5,0.25);
    %\node at(0,0) {$\alpha$};
    \draw[fill=white,draw=none] (0.15,-0.1) rectangle +(0.2,0.2);
    \draw[-] (0.5,-0.25) -- (0,0.25);
\end{scope}
    \end{tikzpicture}, correspond to the variable crossings with $\alpha=-\frac{\pi}{2}$ and $\alpha=\frac{\pi}{2}$, respectively.
We call a code graph with adjustable parameters like this is a \emph{VGQEC code}.
It defines a family of codes, where parameter adjustments enable switching between codes in the family, thereby adapting to various noise models.
In particular, when $\alpha=0$, the variable crossing reduces to the original edge, and this VGQEC code returns to the original repetition code.

To demonstrate the power of this VGQEC code, the amplitude damping noise is examined, and the parameter $\alpha$ is optimized to adapt the noise at different noise rates, as detailed in Supplementary Section D. 
Assuming the full knowledge of the noise model, the optimal recovery map, derived from the semi-definite programming (SDP) approach~\cite{fletcher2007optimum}, can be applied to assess the performance of the encoding map, as given by Eq.~\eqref{eq:optimalfidelity}.
As shown in Fig.~\ref{fig:sdp_opt}(a), the three-qubit repetition code fails to protect information under the noise channel, which performs even worse than the scenario without code protection. 
As a comparison, the optimized three-qubit VGQEC code introduced above significantly increases the channel fidelity.
Specifically, we find that the optimized three-qubit VGQEC code has the following codewords for $\lambda< 0.4$:
\begin{equation}\label{eq:threequbitad}
\begin{aligned}
    &\ket{0}_L=\frac{1}{\sqrt{2}}(\ket{000}+i\ket{011}), \\
    &\ket{1}_L=\frac{1}{\sqrt{2}}(i\ket{100}+\ket{111}).
\end{aligned}
\end{equation}
Compared to quantum codes optimized by other approaches, such as “QVector”~\cite{johnson2017qvector}, a key advantage of the VGQEC code is that it is optimized based on the repetition code, rather than being designed from scratch. As a result, with fewer parameters and optimization steps, the VGQEC code delivers superior error correction performance, as demonstrated by the numerical simulations.

Interestingly, VGQEC codes can bridge the gap between different quantum codes.
In Quon, the logical state of the $[[5,1,3]]$ code~\cite{laflamme1996perfect} can be visualized as a fully connected diagram $K_5$, as shown in Fig.~\ref{fig:513}(b).
By substituting the five positive crossings with variable crossings, a VGQEC code can be constructed, with five parameters $\{\alpha_i\}_{i=1,\dots,5}$ embedded in the crossings.
It turns out that such a VGQEC code achieves the interpolation between the five-qubit repetition code and the $[[5,1,3]]$ code.
Specifically, setting the parameters $\alpha_i=-\frac{\pi}{2}$ for $i=1,\dots,5$ reverts the variable crossings to positive crossings, which yields the five-qubit $[[5,1,3]]$ code, as illustrated in Fig.~\ref{fig:513}(c). 
On the other hand, if setting $\alpha_i = 0$ for $i=1,\dots,5$, the crossings are removed, results in the logical states $\ket{0}_L = \ket{+}^{\otimes 5}$ and $\ket{1}_L = \ket{-}^{\otimes 5}$. In this case, the VGQEC code decays to the five-qubit repetition code in the $\ket{+}$, $\ket{-}$ basis. 
Moreover, based on the Quon graph of the VGQEC code, one can derive the corresponding encoding circuit, as shown in Fig.~\ref{fig:513}(d). 
For example, setting $\alpha_i=-\frac{\pi}{2}$ yields an encoding circuit for the $[[5,1,3]]$ code.

Similar to the three-qubit case, we optimize the parameters $\{\alpha_i\}_{i=1,\dots,5}$ to adapt to a specific noise model. More details can be seen in Supplementary Section D.
The noise model comprises a time-dependent Pauli error channel $\mathcal{N}_1^\eta$, which evolves with time $\eta$, and a fixed correlated error channel $\mathcal{N}_2$.
The correlated noise is modeled as errors occurring simultaneously on adjacent qubits, a phenomenon commonly observed in quantum systems due to cross-talk or interactions between neighboring qubits~\cite{zhao2022quantum,wang2022control}.
As $\eta$ increases, the time-dependent Pauli error channel $\mathcal{N}_1^\eta$ transitions from a dephasing to a depolarizing channel, capturing the shift from primarily phase-flip errors to a more general form of noise. We optimize the VGQEC code to adapt to this shifting noise model.

As shown in Fig.~\ref{fig:sdp_opt}, the optimized VGQEC code initially approximates the five-qubit repetition code when $\eta$ is small. As the noise model evolves, the VGQEC code gradually transforms into the $[[5,1,3]]$ code.
Throughout the evolution of the noise model, the VGQEC code consistently outperforms both the repetition code and the $[[5,1,3]]$ code in terms of information protection.
This adaptability to diverse noise models arises partly from its ability to flexibly switch between multiple quantum codes according to the noise characteristics.
As a result, by harnessing the strengths of different codes, the obtained VGQEC code can achieve superior performance over these original codes.

\section{Hybrid Quantum-Classical Scheme for VGQEC} \label{sec:simulation}

\begin{figure*}[tbp]
    \centering
    \includegraphics[width=\linewidth]{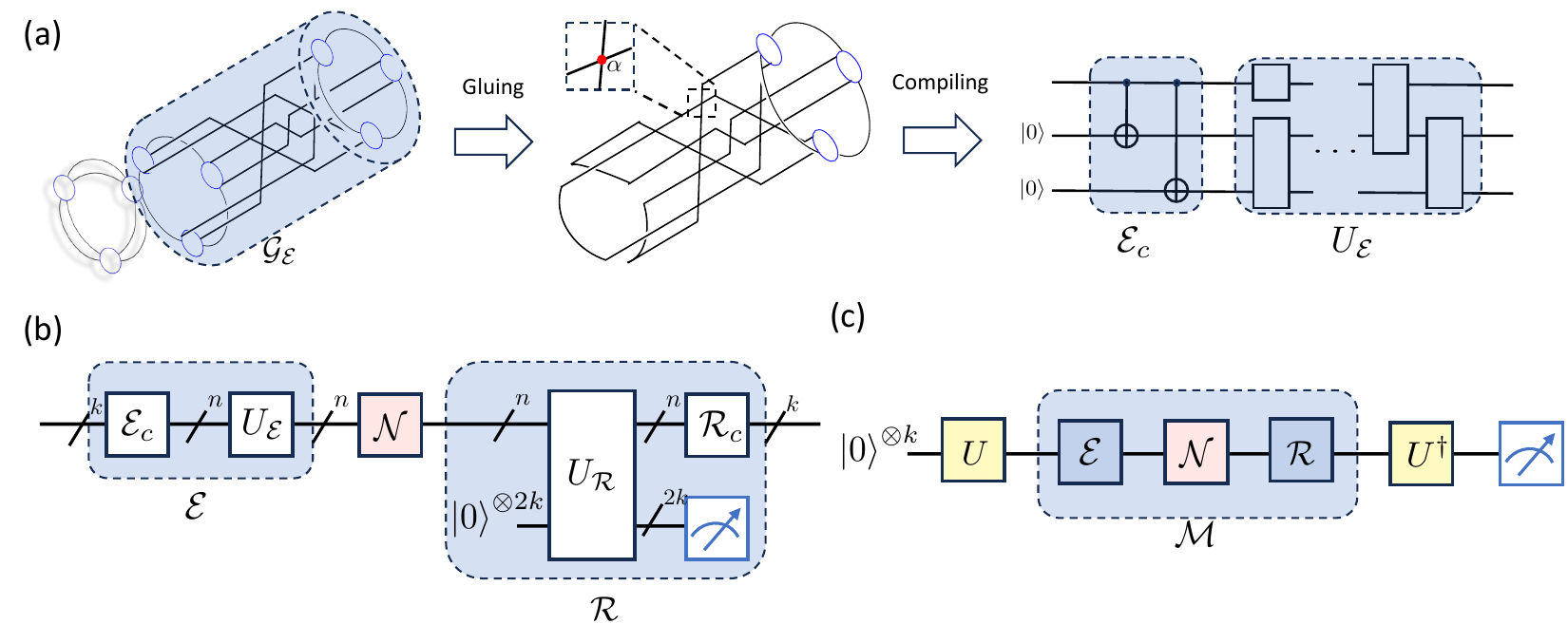}
    \caption{Overview of the construction and the evaluation of VGQEC codes using parameterized quantum circuits, as presented in Sec.~\ref{sec:simulation}. \textbf{(a)} A new VGQEC code is formed by gluing a variable graphical structure $\mathcal{G}_\mathcal{E} $ to the Quon graph of a given code, where all crossings in $\mathcal{G}_\mathcal{E}$ are variable crossings. Its encoding map can be efficiently compiled into the original encoding circuit $\mathcal{E}_c$ alone with a parameterized circuit $U_\mathcal{E}$ associated with $\mathcal{G}_\mathcal{E}$. The explicit form of $U_\mathcal{E}$ is provided in Supplementary Section E. \textbf{(b)} The encoding and the recovery maps are implemented by parameterized quantum circuits, where $\mathcal{N}$ denotes the noise channel. The encoding map $ \mathcal{E} $ is outlined in~(a). In the case of a VGQEC code with $k$ logical qubits and $n$ physical qubits, the recovery map $ \mathcal{R} $ is implemented by introducing $2k$ auxiliary qubits, applying a parameterized circuit $U_\mathcal{R}$, measuring auxiliary qubits, and subsequently applying the original recovery map $\mathcal{R}_c$. \textbf{(c)} The fidelity sampling circuit, where $U$ is sampled from 2-design. The composite channel $\mathcal{M}$ corresponds to the procedure in (b).
    The fidelity is given by the probability of measuring all-zero at the circuit’s outputs. The boxes drawn in blue represent the measurements.}\label{fig:hybrid_method}
\end{figure*}

We have seen that the graphical insight allows us to develop VGQEC codes. However, how to implement their encoding maps with specific quantum circuits poses a significant challenge, especially for those codes with intricate structures.
Additionally, implementing optimal recovery maps~\cite{fletcher2007optimum} or Petz recovery maps~\cite{barnum2002reversing,gilyen2022quantum} for these codes requires thorough characterizations of noise models in quantum devices. 
Besides, the compilations of these recovery maps into quantum circuits also demand substantial computational resources. These difficulties make it highly challenging to deploy VGQEC codes on real-world quantum devices.

To overcome these challenges, we propose a hybrid quantum-classical scheme for implementing VGQEC codes using parameterized quantum circuits, making them compatible with NISQ devices.
Unlike the instances discussed in Sec.~\ref{sec:VMQECC}, which modify the internal structure of the Quon graph associated with an original code, the new approach preserves the internal Quon graph structure, and
the enhancement is achieved by attaching a specially designed variable graphical structure to the physical qubit vertices of the original code, as depicted in Fig.~\ref{fig:hybrid_method}(a).
This variable graphical structure\footnote{This variable graphical structure $\mathcal{G}_\mathcal{E}$ is designed to exhibit ``universal'' properties. For $n$ qubits, it is characterized by $2n$ internal strings that cross each other with variable crossings, which admits all possible connections of $2n$ internal strings. This flexibility facilitates transformations into a wide range of graphical configurations.  The total number of parameters in $\mathcal{G}_\mathcal{E}$ is $n(2n-1)$. For further details, see Supplementary Section E}, denoted as $\mathcal{G}_\mathcal{E}$, introduces significant adaptability to the code, enabling it to address diverse noise models.
For this class of VGQEC codes, the encoding maps $\mathcal{E}$ can be efficiently compiled into a composition of the original code's encoding circuit $\mathcal{E}_c$ and a parameterized quantum circuit $U_\mathcal{E}$ related to $\mathcal{G}_\mathcal{E}$\footnote{The construction of $\mathcal{G}_\mathcal{E}$ enables the straightforward implementation of $U_\mathcal{E}$ using circuits composed of $R_X$, $R_Z$ and $R{ZZ}$ gates. Consequently, it can be easily realized using quantum circuits.}.
Regarding the recovery map $\mathcal{R}$, we employ a combination of parameterized circuits $U_\mathcal{R}$ and the original code's recovery maps $\mathcal{R}_c$ with the help of necessary auxiliary qubits~\footnote{The intuition of this design is to ensure that the recovery map can implement the original code's recovery process and retain the flexibility for further optimization, with auxiliary qubits extending its degrees of freedom, see Supplementary Section H}, as depicted in Fig.~\ref{fig:hybrid_method}(b).
More details on the structures of $\mathcal{G}_\mathcal{E}$ and the circuits $U_\mathcal{E},U_\mathcal{R}$ are provided in Supplementary Section E.

To optimize the parameters within the VGQEC codes along with their associated recovery circuits, we choose the objective function of optimization to be the average entanglement fidelity over the Haar random state (see Methods),
which is evaluated using the procedure introduced in Ref.~\cite{johnson2017qvector}.
The fidelity sampling circuit, denoted as $U^\dagger \circ \mathcal{M} \circ U$, is repeatedly measured on the computational basis. 
Here, $\mathcal{M}= \mathcal{R}\circ\mathcal{N}\circ\mathcal{E}$ represents the noisy channel protected by the VGQEC code, and the unitary operator $U$ is randomly selected from a 2-design distribution, as depicted in Fig.~\ref{fig:hybrid_method}(c).
Through iterative sampling and measurement on the outputs, the probability of obtaining all-0 outcomes provides an unbiased estimate of the desired fidelity.
The process is shown in Fig.~\ref{fig:hybrid_method}(d).
For $N$ repetitions, the estimated fidelity has a standard deviation of order $\mathcal{O}(\frac{1}{\sqrt{N}})$.
Classical optimization algorithms are then employed to maximize the fidelity. 
Further details can be found in Supplementary Section F.

We numerically investigate the performance of the proposed scheme, for which we focus on the three-qubit and the five-qubit VGQEC codes derived from the three-qubit repetition code and the $[[5,1,3]]$ code respectively\footnote{To avoid ambiguity, we clarify that the VGQEC codes in this section differ from the structure depicted in Fig.~\ref{fig:513}. Instead, they are constructed by gluing together $\mathcal{G}_\mathcal{E}$, shown in Fig.~\ref{fig:hybrid_method}(a).}, and the noise models we choose are the amplitude damping noise and the thermal relaxation process.  
In the amplitude damping case, we assume the noise intensity to be uniform across all qubits, whereas in the thermal relaxation case, we assume the noise intensity to vary among the qubits.

\begin{figure*}[!htb]
    \subfigure[]{
      \includegraphics[width=0.45\textwidth]{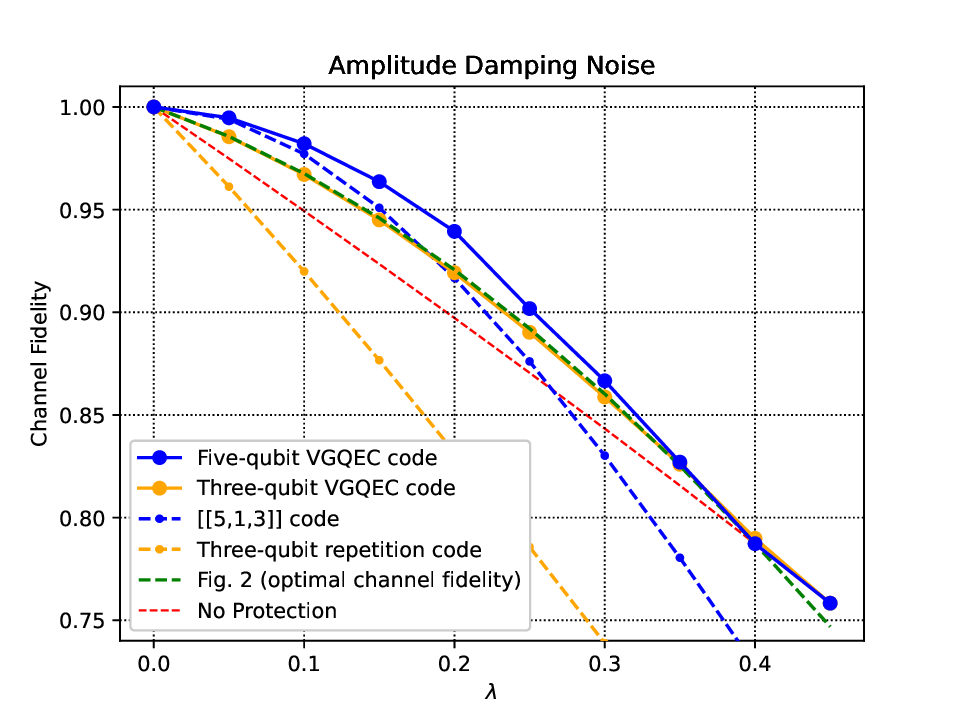}
      \label{fig:ad_three}}
    \subfigure[]{
      \includegraphics[width=0.45\textwidth]{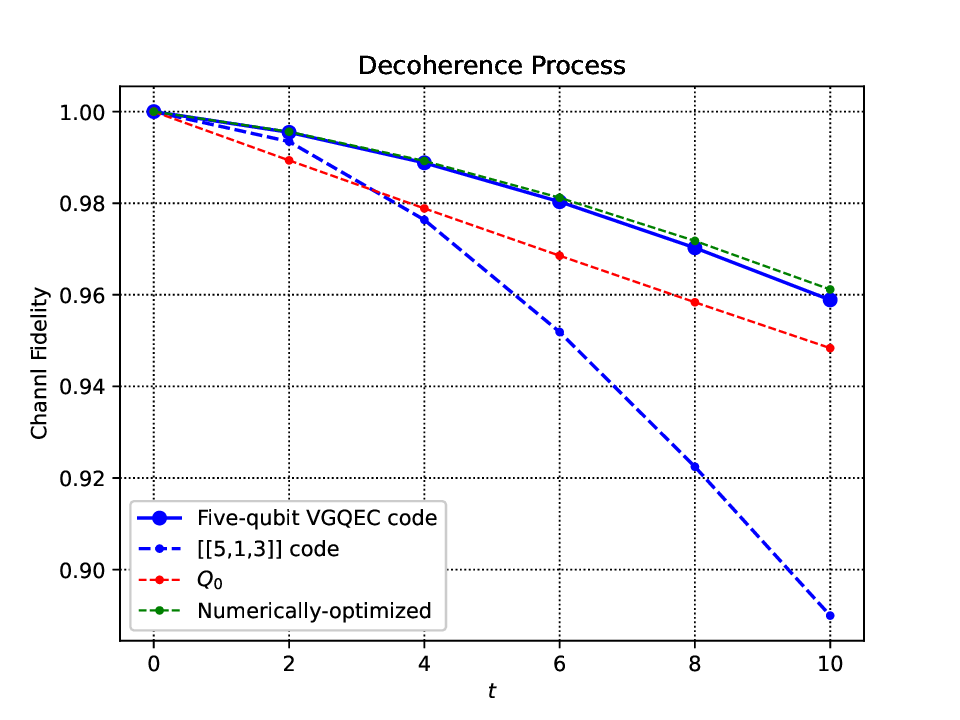}
      \label{fig:pad_five}
    }
    \caption{Performance of the VGQEC codes implemented using parameterized quantum circuits. The three-qubit VGQEC code is derived from the three-qubit repetition code, and the five-qubit VGQEC code is derived from the $[[5,1,3]]$ code by gluing a variable graphical structure $\mathcal{G}_\mathcal{E}$, as illustrated in Fig.~\ref{fig:hybrid_method}.
    \textbf{(a)} Performances under the amplitude damping channel. The vertical axis represents the channel fidelity. 
    The data labeled ``Fig.~2 (optimal channel fidelity)" depicts the optimal channel fidelity of the three-qubit VGQEC code shown in Fig.~\ref{fig:sdp_opt}. \textbf{(b)} Performance of the five-qubit VGQEC code and the $[[5,1,3]]$ code under asymmetric noise settings. The noise is modeled as the thermal relaxation error, which affects each qubit with uneven intensities. These intensities are chosen to match the coherence time of the real \textit{IBM-LIMA} machine, as shown in Methods.
    $Q_0$ represents the channel fidelity of the unprotected channel corresponding to the first qubit in the five-qubit system. The numerically-optimized results are obtained using an iterated convex optimization~\cite{kosut2009quantum}.}
  \end{figure*}

As depicted in Fig.~\ref{fig:ad_three}, the proposed scheme optimizes the three-qubit VGQEC code, achieving the optimal channel fidelity of the code in Eq.~\eqref{eq:threequbitad}. 
Notably, it can be seen that the performance of the three-qubit VGQEC code surpasses that of the $[[5,1,3]]$ code when the damping parameter $\lambda\geq 0.2$, even with fewer physical qubits used.
Furthermore, the performance of the five-qubit VGQEC code also exceeds that of the $[[5,1,3]]$ code, for all values of $\lambda$.
In comparison to the three-qubit scenario, the five-qubit VGQEC code effectively utilizes redundant information in the additional qubits to further enhance channel fidelity. This outcome highlights the scheme's efficacy in tailoring VGQEC codes to adapt quantum noise.

We further examine a thermal relaxation process with duration $t$ occurring in a five-qubit quantum system, which is characterized by the $T_1$ and $T_2$ times matching those of the \textit{IBM-LIMA} device.
The $[[5,1,3]]$ code is specifically designed to correct arbitrary errors on individual qubits. 
However, when the qubits exhibit different error rates, the symmetrically designed $[[5,1,3]]$ struggles to perform effectively.
As shown in Fig.~\ref{fig:pad_five}, when $t\geq 2.5$, the channel protected by the $[[5,1,3]]$ code exhibits even a lower fidelity compared with the case without error correction ($Q_0$ in Fig.~\ref{fig:pad_five}), which indicates the $[[5,1,3]]$ code is ill-suited to asymmetric noise environments.
In contrast, our proposed approach demonstrates a significant performance enhancement compared to the $[[5,1,3]]$ code, which almost achieves the numerically optimized results obtained through the iterated convex optimization method~\cite{kosut2009quantum}.
In this case, our scheme successfully adapts the original code to the case that qubits have uneven qualities, thus enhancing its performance on asymmetric quantum noise.

\section{Photonic Realization of Three-qubit VGQEC Code}\label{sec:experiment}
\begin{figure*}[htbp]
    \centering
    \includegraphics[width=0.95\linewidth]{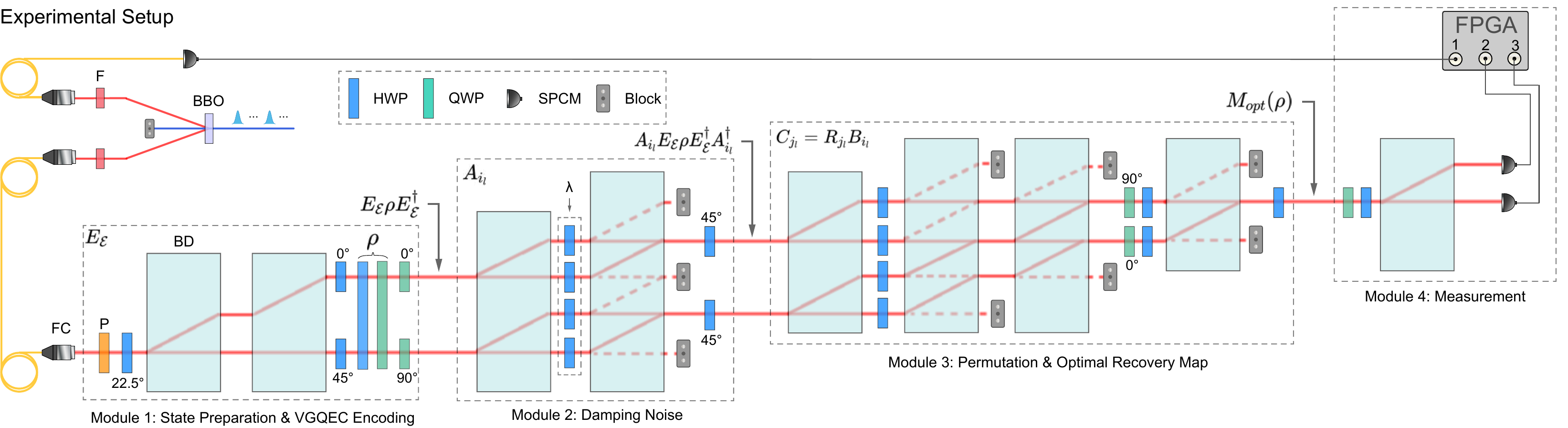}
    \caption{Experimental setup. The single photons, generated in a heralded manner through the parametric down-conversion process in a $\beta-$Barium borate (BBO) crystal, are initially injected into the optical circuit. Initial states are prepared by a half-wave plate (HWP) and a quarter-wave plate (QWP), while Module-1 encodes polarization and path modes of single photons, realizing the three-qubit VGQEC code. Module-2 implements diagonal damping operators with corresponding probability, and simulates part of amplitude damping channel. Module-3 implements permutation of the damping noise and the optimal recovery map. Module-4 performs projective measurement onto the initial logical state through a group of wave-plates and a beam displacer (BD) followed by two single-photon counting modules (SPCMs). A field-programmable gate array (FPGA) registers the clicks of the two detectors. F: 3nm interference filter, FC: fiber connector, P: polarizer.}
    \label{fig_exp_schematic}
\end{figure*}

To experimentally demonstrate the performance of the optimized three-qubit VGQEC code under the amplitude damping channel, we devise optical circuits to realize the error correction encoding map $\mathcal{E}$ in Eq.~\eqref{eq:threequbitad}, simulate the three-qubit amplitude damping channel $\mathcal{N}$, and perform the optimal recovery map $\mathcal{R}_{opt}$ derived from Section~\ref{sec:simulation}, respectively. Using a fidelity sampling circuit, we experimentally measure the channel fidelity of the composite channel $\mathcal{M}_{opt}=\mathcal{R}_{opt}\circ\mathcal{N}\circ\mathcal{E}$.
We now give a brief description for the composite channel $M_{opt}$ and the quantum measurements we conduct. More details can be seen in Supplementary Section G.

The experimental setup is shown in Fig.~\ref{fig_exp_schematic}.
Heralded single photons, generated via a parametric down-conversion process, are injected into a linear optical circuit. 
Harnessing the three-qubit VGQEC code, the logical qubit is encoded in the path and the polarization degrees of freedom (DoFs) of photons in the state preparation and the encoding module (Module-1).  
A half-wave plate (HWP) and a quarter-wave plate (QWP) prepare the initial single-qubit state. 
Generally, three qubits can be encoded with eight modes of single photons.
Since the subspace, which supports the encoding space of $\mathcal{E}$, is four-dimensional, the three-quibt VGQEC code can equivalently be implemented by two path states ($\ket{s_0}$ and $\ket{s_1})$ and two polarization states ($\ket{H}$ and $\ket{V}$), allowing the logical qubits to be represented as $\ket{0}_L = 1/\sqrt{2}(\ket{H}\ket{s_0}+i\ket{H}\ket{s_1})$ and $\ket{1}_L = 1/\sqrt{2}(i\ket{V}\ket{s_0}+\ket{V}\ket{s_1})$.
Here, $\{\ket{H},\ket{V}\}$ corresponds to the basis states $\{\ket{0}_1,\ket{1}_1\}$ of the first qubit, and $\{\ket{s_0},\ket{s_1}\}$ corresponds to two out of four basis states $\{\ket{0}_2\ket{0}_{3},\ket{1}_2\ket{1}_{3}\}$ of the second and third qubits, respectively.

To implement the amplitude damping channel $\mathcal{N}$ and the optimal recovery map $\mathcal{R}_{opt}$, we decompose these into Kraus operators and apply each operator with its corresponding probability~\cite{PhysRevA.84.032304}  (see Methods).
The Kraus operators $N_i$ ($i=1,\dots,8$) of $\mathcal{N}$ are decomposed into diagonal damping operators $A_i$ and unitary permutations $B_i$.
The optimal recovery map $\mathcal{R}_{opt}$, derived using SDP and characterized by a rank-5 Choi matrix, is implemented using five Kraus operators $R_j$ (where $j=1,\dots,5$).
The damping operations $A_i$ are implemented in the damping noise module (Module-2, Fig.~\ref{fig_exp_schematic}), while the permutation operators $B_i$ along with  ${R}_{j}$, denoted by $C_{k}=B_iR_j$, are implemented assembly in the permutation and recovery module (Module-3). 
The effect of the composite channel $\mathcal{M}_{opt}(\rho)= \mathcal{R}_{opt}\circ\mathcal{N}\circ\mathcal{E}(\rho)$ is then given by $\mathcal{M}_{opt}(\rho) = \sum_{i,j}R_j N_i E_{\mathcal{E}}\rho E_{\mathcal{E}} N_i^{\dagger} R_j^{\dagger}$, involving a total of 40 Kraus operators. 
By eliminating zero terms in $R_j N_i E_{\mathcal{E}}$ and rearranging the non-zero terms, we obtain 14 non-zero Kraus operators $K_l$ such that $\mathcal{M}_{opt}(\rho) = \sum_{l=1}^{14}K_l\rho K_l^{\dagger}$, with $K_l = R_{j_l}N_{i_l}E_{\mathcal{E}} = C_{j_l}A_{i_l}E_{\mathcal{E}}$.

Measurements (module-4) are performed by a HWP and a QWP followed by a beam displacer. The wave-plates are configured to implement measurements in Mutually Unbiased Bases (MUBs). Two single photon counting modules register the measurement outcomes, and a field-programmable gate array (FPGA) system records the outcome probabilities. 
The fidelity between the optimal composite channel $\mathcal{M}_{opt}$ and the identity channel is estimated by the effect of $M_{opt}$ on MUBs. 
Initial states are prepared from the set $S=\{\ket{H},\ket{V},\ket{D},\ket{A},\ket{L},\ket{R}\}$. After each state passes through $\mathcal{M}_{opt}$, the output state is projected onto the corresponding initial state. The average entanglement fidelity is then calculated as the average success probability of these projections and subsequently converted to the channel fidelity (see Methods) .

\begin{figure}
    \centering
    \includegraphics[width=\linewidth]{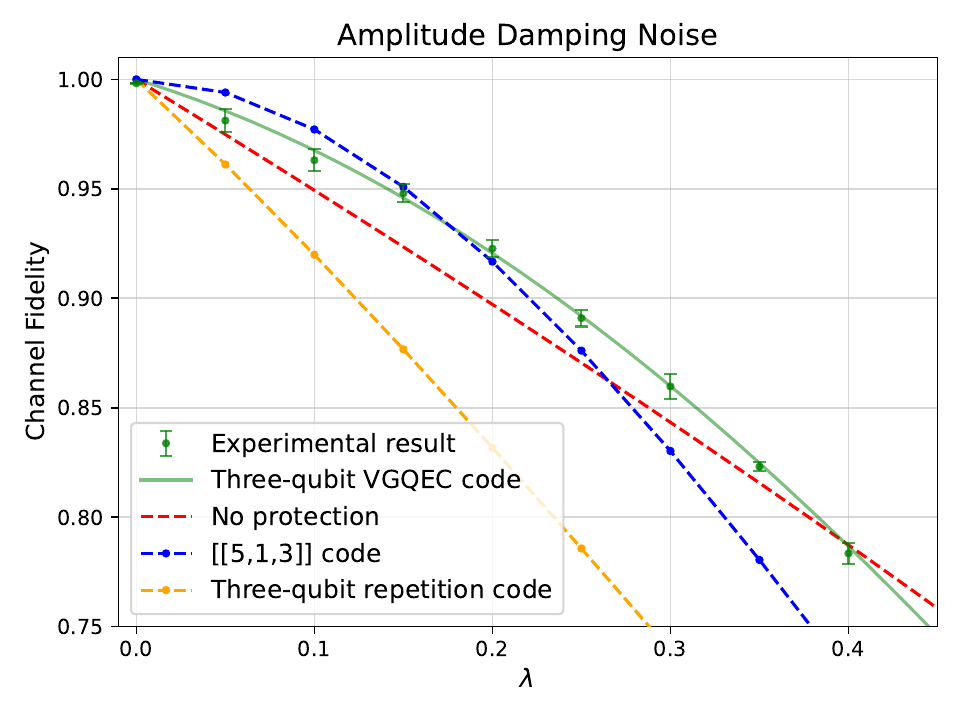}
    \caption{Experimentally measured channel fidelity as a function of the amplitude damping strength $\lambda$. The performance of the three-qubit VGQEC encoding (the data points) closely matches the theoretical predictions (the green line). Error bars represent uncertainties arising from laser power and environmental fluctuations. The performance of an unprotected qubit is included for comparison (the red dashed line).}
    \label{fig_exp_results}
\end{figure}

Fig.~\ref{fig_exp_results} presents the experimental results. To suppress statistical fluctuations, the success probability of projective measurement for each state is estimated from approximately 18,000 photon detection events. Error bars represent the standard deviation obtained from seven experimental repetitions.
In the low-to-medium damping noise regime, the three-qubit VGQEC code effectively enhances the channel fidelity compared to the unprotected qubit.
Compared with the [[5,1,3]] code, the code gradually show superiority as the damping parameter $\lambda$ exceeds $0.2$.
Since the three-qubit repetition code is terrible even with its optimal Recovery map, our results further highlight the error correction advantage of VGQEC codes.
All the experimental data points agree with the theoretical predictions within their respective error margins.

To verify the faithful implementation of the Kraus operators constituting the composite channel $\mathcal{M}_{opt}$, we evaluate the classical fidelity of the measured probability distributions from realizing Kraus operators individually. For a given initial state in the set $S$, the classical fidelity is defined as
\begin{equation}
f_c=\left(\sum_{l=1}^{14}\left[\sqrt{p_s(K_l)q_s(K_l)}+ \sqrt{p_f(K_l)q_f(K_l)}\right]\right)^2,
\end{equation}
where $q_s(K_l)$ and $q_f(K_l)$ represent theoretical probabilities of success and failure for projective measurement onto the initial logical state associated with Kraus operator $K_l$. $p_s$ and $p_f$ are the corresponding experimentally measured probabilities. The average classical fidelity $\overline{f}_c$ is calculated by averaging $f_c$ over different initial states in $S$ and subsequently across 9 different damping noise strengths.
The high value of $\overline{f}_c= 0.9969(16)$ confirms the reliability of our experimental implementation of $\mathcal{M}_{opt}$. Using a photonic system, we have thus experimentally demonstrated the effectiveness of the three-qubit VGQEC code in mitigating the amplitude damping noise across the low-to-medium noise regime.

\section{Discussion} \label{sec:discussion}

Tailoring quantum codes to adapt to specific noise models is a crucial yet challenging task. In this work, we introduce a new approach for adaptively constructing noise-tailored quantum codes, which we refer to as VGQEC codes. 
Notably, the VGQEC codes we construct demonstrate exceptional adaptability across various noise models, along with significantly improved error correction performance.

Yet to fully realize its potential requires further efforts.
First, the surgeries on graphs, particularly the placement of embedded parameters, offer considerable flexibility. Despite the existence of a few basic heuristics, there is a lack of theoretical guidance regarding the optimal embedding schemes.
Second, acquiring the recovery maps for VGQEC codes poses a challenge.
Although options such as optimal recovery maps derived from SDP, Petz recovery maps, and parameterized quantum circuits, are available, these approaches often necessitate noise modeling or entail substantial training costs.
Therefore, further investigations into the recovery maps of VGQEC codes are warranted.
Third, we approach codes from the perspective of logical states, and a natural extension is to directly modify their encoding maps. This modification could offer advantages in terms of circuit complexity. Achieving this may require a new mathematical tool, presented in Ref.~\cite{liu2024functional}, which will be discussed in future work.

Overall, we have shown that adaptive quantum codes can greatly improve error correction performance, enabling quantum systems to better withstand noise.
The proposed VGQEC codes, along with the hybrid quantum-classical scheme, provide a promising approach for designing device-tailored quantum codes, which we hope leads to further studies and implementations of more robust codes in the future.

\paragraph{Acknowledgements}
We thank Song Cheng, Fan Lu, and Ningfeng Wang for useful discussions.
Zhaohui Wei and Zhengwei Liu were supported by Beijing Natural Science Foundation under Grant Z220002.
Yuguo Shao, Fuchuan Wei, and Zhengwei Liu were supported by BMSTC and ACZSP (Grant No.~Z221100002722017).
Zhaohui Wei was supported in part by the National Natural Science
Foundation of China under Grant 62272259 and Grant 62332009.
Yong-chang Li, Hao Zhan, Ben Wang and Lijian Zhang were supported by the National Key Research and Development Program of China (Grants No. 2023YFC2205802), National Natural Science Foundation of China (Grants No. 12347104, No. U24A2017, No. 12461160276) and Natural Science Foundation of Jiangsu Province (Grants No. BK20243060, BK20233001).

%%%%%%%%%%%%%%%%%%%%%%%%%%%%%%%%%%%%%%%%%%%%%%%%

\section*{Methods}
%%%%%%%%%%%%%%%%%%%%%%%%%%%%%%%%%%%%%%%%%%%%%%%%
\section{Preliminaries}\label{sec:preliminaries}

The main idea of QEC is to encode logical qubits into noisy physical qubits. 
Within the framework of QEC, the correction process can be divided into two steps: encoding map and recovery map.
The encoding map, denoted as $\mathcal{E}$, transforms the logical qubits into physical qubits by adding redundant information to safeguard the logical information.
After the physical qubits experience a noise channel $\mathcal{N}$, the recovery map $\mathcal{R}$ is employed to extract the logical information from the noisy physical qubits.
To better protect quantum information, the encoding and the recovery maps should be chosen in such a way that the composite channel approximates the identity channel, i.e., $\mathcal{R}\circ\mathcal{N}\circ\mathcal{E}\approx \mathcal{I}$.

The disparity between the identity channel $\mathcal{I}$ and an arbitrary channel $\mathcal{M}$ can be quantified using the concept of \textit{Entanglement Fidelity}~\cite{schumacher1996sending}.
For a given state $\rho$, the entanglement fidelity of a channel $\mathcal{M}$ is defined as
\begin{equation}\label{eq:entanglementfidelity}
    F_e(\rho,\mathcal{M})\coloneqq f(\ketbra{\psi}{\psi},(\mathcal{M}\otimes \mathcal{I})(\ketbra{\psi}{\psi})),
\end{equation}
where $\ket{\psi}$ is a purification of $\rho$, and $f(\rho,\sigma)\coloneqq (\tr{\sqrt{\sqrt{\rho}\sigma\sqrt{\rho}}})^2$ denotes the state fidelity between two states $\rho$ and $\sigma$.
In particular, when the reference state $\rho$ is chosen as the maximally mixed state $\rho=I/d$, where $d$ represents the dimension of the Hilbert space, Eq.~\eqref{eq:entanglementfidelity} yields the value of \textit{Channel fidelity}~\cite{reimpell2005iterative}:
\begin{equation}\label{eq:channelfidelity}
    F_C(\mathcal{M})\coloneqq F_e(I/d,\mathcal{M}).
\end{equation}
By selecting $\mathcal{M}= \mathcal{R}\circ\mathcal{N}\circ\mathcal{E}$, these fidelity metrics evaluate the effectiveness of the code in safeguarding quantum information against noise.
The optimal channel fidelity of the code, denoted $ F_C^{\text{opt}} $, is defined as the maximum channel fidelity across all recovery maps:
\begin{equation}\label{eq:optimalfidelity}
    F^{opt}_C\coloneqq\underset{\mathcal{R}}{\max}~F_C(\mathcal{R}\circ\mathcal{N}\circ\mathcal{E})=F_C(\mathcal{R}_{opt}\circ\mathcal{N}\circ\mathcal{E}),
\end{equation}
where $\mathcal{R}_{opt}$ is the optimal recovery map that achieves the maximum channel fidelity $F^{opt}_C$.
Since the channel fidelity depends linearly on the Choi matrix of the recovery map, the optimal recovery map $ \mathcal{R}_{\text{opt}} $ can be efficiently determined by solving a semi-definite programming (SDP) problem, as described in~\cite{fletcher2007optimum}, which is briefly reviewed in Supplementary Section C.

Directly measuring channel fidelity in experiments is challenging. 
To experimentally assess a code's performance, one can utilize the average entanglement fidelity over the Haar random states, defined as
\begin{equation}\label{eq:AEF}
\begin{aligned}
    \overline{F}_e(\mathcal{M})\coloneqq &{\mathbb{E}}_{\ket{\psi}\sim\mu_H} F_e(\ket{\psi}\bra{\psi},\mathcal{M})\\
    =&{\mathbb{E}}_{\ket{\psi}\sim\mu_H}\bra{\psi}\mathcal{M}(\ket{\psi}\bra{\psi})\ket{\psi},
\end{aligned}
\end{equation}
where $\mu_H$ represents the Haar measure~\cite{mele2024introduction} on the Hilbert space. 
Since the average entanglement fidelity depends only on second-order moments, the Haar distribution can be substituted with 2-design distribution~\cite{dankert2009exact,Harper_2017}.
Examples of 2-designs include mutually unbiased bases (MUBs)~\cite{PhysRevA.80.012304,10.1117/12.615759} and symmetric informationally complete positive operator-valued measures (SIC-POVMs)~\cite{renes2004symmetric,scott2006tight}, which form complex projective 2-design and enable efficiently estimation of the average entanglement fidelity.
Additionally, there exists a relationship between the average entanglement fidelity $\overline{F}_e$ and the channel fidelity $F_C$, which is given by~\cite{nielsen2002simple}
\begin{equation}
    \overline{F}_e(\mathcal{M})=\frac{d F_C(\mathcal{M})+1}{d+1}.
\label{eq:average_fidelity}
\end{equation}

\section{Quon language}

The Quon language \cite{liu2019quantized,liu2017quon,jaffe2018mathematical} provides a pictorial formalism for studying quantum information. It uses braided charged strings in three-dimensional space to represent quantum states and quantum operations. These diagrams serve as a visual tool for understanding quantum information concepts, such as quantum error correction codes.

In this section, we introduce the fundamental computational rules of the Quon language.
In Quon, the 1-qubit computational basis states are represented by diagrams in the hemisphere. For the state $\ket{0}$, the corresponding diagram is as follows:
\begin{equation}\label{eq:quon_state_0}
\begin{tikzpicture}
\node at(-1.25,0.125) {$\sqrt{2} \ket{0}=$};
\begin{scope}
    \clip (-0.25,0) rectangle (0.25,0.25);
    \draw (0,0) circle(0.25);
\end{scope}
\begin{scope}
    \clip (-0.25+1,0) rectangle (0.25+1,0.25);
    \draw (0+1,0) circle(0.25);
\end{scope}
\node at(1.35,0) {. };
\end{tikzpicture}
\end{equation}
The state $\ket{1}$ is represented by adding a pair of charges to the two strings in the diagram for $\ket{0}$, as shown below:
\begin{equation}\label{eq:quon_state_1}
    \begin{tikzpicture}
        \node at(-1.25,0.125) {$\sqrt{2} \ket{1}=$};
        \begin{scope}
            \clip (-0.25,0) rectangle (0.25,0.25);
            \draw (0,0) circle(0.25);
            \fill (-0.18,0.19) circle (1.5pt);
        \end{scope}
        \begin{scope}
            \clip (-0.25+1,0) rectangle (0.25+1,0.25);
            \draw (0+1,0) circle(0.25);
            \fill (0.18+1,0.19) circle (1.5pt);
        \end{scope}
        \node at(1.35,0) {, };
    \end{tikzpicture}
\end{equation}
The braid crossings in Quon are defined by the following equations:
\begin{equation}\label{eq:quon_crossing}
    \begin{array}{cc}
    \begin{tikzpicture}
    \begin{scope}[shift={(0,0)}]
        \draw[fill=white,draw=none,opacity=0] (0,0) rectangle (0.01,0.01);
    \end{scope}
    \begin{scope}[shift={(0.1,0)}]
        \draw[-] (0.5,-0.5) -- (0,0.5);
        %\node at(0,0) {$\alpha$};
        \draw[fill=white,draw=none] (0.15,-0.1) rectangle +(0.2,0.2);
        \draw[-] (0,-0.5) -- (0.5,0.5);
    \end{scope}
    \node at(1.25,0) {$\coloneqq \omega^{-\frac{1}{2}}$};
    \node at(2.1,0) {$(\frac{1}{\sqrt{2}}$};
    \begin{scope}[shift={(2.5,0)}]
        \draw[-] (0,-0.5) -- (0,0.5);
        %\draw[fill=white,draw=none] (0.4,-0.1) rectangle +(0.2,0.2);
        \draw[-] (0.5,-0.5) -- (0.5,0.5);
        %\fill (0.25,0.25) circle (1.5pt);
    \end{scope}
    \node at(3.5,0) {$+\frac{i}{\sqrt{2}}$};
    \begin{scope}[shift={(4.4,0)}]
        \draw[-] (0,-0.5) -- (0,0.5);
        %\draw[fill=white,draw=none] (0.4,-0.1) rectangle +(0.2,0.2);
        \draw[-] (0.5,-0.5) -- (0.5,0.5);
        \fill (0,0) circle (1.5pt);
        \fill (0.5,0) circle (1.5pt);
    \end{scope}
    \node at(5.1,0) {$),$};
    %\node at(2.6,-0.25) {$,$};
    \begin{scope}[shift={(6.1,0)}]
        \draw[fill=white,draw=none,opacity=0] (0,0) rectangle (0.01,0.01);
    \end{scope}
    \end{tikzpicture}
    \\
    \begin{tikzpicture}
        \begin{scope}[shift={(0,0)}]
            \draw[fill=white,draw=none,opacity=0] (0,0) rectangle (0.01,0.01);
        \end{scope}
    \begin{scope}[shift={(0.1,0)}]
        \draw[-] (0,-0.5) -- (0.5,0.5);
        %\node at(0,0) {$\alpha$};
        \draw[fill=white,draw=none] (0.15,-0.1) rectangle +(0.2,0.2);
        \draw[-] (0.5,-0.5) -- (0,0.5);
    \end{scope}
    \node at(1.25,0) {$\coloneqq \omega^{-\frac{1}{2}}$};
    \node at(2.1,0) {$(\frac{1}{\sqrt{2}}$};
    \begin{scope}[shift={(2.5,0)}]
        \draw[-] (0,-0.5) -- (0,0.5);
        %\draw[fill=white,draw=none] (0.4,-0.1) rectangle +(0.2,0.2);
        \draw[-] (0.5,-0.5) -- (0.5,0.5);
        %\fill (0.25,0.25) circle (1.5pt);
    \end{scope}
    \node at(3.5,0) {$-\frac{i}{\sqrt{2}}$};
    \begin{scope}[shift={(4.4,0)}]
        \draw[-] (0,-0.5) -- (0,0.5);
        %\draw[fill=white,draw=none] (0.4,-0.1) rectangle +(0.2,0.2);
        \draw[-] (0.5,-0.5) -- (0.5,0.5);
        \fill (0,0) circle (1.5pt);
        \fill (0.5,0) circle (1.5pt);
    \end{scope}
    \node at(5.1,0) {$).$};
    \begin{scope}[shift={(6.1,0)}]
        \draw[fill=white,draw=none,opacity=0] (0,0) rectangle (0.01,0.01);
    \end{scope}
    %\node at(2.6,-0.25) {$,$};
    \end{tikzpicture}
\end{array}
    %\caption{Graphical interpretation of braids crossings: (a) Variable braids crossing with parameter $\alpha$. }
    %\label{fig:crossing}
\end{equation}
A variable braid crossing is defined as follows:
\begin{equation}\label{eq:quon_crossing_var}
    \begin{tikzpicture}
        \begin{scope}[shift={(0,0)}]
            \draw[fill=white,draw=none,opacity=0] (0,0) rectangle (0.01,0.01);
        \end{scope}
        \begin{scope}
            \draw[-] (0.5,-0.5) -- (0,0.5);
            \node at(0.1,0) {$\alpha$};
            %\draw[fill=white,draw=none] (0.4,-0.1) rectangle +(0.2,0.2);
            \draw[-] (0,-0.5) -- (0.5,0.5);
        \end{scope}
        \node at(0.8,0) {$\coloneqq$};
        \node at(1.6,0) {$\frac{1+e^{i\alpha}}{2}$};
        \begin{scope}[shift={(2.4,0)}]
            \draw[-] (0,-0.5) -- (0,0.5);
            %\draw[fill=white,draw=none] (0.4,-0.1) rectangle +(0.2,0.2);
            \draw[-] (0.5,-0.5) -- (0.5,0.5);
            %\fill (0.25,0.25) circle (1.5pt);
        \end{scope}
        \node at(3.6,0) {$+\frac{1-e^{i\alpha}}{2}$};
        \begin{scope}[shift={(4.3,0)}]
            \draw[-] (0,-0.5) -- (0,0.5);
            %\draw[fill=white,draw=none] (0.4,-0.1) rectangle +(0.2,0.2);
            \draw[-] (0.5,-0.5) -- (0.5,0.5);
            \fill (0,0) circle (1.5pt);
            \fill (0.5,0) circle (1.5pt);
        \end{scope}
        \node at(5,-0.25) {$.$};
        \begin{scope}[shift={(6,0)}]
            \draw[fill=white,draw=none,opacity=0] (0,0) rectangle (0.01,0.01);
        \end{scope}
    \end{tikzpicture}
\end{equation}
When the crossing parameter $\alpha=0$, the crossed strings will be transformed into parallel strings. The positive and the negative braids crossings in Eq.~\eqref{eq:quon_crossing} differ from variable braids crossings with parameters $-\frac{\pi}{2}$ and $\frac{\pi}{2}$ by a global phase $\omega^{-\frac{1}{2}}=(\frac{1+i}{\sqrt{2}})^{-\frac{1}{2}}$, respectively.

The charges, represented by dots ``$\bullet$", behave similarly to Majorana fermions, and the braid satisfies Reidemeister moves of type I, II, III~\cite{reidemeister1927elementare}. The essential properties of the quon language can be summarized pictorially as follows:
\begin{equation}\label{eq:quon_relation}
\begin{array}{cc}
\begin{tikzpicture}
\begin{scope}
    \draw (0,0) circle(0.25);
    \node at(1,0) {$=\sqrt{2}$, };
\end{scope}
\end{tikzpicture}
\begin{tikzpicture}
\begin{scope}
    \draw (0,0) circle(0.25);
    \fill (-0.25,0) circle (1.5pt);
    \node at(1,0) {$=0$, };
\end{scope}
\end{tikzpicture}
\begin{tikzpicture}
\begin{scope}
    \draw[-] (0,-0.5) -- (0,0.5);
    \fill (0,0.25) circle (1.5pt);
    \fill (0,-0.25) circle (1.5pt);
    \node at(0.5,0) {$=$};
    \draw[-] (1,-0.5) -- (1,0.5);
    \node at(1.25,-0.25) {$,$};
\end{scope}
\end{tikzpicture}\\
\begin{tikzpicture}
\begin{scope}
    \fill (-0.125,0.2) circle (1.5pt);
    \clip (-0.25,0) rectangle (0.25,0.25);
    \draw (0,0) circle(0.25);
\end{scope}
\node at(0.75,0.125) {$= i$};
\begin{scope}[shift={(1.5,0)}]
    \fill (0.125,0.2) circle (1.5pt);
    \node at(0.5,0) {$,$};
    \clip (-0.25,0) rectangle (0.25,0.25);
    \draw (0,0) circle(0.25);
\end{scope}
\end{tikzpicture}
\begin{tikzpicture}
\begin{scope}
    \draw[-] (0,-0.5) -- (0,0.5);
    \draw[-] (0.25,-0.5) -- (0.25,0.5);
    \fill (0,0.25) circle (1.5pt);
    \fill (0.25,-0.25) circle (1.5pt);
    \node at(0.75,0) {$=-$};
    \draw[-] (1.25,-0.5) -- (1.25,0.5);
    \draw[-] (1.5,-0.5) -- (1.5,0.5);
    \fill (1.25,-0.25) circle (1.5pt);
    \fill (1.5,0.25) circle (1.5pt);
    \node at(2,0) {$=-i$};
    \draw[-] (2.5,-0.5) -- (2.5,0.5);
    \draw[-] (2.75,-0.5) -- (2.75,0.5);
    \fill (2.5,0) circle (1.5pt);
    \fill (2.75,0) circle (1.5pt);
    \node at(3,-0.25) {$,$};
\end{scope}
\end{tikzpicture}\\
\begin{tikzpicture}
\begin{scope}
    \draw[-] (1,-0.5) -- (0,0.5);
    \draw[fill=white,draw=none] (0.4,-0.1) rectangle +(0.2,0.2);
    \draw[-] (0,-0.5) -- (1,0.5);
    \fill (0.75,-0.25) circle (1.5pt);
\end{scope}
\node at(1.25,0) {$=i$};
\begin{scope}[shift={(1.5,0)}]
    \draw[-] (1,-0.5) -- (0,0.5);
    \draw[fill=white,draw=none] (0.4,-0.1) rectangle +(0.2,0.2);
    \draw[-] (0,-0.5) -- (1,0.5);
    \fill (0.25,0.25) circle (1.5pt);
\end{scope}
\node at(2.6,-0.25) {$,$};
\end{tikzpicture}
\begin{tikzpicture}
    \begin{scope}
        \draw[-] (0,-0.5) -- (1,0.5);
        \draw[fill=white,draw=none] (0.4,-0.1) rectangle +(0.2,0.2);
        \draw[-] (1,-0.5) -- (0,0.5);
    \end{scope}
    \node at(1.25,0) {$=$};
    \begin{scope}[shift={(1.5,0)}]
        \draw[-] (0.75,-0.25) -- (0.25,0.25);
        \draw[fill=white,draw=none] (0.4,-0.1) rectangle +(0.2,0.2);
        \draw[-] (0.25,-0.25) -- (0.75,0.25);
        \draw[-] (0.75,0.25) -- (1,-0.5);
        \draw[-] (0.25,-0.25) -- (0,0.5);
        \draw[-] (1,0.5) -- (0.25,0.25);
        \draw[-] (0,-0.5) -- (0.75,-0.25);
    \end{scope}
    \node at(2.6,-0.25) {$,$};
\end{tikzpicture}
\\
\begin{tikzpicture}
\begin{scope}
    \draw[-] (0,-0.5) -- (1,0.5);
    \draw[fill=white,draw=none] (0.4,-0.1) rectangle +(0.2,0.2);
    \draw[-] (1,-0.5) -- (0,0.5);
\end{scope}
\node at(1.25,0) {$=$};
\begin{scope}[shift={(1.5,0)}]
    \draw[-] (1,-0.5) -- (0,0.5);
    \draw[fill=white,draw=none] (0.4,-0.1) rectangle +(0.2,0.2);
    \draw[-] (0,-0.5) -- (1,0.5);
    \fill (0.75,-0.25) circle (1.5pt);
    \fill (0.25,-0.25) circle (1.5pt);
\end{scope}
\node at(2.6,-0.25) {$,$};
\end{tikzpicture}
\begin{tikzpicture}
    \begin{scope}
        \draw[-] (1,-0.5) -- (0,0.5);
        \draw[fill=white,draw=none] (0.4,-0.1) rectangle +(0.2,0.2);
        \draw[-] (0,-0.5) -- (1,0.5);
        \fill (0.25,-0.25) circle (1.5pt);
    \end{scope}
    \node at(1.25,0) {$=-$};
    \begin{scope}[shift={(1.5,0)}]
        \draw[-] (1,-0.5) -- (0,0.5);
        \draw[fill=white,draw=none] (0.4,-0.1) rectangle +(0.2,0.2);
        \draw[-] (0,-0.5) -- (1,0.5);
        \fill (0.75,0.25) circle (1.5pt);
    \end{scope}
    \node at(2.6,-0.25) {$.$};
\end{tikzpicture}
\end{array}
\end{equation}
The strings and charges in the Quon language are free to move within three-dimensional space, without affecting the global phase. Modulo the charges, the layers of the braid can be interchanged freely.

For example, the equality $\bra{0}\ket{1}=0$ can be represented by the following diagram:
\begin{equation}
\begin{tikzpicture}
\begin{scope}
    \draw[dashed] (-0.5,0.3) rectangle (1.5,0.8) node[anchor=east,xshift=-55,yshift=-8] {$\sqrt{2}\ket{1}$}; 
    \draw[dashed] (-0.5,-0.3) rectangle (1.5,-0.8) node[anchor=east,xshift=-55,yshift=8] {$\sqrt{2}\bra{0}$}; 
    \begin{scope}[shift={(0,0.5)}]
        \clip (-0.25,0) rectangle (0.25,0.25);
        \draw (0,0) circle(0.25);
        \fill (-0.18,0.19) circle (1.5pt);
    \end{scope}
    \begin{scope}[shift={(0,0.5)}]
        \clip (-0.25+1,0) rectangle (0.25+1,0.25);
        \draw (0+1,0) circle(0.25);
        \fill (0.18+1,0.19) circle (1.5pt);
    \end{scope}
    \begin{scope}
        \draw[-] (-0.25,-0.5) -- (-0.25,0.5);
        \draw[-] (0.25,-0.5) -- (0.25,0.5);
        \draw[-] (0.75,-0.5) -- (0.75,0.5);
        \draw[-] (1.25,-0.5) -- (1.25,0.5);
    \end{scope}
    \begin{scope}[shift={(0,-0.5)}]
        \clip (-0.25,0) rectangle (0.25,-0.25);
        \draw (0,0) circle(0.25);
    \end{scope}
    \begin{scope}[shift={(0,-0.5)}]
        \clip (-0.25+1,0) rectangle (0.25+1,-0.25);
        \draw (0+1,0) circle(0.25);
    \end{scope}
\end{scope}
\end{tikzpicture}
=-i
\begin{tikzpicture}
    \begin{scope}
        \draw (0,0) circle(0.25);
        \fill (-0.25,0) circle (1.5pt);
    \end{scope}
    \begin{scope}[shift={(0.6,0)}]
        \draw (0,0) circle(0.25);
        \fill (-0.25,0) circle (1.5pt);
    \end{scope}
\end{tikzpicture}
=0,
\end{equation}
where $\begin{tikzpicture}\begin{scope}[shift={(0.6,0)}]
    \draw (0,0) circle(0.25);
    \fill (-0.25,0) circle (1.5pt);
\end{scope}\end{tikzpicture}$ is given in Eq.~\eqref{eq:quon_relation}.

We now illustrate the relationship between Fig.~\ref{fig:513}(a) and the three-qubit repetition code in Quon language.
The figure represents the state $\ket{\phi}$, defined as:
\begin{equation}\label{ap:eq:quon_3_code}
    \ket{\phi}:=
    \begin{tikzpicture}
        \begin{scope}
            %\draw (0,0) circle(1);
            \node[circle,draw=blue] (1) at(-1,0) {2};
            \draw[->,thick,blue] (-1.3,0)--(-1.3,0.1);
            \node[circle,draw=blue] (2) at(0.5,0.866) {1};
            \draw[->,thick,blue] (0.8,0.866)--(0.8,0.766);
            \node[circle,draw=blue] (3) at(0.5,-0.866) {3};
            \draw[->,thick,blue] (0.8,-0.866)--(0.8,-0.966);
            \draw [-,black] (1) to [out=90,in=150] (2);
            \draw [-,black] (1) to [out=60,in=180] (2);
            \draw [-,black] (1) to [out=270,in=210] (3);
            \draw [-,black] (1) to [out=300,in=180] (3);
            \draw [-,black] (2) to [out=330,in=30] (3);
            \draw [-,black] (2) to [out=300,in=60] (3);
        \end{scope}
    \end{tikzpicture}.%=(\ket{000}+\ket{111})=\sqrt{2}\ket{+}_L.
\end{equation}
The three discs represent the three qubits, and the arrows indicate the starting points of the strings. The first string, pointed to by the arrow, is the first, and so on.

In these discs, Eq.~\eqref{eq:quon_state_0} and \eqref{eq:quon_state_1} can be represented by the following diagram:
\begin{equation}\label{eq:quon_basis}
    \begin{tikzpicture}
        \node[circle,draw=blue,minimum size=18] (1) at(-1,0) {};
        \draw[->,thick,blue] (-1.3,0)--(-1.3,0.1);
        \draw [-,black] (1) to [out=120,in=180] (-1,0.05);
        \draw [-,black] (1) to [out=60,in=0] (-1,0.05);
        \draw [-,black] (1) to [out=240,in=180] (-1,-0.05);
        \draw [-,black] (1) to [out=300,in=0] (-1,-0.05);
    \end{tikzpicture}
    =\sqrt{2}\ket{0}, \quad
    \begin{tikzpicture}
        \node[circle,draw=blue,minimum size=18] (1) at(-1,0) {};
        \draw[->,thick,blue] (-1.3,0)--(-1.3,0.1);
        \draw [-,black] (1) to [out=120,in=180] (-1,0.05);
        \draw [-,black] (1) to [out=60,in=0] (-1,0.05);
        \draw [-,black] (1) to [out=240,in=180] (-1,-0.05);
        \draw [-,black] (1) to [out=300,in=0] (-1,-0.05);
        \fill (-1.1,0.15) circle (1.5pt);
        \fill (-1.1,-0.15) circle (1.5pt);
    \end{tikzpicture}
    =\sqrt{2}\ket{1}
\end{equation}

The amplitude of the computational basis states can be calculated by inserting the computational basis states from Eq.~\eqref{eq:quon_basis} into the graph.
For example, the amplitude of $\ket{000}$ is given by $\braket{000}{\phi}$, which can be calculated as follows:
\begin{equation}
    \braket{000}{\phi}=
    \frac{1}{(\sqrt{2})^3}\begin{tikzpicture}
        \begin{scope}
            %\draw (0,0) circle(1);
            \node[circle,draw=blue,minimum size=18] (1) at(-1,0) {};
            \draw[->,thick,blue] (-1.3,0)--(-1.3,0.1);
            \node[circle,draw=blue,minimum size=18] (2) at(0.5,0.866) {};
            \draw[->,thick,blue] (0.8,0.866)--(0.8,0.766);
            \node[circle,draw=blue,minimum size=18] (3) at(0.5,-0.866) {};
            \draw[->,thick,blue] (0.8,-0.866)--(0.8,-0.966);
            \draw [-,black] (1) to [out=90,in=150] (2);
            \draw [-,black] (1) to [out=60,in=180] (2);
            \draw [-,black] (1) to [out=270,in=210] (3);
            \draw [-,black] (1) to [out=300,in=180] (3);
            \draw [-,black] (2) to [out=330,in=30] (3);
            \draw [-,black] (2) to [out=300,in=60] (3);
            \draw [-,black] (1) to [out=90,in=150] (-1,0);
            \draw [-,black] (1) to [out=60,in=330] (-1,0);
            \draw [-,black] (1) to [out=270,in=210] (-1,0);
            \draw [-,black] (1) to [out=300,in=30] (-1,0);
            \draw [-,black] (2) to [out=330,in=30] (0.5,0.866);
            \draw [-,black] (2) to [out=300,in=210] (0.5,0.866);
            \draw [-,black] (2) to [out=150,in=90] (0.5,0.866);
            \draw [-,black] (2) to [out=180,in=270] (0.5,0.866);
            \draw [-,black] (3) to [out=210,in=330] (0.5,-0.866);
            \draw [-,black] (3) to [out=180,in=150] (0.5,-0.866);
            \draw [-,black] (3) to [out=30,in=270] (0.5,-0.866);
            \draw [-,black] (3) to [out=60,in=90] (0.5,-0.866);
        \end{scope}
    \end{tikzpicture}
    =\frac{1}{(\sqrt{2})^3}
    \begin{tikzpicture}
    \begin{scope}
        \draw (0,0) circle(0.25);
    \end{scope}
    \begin{scope}[shift={(0.6,0)}]
        \draw (0,0) circle(0.25);
    \end{scope}
    \begin{scope}[shift={(1.2,0)}]
        \draw (0,0) circle(0.25);
    \end{scope}
    \end{tikzpicture}
    =1.
\end{equation}
Similarly, $\braket{111}{\phi}=1$, and for any other computational basis $\ket{\psi}\neq\ket{000},\ket{111}$, $\braket{\psi}{\phi}=0$.
Therefore, the state $\ket{\phi}=(\ket{000}+\ket{111})=\sqrt{2}\ket{+}_L$ is the logical state of the three-qubit repetition code.

Further information can be found in Supplementary Section B.

\section{Noise Models}
\subsection{Amplitude Damping Noise}
The single-qubit amplitude damping noise channel is given by
\begin{equation}
\begin{aligned}
    \mathcal{N}(\rho)&=\sum_{k=0,1}E_k\rho E_k^\dagger,\\
   \text{where } E_0&=\left[
  \begin{array}{cc}
    1 & 0 \\
    0 & \sqrt{1-\lambda}
  \end{array}
\right],\quad
E_1=\left[
  \begin{array}{cc}
    0 & \sqrt{\lambda} \\
    0 & 0
  \end{array}
\right] ,
\end{aligned}
\end{equation}
with $\lambda\in[0,1]$.

\subsection{Thermal Relaxation Process}

The thermal relaxation process of a single qubit corresponding to duration $t$ can be described by the following map :
\begin{equation}
    \rho=\begin{bmatrix}
1-\rho_{11} & \rho_{01} \\
\bar{\rho}_{01} & \rho_{11}
\end{bmatrix} 
\longrightarrow 
\begin{bmatrix}
1-\rho_{11}e^{-\frac{t}{T_1}} & \rho_{01}e^{-\frac{t}{2T_1}-\frac{t}{T_\phi}} \\
\bar{\rho}_{01}e^{-\frac{t}{2T_1}-\frac{t}{T_\phi}} & \rho_{11}e^{-\frac{t}{T_1}}
\end{bmatrix},
\end{equation}
where $\frac{1}{T_\phi}=\frac{1}{T_2}-\frac{1}{2T_1}$ and $T_2\leq 2T_1$.
This process has the following Kraus representation:
\begin{align*}
    \mathcal{N}(\rho)&=\sum_{k=1,2,3}A_k\rho A_k^\dagger,\\
    \text{where }A_1=&\left[
  \begin{array}{cc}
    1 & 0 \\
    0 & \sqrt{1-\gamma-\lambda}
  \end{array}
\right],
A_2=\left[
  \begin{array}{cc}
    0 & \sqrt{\gamma} \\
    0 & 0
  \end{array}
\right],
A_3=\left[
  \begin{array}{cc}
    0 & 0 \\
    0 & \sqrt{\lambda}
  \end{array}
\right],
\end{align*}
$\gamma=1-e^{-\frac{t}{T_1}}$, and $\lambda=e^{-\frac{t}{T_1}}-e^{-\frac{t}{2T_2}}$.
Such a CPTP map describes a Phase Amplitude Damping channel.

As reported in Ref.~\cite{IBM}, the publicly available coherence time data for the \textit{IBMQ-LIMA} (5-qubit) device is presented in Table~\ref{table:ibm}.
It is worth noting that IBM devices are calibrated nearly every day, with error reports updated after each calibration cycle; therefore, the data may not be consistent with the current state.
The data shows a significant variability in noise intensity across the qubits, with $Q_4$ exhibiting the shortest coherence time, suggesting that it is subject to the most intense noise.
Among the five qubits, $ Q_0 $ demonstrates the highest quality and is thus used as the baseline for comparison.

\begin{table}[htbp]
\centering
\begin{tabular}{|l|c|c|c|c|c|}
\hline 
Qubit &$Q_0$&$Q_1$&$Q_2$&$Q_3$&$Q_4$\\
\hline  
$T_1$&97.51 $\mu$ s&127.61 $\mu$ s&92.68 $\mu$ s&79.36 $\mu$ s&19.76 $\mu$ s\\
\hline 
$T_2$&178.3 $\mu$ s&109.28 $\mu$ s&120.95 $\mu$ s&35.71 $\mu$ s&19.4 $\mu$ s\\
\hline 
\end{tabular}
\caption{Coherence time data of IBMQ-LIMA device. The five qubits in the device are labeled by $Q_0,\dots,Q_4$.}\label{table:ibm}
\end{table}
\section{Composite channel design}

Our experiment implements the composite channel $M_{opt}$, which consists of the error correction encoding map $\mathcal{E}$, the amplitude damping channel $\mathcal{N}$, and the optimal recovery map $R_{opt}$. Corresponding projective measurements are performed to evaluate the composite channel.
The quantum operations are executed via their respective Kraus operators, each weighted by probabilities governing their contributions.

The error correction encoding map $\mathcal{E}$ maps the physical qubits into the logical qubit and is represented as $E_{\mathcal{E}}\rho E_{\mathcal{E}}^{\dagger}$. 

The amplitude damping channel $\mathcal{N}$ is modeled as a tensor product of single-qubit amplitude damping operators.
Each single-qubit operator is decomposed into a diagonal damping operator and a permutation operator:
\begin{equation}\label{eq:dampingoperator}
\begin{split}
E_0  = \begin{bmatrix}
    1 & 0 \\
    0 & 1
\end{bmatrix}
\begin{bmatrix}
1 & 0 \\
0 & \sqrt{1-\lambda}
\end{bmatrix},\ 
E_1  = \begin{bmatrix}
    0 & 1 \\
    1 & 0
\end{bmatrix}
\begin{bmatrix}
0 & 0 \\
0 & \sqrt{\lambda}
\end{bmatrix}.
\end{split}
\end{equation}
For a three physical qubit system, the three-quibt amplitude damping channel $\mathcal{N}$ is defined using eight Kraus operator $N_i$ as follows:
\begin{equation}
    \begin{aligned}
        \mathcal{N}(\rho)&=\sum_{a,b,c\in\{0,1\}}(E_{a}\otimes E_{b}\otimes E_{c})\rho(E_{a}^{\dagger}\otimes E_{b}^{\dagger}\otimes E_{c}^{\dagger})\\
        &=\sum_{i=1}^{8}N_i\rho N_i^{\dagger}=\sum_{i=1}^{8}B_iA_i\rho A_i^{\dagger}B_i^{\dagger},
    \end{aligned}
\end{equation}
where $A_i$ and $B_i$ are third-order tensor products of diagonal damping operator and permutation operators, respectively.
The diagonal damping operator $A_i$ is implemented in the damping noise module, while the permutation operator $B_i$ is realized in the subsequent module.

The optimal recovery map $R_{opt}$, obtained by solving a semi-definite programming (SDP) problem, is represented by a 16 × 16 Choi matrix with rank 5, allowing the implementation using five Kraus operators $R_j$.
The permutation operator $B_i$ and the optimal recovery operator $R_j$ are implemented in the permutation and recovery module.

Since the composition of the noise channel and the recovery map has 40 different combinations of Kraus operators (8 $N_i$ and 5 $R_j$), we need to execute 40 corresponding wave plates degree settings and {corresponding} projective measurements. 
However, because the error correction encoding map $\mathcal{E}$ assigns physical qubits to a specific logical subspace, non-zero contributions arise only when Kraus operators act within that subspace.
By eliminating zero terms in $R_j N_i E_{\mathcal{E}}$ and rearranging the non-zero terms, we define a new group of Kraus operators $K_l = C_{j_l}A_{i_l} E_{\mathcal{E}}$ with $C_{j_l} = R_{j_l}B_{i_l}$, where $l$ denotes the index of non-zero Kraus operators for $\mathcal{M}_{opt}$. The optimal composite channel naturally becomes
\begin{equation}
\begin{aligned}
\mathcal{M}_{opt} (\rho)
=\sum_{l=1}^{14} K_l \mathcal{E}(\rho) K_l^\dagger =\sum_{l=1}^{14} C_{j_l}A_{i_l} E_{\mathcal{E}} \rho E_{\mathcal{E}}^\dagger C_{j_l}^\dagger A_{i_l}^\dagger. \\
\end{aligned}
\end{equation}
Here $E_{\mathcal{E}}$, $A_{i_l}$ and $C_{j_l}$ are implemented  in Module-1, Module-2 and Module-3 of the experimental setup, respectively.

\textit{Note on Ref.~\cite{dutta2024smallest}:} A few days before submitting this paper to arXiv, we became aware of the work~\cite{dutta2024smallest}, which presents a 3-qubit code designed for amplitude damping noise. It is worth noting that the code proposed in this paper differs from the one in Ref.~\cite{dutta2024smallest}. Specifically, the code in Ref.~\cite{dutta2024smallest} is a probabilistic one, necessitating a post-selection process to filter out protocol failures. 
The fidelity data presented in Ref.~\cite{dutta2024smallest} illustrates the code's performance after this filtering step.
On the other hand, the VGQEC code in this paper is deterministic and does not require post-selections.

\bibliographystyle{ieeetr}
\bibliography{main}

\clearpage
\widetext

\appendix

\section{Quon Language}\label{ap:quon}
The quon language \cite{liu2019quantized,liu2017quon,jaffe2018mathematical} provides a mathematical picture language to study quantum information.  The pictures are given by braided charged strings in three-dimensional space. 
In the following, we introduce the basic computational rules of Quon language.
% and show how to represent quantum error correction codes in the Quon language.

In Quon, the 1-qubit XYZ computational bases are represented by the following diagrams in a hemisphere:
\begin{equation}
\begin{array}{cc}\label{eq:quon_basis}
\begin{tikzpicture}
\node at(-2.25,0.125) {$\sqrt{2} \ket{0}_Z=\sqrt{2} \ket{0}=$};
\begin{scope}
    \clip (-0.25,0) rectangle (0.25,0.25);
    \draw (0,0) circle(0.25);
\end{scope}
\begin{scope}
    \clip (-0.25+1,0) rectangle (0.25+1,0.25);
    \draw (0+1,0) circle(0.25);
\end{scope}
\node at(1.35,0) {, };
\end{tikzpicture}\\
\begin{tikzpicture}
\node at(-2,0.125) {$\sqrt{2} \ket{0}_Y=\ket{0}+i\ket{1}=$};
\begin{scope}
    \clip (-0.25,0) rectangle (1,0.5);
    \draw (0.25,0) circle(0.5);
    \draw[fill=white,draw=none] (0.4,0.35) rectangle +(0.2,0.2);
\end{scope}
\begin{scope}
    \clip (-0.25+0.5,0) rectangle (1.25,0.5);
    \draw (0.75,0) circle(0.5);
\end{scope}
\node at(1.35,0) {, };
\end{tikzpicture}\\
\begin{tikzpicture}
\node at(-2,0.125) {$\sqrt{2} \ket{0}_X=\ket{0}+\ket{1}=$};
\begin{scope}
    \clip (-0.25,0) rectangle (1.25,0.8);
    \draw (0.5,0) circle(0.75);
\end{scope}
\begin{scope}
    \clip (-0.25+0.5,0) rectangle (1.25,0.5);
    \draw (0.5,0) circle(0.25);
\end{scope}
\node at(1.35,0) {.};
\end{tikzpicture}
\end{array}
\end{equation}

Correspondingly, $\ket{1}_Z$, $\ket{1}_Y$, $\ket{1}_X$ are obtained by adding a pair of charges separately to the two strings in $\ket{0}_Z$, $\ket{0}_Y$, $\ket{0}_X$, respectively. There is an example for $\ket{1}_Z$:
\begin{equation}
    \begin{tikzpicture}
        \node at(-1.25,0.125) {$\sqrt{2} \ket{1}=$};
        \begin{scope}
            \clip (-0.25,0) rectangle (0.25,0.25);
            \draw (0,0) circle(0.25);
            \fill (-0.18,0.19) circle (1.5pt);
        \end{scope}
        \begin{scope}
            \clip (-0.25+1,0) rectangle (0.25+1,0.25);
            \draw (0+1,0) circle(0.25);
            \fill (0.18+1,0.19) circle (1.5pt);
        \end{scope}
        \node at(1.35,0) {, };
    \end{tikzpicture}
\end{equation}
Pauli $X$, $Y$, and $Z$ gates are represented by the following diagrams in a cylinder:
\begin{equation}\label{eq:quon_pauli}
\begin{array}{cc}
\begin{tikzpicture}
\node at(-0.5,0) {$I=$};
\begin{scope}
\draw[-] (0,-0.5) -- (0,0.5);
\draw[-] (0.25,-0.5) -- (0.25,0.5);
\draw[-] (0.5,-0.5) -- (0.5,0.5);
\draw[-] (0.75,-0.5) -- (0.75,0.5);
\end{scope}
\node at(1,0) {$=$};
\begin{scope}[shift={(1.5,0)}]
\draw[-] (0,-0.5) -- (0,0.5);
\draw[-] (0.25,-0.5) -- (0.25,0.5);
\draw[-] (0.5,-0.5) -- (0.5,0.5);
\draw[-] (0.75,-0.5) -- (0.75,0.5);
\fill (0,0) circle (1.5pt);
\fill (0.25,0) circle (1.5pt);
\fill (0.5,0) circle (1.5pt);
\fill (0.75,0) circle (1.5pt);
\end{scope}
\node at(2.5,-0.35) {,};
\end{tikzpicture}\\
\begin{tikzpicture}
\node at(-0.5,0) {$Z=$};
\begin{scope}
\draw[-] (0,-0.5) -- (0,0.5);
\draw[-] (0.25,-0.5) -- (0.25,0.5);
\draw[-] (0.5,-0.5) -- (0.5,0.5);
\draw[-] (0.75,-0.5) -- (0.75,0.5);
\fill (0,0) circle (1.5pt);
\fill (0.25,0) circle (1.5pt);
\end{scope}
\node at(1,0) {$=$};
\begin{scope}[shift={(1.5,0)}]
\draw[-] (0,-0.5) -- (0,0.5);
\draw[-] (0.25,-0.5) -- (0.25,0.5);
\draw[-] (0.5,-0.5) -- (0.5,0.5);
\draw[-] (0.75,-0.5) -- (0.75,0.5);
\fill (0.5,0) circle (1.5pt);
\fill (0.75,0) circle (1.5pt);
\end{scope}
\node at(2.5,-0.35) {,};
\end{tikzpicture}\\
\begin{tikzpicture}
\node at(-0.5,0) {$Y=$};
\begin{scope}
\draw[-] (0,-0.5) -- (0,0.5);
\draw[-] (0.25,-0.5) -- (0.25,0.5);
\draw[-] (0.5,-0.5) -- (0.5,0.5);
\draw[-] (0.75,-0.5) -- (0.75,0.5);
\fill (0,0) circle (1.5pt);
\fill (0.5,0) circle (1.5pt);
\end{scope}
\node at(1,0) {$=$};
\begin{scope}[shift={(1.5,0)}]
\draw[-] (0,-0.5) -- (0,0.5);
\draw[-] (0.25,-0.5) -- (0.25,0.5);
\draw[-] (0.5,-0.5) -- (0.5,0.5);
\draw[-] (0.75,-0.5) -- (0.75,0.5);
\fill (0.25,0) circle (1.5pt);
\fill (0.75,0) circle (1.5pt);
\end{scope}
\node at(2.5,-0.35) {,};
\end{tikzpicture}\\
\begin{tikzpicture}
\node at(-0.5,0) {$X=$};
\begin{scope}
\draw[-] (0,-0.5) -- (0,0.5);
\draw[-] (0.25,-0.5) -- (0.25,0.5);
\draw[-] (0.5,-0.5) -- (0.5,0.5);
\draw[-] (0.75,-0.5) -- (0.75,0.5);
\fill (0,0) circle (1.5pt);
\fill (0.75,0) circle (1.5pt);
\end{scope}
\node at(1,0) {$=$};
\begin{scope}[shift={(1.5,0)}]
\draw[-] (0,-0.5) -- (0,0.5);
\draw[-] (0.25,-0.5) -- (0.25,0.5);
\draw[-] (0.5,-0.5) -- (0.5,0.5);
\draw[-] (0.75,-0.5) -- (0.75,0.5);
\fill (0.5,0) circle (1.5pt);
\fill (0.25,0) circle (1.5pt);
\end{scope}
\node at(2.5,-0.35) {.};
\end{tikzpicture}
\end{array}
\end{equation}

The braid crossings in Quon are defined as follows:

\begin{equation}\label{eq:quon_crossing}
    \begin{array}{cc}
    \begin{tikzpicture}
    \begin{scope}[shift={(0,0)}]
        \draw[fill=white,draw=none,opacity=0] (0,0) rectangle (0.01,0.01);
    \end{scope}
    \begin{scope}[shift={(0.1,0)}]
        \draw[-] (0.5,-0.5) -- (0,0.5);
        %\node at(0,0) {$\alpha$};
        \draw[fill=white,draw=none] (0.15,-0.1) rectangle +(0.2,0.2);
        \draw[-] (0,-0.5) -- (0.5,0.5);
    \end{scope}
    \node at(1.25,0) {$\coloneqq \omega^{-\frac{1}{2}}$};
    \node at(2.1,0) {$(\frac{1}{\sqrt{2}}$};
    \begin{scope}[shift={(2.5,0)}]
        \draw[-] (0,-0.5) -- (0,0.5);
        %\draw[fill=white,draw=none] (0.4,-0.1) rectangle +(0.2,0.2);
        \draw[-] (0.5,-0.5) -- (0.5,0.5);
        %\fill (0.25,0.25) circle (1.5pt);
    \end{scope}
    \node at(3.5,0) {$+\frac{i}{\sqrt{2}}$};
    \begin{scope}[shift={(4.4,0)}]
        \draw[-] (0,-0.5) -- (0,0.5);
        %\draw[fill=white,draw=none] (0.4,-0.1) rectangle +(0.2,0.2);
        \draw[-] (0.5,-0.5) -- (0.5,0.5);
        \fill (0,0) circle (1.5pt);
        \fill (0.5,0) circle (1.5pt);
    \end{scope}
    \node at(5.1,0) {$),$};
    %\node at(2.6,-0.25) {$,$};
    \begin{scope}[shift={(6.1,0)}]
        \draw[fill=white,draw=none,opacity=0] (0,0) rectangle (0.01,0.01);
    \end{scope}
    \end{tikzpicture}
    \\
    \begin{tikzpicture}
        \begin{scope}[shift={(0,0)}]
            \draw[fill=white,draw=none,opacity=0] (0,0) rectangle (0.01,0.01);
        \end{scope}
    \begin{scope}[shift={(0.1,0)}]
        \draw[-] (0,-0.5) -- (0.5,0.5);
        %\node at(0,0) {$\alpha$};
        \draw[fill=white,draw=none] (0.15,-0.1) rectangle +(0.2,0.2);
        \draw[-] (0.5,-0.5) -- (0,0.5);
    \end{scope}
    \node at(1.25,0) {$\coloneqq \omega^{-\frac{1}{2}}$};
    \node at(2.1,0) {$(\frac{1}{\sqrt{2}}$};
    \begin{scope}[shift={(2.5,0)}]
        \draw[-] (0,-0.5) -- (0,0.5);
        %\draw[fill=white,draw=none] (0.4,-0.1) rectangle +(0.2,0.2);
        \draw[-] (0.5,-0.5) -- (0.5,0.5);
        %\fill (0.25,0.25) circle (1.5pt);
    \end{scope}
    \node at(3.5,0) {$-\frac{i}{\sqrt{2}}$};
    \begin{scope}[shift={(4.4,0)}]
        \draw[-] (0,-0.5) -- (0,0.5);
        %\draw[fill=white,draw=none] (0.4,-0.1) rectangle +(0.2,0.2);
        \draw[-] (0.5,-0.5) -- (0.5,0.5);
        \fill (0,0) circle (1.5pt);
        \fill (0.5,0) circle (1.5pt);
    \end{scope}
    \node at(5.1,0) {$).$};
    \begin{scope}[shift={(6.1,0)}]
        \draw[fill=white,draw=none,opacity=0] (0,0) rectangle (0.01,0.01);
    \end{scope}
    %\node at(2.6,-0.25) {$,$};
    \end{tikzpicture}
\end{array}
    %\caption{Graphical interpretation of braids crossings: (a) Variable braids crossing with parameter $\alpha$. }
    %\label{fig:crossing}
\end{equation}

In addition, we define the variable braids crossing with parameters as follows:
\begin{equation}\label{eq:quon_crossing_var}
    \begin{tikzpicture}
        \begin{scope}[shift={(0,0)}]
            \draw[fill=white,draw=none,opacity=0] (0,0) rectangle (0.01,0.01);
        \end{scope}
        \begin{scope}
            \draw[-] (0.5,-0.5) -- (0,0.5);
            \node at(0.1,0) {$\alpha$};
            %\draw[fill=white,draw=none] (0.4,-0.1) rectangle +(0.2,0.2);
            \draw[-] (0,-0.5) -- (0.5,0.5);
        \end{scope}
        \node at(0.8,0) {$\coloneqq$};
        \node at(1.6,0) {$\frac{1+e^{i\alpha}}{2}$};
        \begin{scope}[shift={(2.4,0)}]
            \draw[-] (0,-0.5) -- (0,0.5);
            %\draw[fill=white,draw=none] (0.4,-0.1) rectangle +(0.2,0.2);
            \draw[-] (0.5,-0.5) -- (0.5,0.5);
            %\fill (0.25,0.25) circle (1.5pt);
        \end{scope}
        \node at(3.6,0) {$+\frac{1-e^{i\alpha}}{2}$};
        \begin{scope}[shift={(4.3,0)}]
            \draw[-] (0,-0.5) -- (0,0.5);
            %\draw[fill=white,draw=none] (0.4,-0.1) rectangle +(0.2,0.2);
            \draw[-] (0.5,-0.5) -- (0.5,0.5);
            \fill (0,0) circle (1.5pt);
            \fill (0.5,0) circle (1.5pt);
        \end{scope}
        \node at(5,-0.25) {$.$};
        \begin{scope}[shift={(6,0)}]
            \draw[fill=white,draw=none,opacity=0] (0,0) rectangle (0.01,0.01);
        \end{scope}
    \end{tikzpicture}
\end{equation}
When the crossing parameter $\alpha=0$, the crossed strings will be transformed into parallel strings. The positive and negative braids crossings in Eq.~\eqref{eq:quon_crossing} differ from variable braids crossings with parameters $-\frac{\pi}{2}$ and $\frac{\pi}{2}$ by a global phase $\omega^{-\frac{1}{2}}=(\frac{1+i}{\sqrt{2}})^{-\frac{1}{2}}$, respectively.

The charges represented by dots ``$\bullet$" exhibit similar behavior to Majorana fermions, and the braid satisfies Reidemeister moves of type I, II, III. The pertinent properties of the quon language can be summarized pictorially as follows:
\begin{equation}\label{eq:quon_relation}
\begin{array}{cc}
\begin{tikzpicture}
\begin{scope}
    \draw (0,0) circle(0.25);
    \node at(1,0) {$=\sqrt{2}$, };
\end{scope}
\end{tikzpicture}
\begin{tikzpicture}
\begin{scope}
    \draw (0,0) circle(0.25);
    \fill (-0.25,0) circle (1.5pt);
    \node at(1,0) {$=0$, };
\end{scope}
\end{tikzpicture}
\begin{tikzpicture}
\begin{scope}
    \draw[-] (0,-0.5) -- (0,0.5);
    \fill (0,0.25) circle (1.5pt);
    \fill (0,-0.25) circle (1.5pt);
    \node at(0.5,0) {$=$};
    \draw[-] (1,-0.5) -- (1,0.5);
    \node at(1.25,-0.25) {$,$};
\end{scope}
\end{tikzpicture}\\
\begin{tikzpicture}
\begin{scope}
    \fill (-0.125,0.2) circle (1.5pt);
    \clip (-0.25,0) rectangle (0.25,0.25);
    \draw (0,0) circle(0.25);
\end{scope}
\node at(0.75,0.125) {$= i$};
\begin{scope}[shift={(1.5,0)}]
    \fill (0.125,0.2) circle (1.5pt);
    \node at(0.5,0) {$,$};
    \clip (-0.25,0) rectangle (0.25,0.25);
    \draw (0,0) circle(0.25);
\end{scope}
\end{tikzpicture}
\begin{tikzpicture}
\begin{scope}
    \draw[-] (0,-0.5) -- (0,0.5);
    \draw[-] (0.25,-0.5) -- (0.25,0.5);
    \fill (0,0.25) circle (1.5pt);
    \fill (0.25,-0.25) circle (1.5pt);
    \node at(0.75,0) {$=-$};
    \draw[-] (1.25,-0.5) -- (1.25,0.5);
    \draw[-] (1.5,-0.5) -- (1.5,0.5);
    \fill (1.25,-0.25) circle (1.5pt);
    \fill (1.5,0.25) circle (1.5pt);
    \node at(2,0) {$=-i$};
    \draw[-] (2.5,-0.5) -- (2.5,0.5);
    \draw[-] (2.75,-0.5) -- (2.75,0.5);
    \fill (2.5,0) circle (1.5pt);
    \fill (2.75,0) circle (1.5pt);
    \node at(3,-0.25) {$,$};
\end{scope}
\end{tikzpicture}\\
\begin{tikzpicture}
\begin{scope}
    \draw[-] (1,-0.5) -- (0,0.5);
    \draw[fill=white,draw=none] (0.4,-0.1) rectangle +(0.2,0.2);
    \draw[-] (0,-0.5) -- (1,0.5);
    \fill (0.75,-0.25) circle (1.5pt);
\end{scope}
\node at(1.25,0) {$=i$};
\begin{scope}[shift={(1.5,0)}]
    \draw[-] (1,-0.5) -- (0,0.5);
    \draw[fill=white,draw=none] (0.4,-0.1) rectangle +(0.2,0.2);
    \draw[-] (0,-0.5) -- (1,0.5);
    \fill (0.25,0.25) circle (1.5pt);
\end{scope}
\node at(2.6,-0.25) {$,$};
\end{tikzpicture}
\begin{tikzpicture}
    \begin{scope}
        \draw[-] (0,-0.5) -- (1,0.5);
        \draw[fill=white,draw=none] (0.4,-0.1) rectangle +(0.2,0.2);
        \draw[-] (1,-0.5) -- (0,0.5);
    \end{scope}
    \node at(1.25,0) {$=$};
    \begin{scope}[shift={(1.5,0)}]
        \draw[-] (0.75,-0.25) -- (0.25,0.25);
        \draw[fill=white,draw=none] (0.4,-0.1) rectangle +(0.2,0.2);
        \draw[-] (0.25,-0.25) -- (0.75,0.25);
        \draw[-] (0.75,0.25) -- (1,-0.5);
        \draw[-] (0.25,-0.25) -- (0,0.5);
        \draw[-] (1,0.5) -- (0.25,0.25);
        \draw[-] (0,-0.5) -- (0.75,-0.25);
    \end{scope}
    \node at(2.6,-0.25) {$,$};
\end{tikzpicture}
\\
\begin{tikzpicture}
\begin{scope}
    \draw[-] (0,-0.5) -- (1,0.5);
    \draw[fill=white,draw=none] (0.4,-0.1) rectangle +(0.2,0.2);
    \draw[-] (1,-0.5) -- (0,0.5);
\end{scope}
\node at(1.25,0) {$=$};
\begin{scope}[shift={(1.5,0)}]
    \draw[-] (1,-0.5) -- (0,0.5);
    \draw[fill=white,draw=none] (0.4,-0.1) rectangle +(0.2,0.2);
    \draw[-] (0,-0.5) -- (1,0.5);
    \fill (0.75,-0.25) circle (1.5pt);
    \fill (0.25,-0.25) circle (1.5pt);
\end{scope}
\node at(2.6,-0.25) {$,$};
\end{tikzpicture}
\begin{tikzpicture}
    \begin{scope}
        \draw[-] (1,-0.5) -- (0,0.5);
        \draw[fill=white,draw=none] (0.4,-0.1) rectangle +(0.2,0.2);
        \draw[-] (0,-0.5) -- (1,0.5);
        \fill (0.25,-0.25) circle (1.5pt);
    \end{scope}
    \node at(1.25,0) {$=-$};
    \begin{scope}[shift={(1.5,0)}]
        \draw[-] (1,-0.5) -- (0,0.5);
        \draw[fill=white,draw=none] (0.4,-0.1) rectangle +(0.2,0.2);
        \draw[-] (0,-0.5) -- (1,0.5);
        \fill (0.75,0.25) circle (1.5pt);
    \end{scope}
    \node at(2.6,-0.25) {$.$};
\end{tikzpicture}
\end{array}
\end{equation}
In fact, the first three rows can be generalized in qudit case~\cite{liu2017quon}.
The last two new relations hold only for qubit case.
The strings and charges are able to move freely in the 3D space, disregarding the global phase. Modulo the charges, the layers of the braid can be freely interchanged. 
%This is important to do pictorial computation efficiently, while it is difficult to evaluate a link in polynomial time in general.
%Furthermore, this property implies that our construction of the graphical QECC is essentially independent of the choice of the layers of the braids.

The following string-genus relation removes a hole surrounded by a string. While it still has no physical interpretation, its pictorial representation is:
\begin{equation}\label{eq:quon_genus}
    \begin{tikzpicture}
        \begin{scope}
            \draw (0,0) circle(0.5);
            \clip (-0.25,0.1) rectangle (0.25,-0.4);
            \draw[draw=blue] (0,0.1) circle(0.25);
            \clip (-0.125,0) rectangle (0.125,0.125);
            \draw[draw=blue] (0,-0.05) circle(0.125);
        \end{scope}
        \node at(1.25,0) {$=\frac{1}{\sqrt{2}}$, };
    \end{tikzpicture}
\end{equation}

The pictures in the bulk represent gates projectively and the pictures on the boundary represent states linearly.
For example, the equality $\bra{0}X\ket{0}=0$ can be represented by the following diagram:
\begin{equation}
\begin{tikzpicture}
\begin{scope}
    \draw[dashed] (-0.5,0.3) rectangle (1.5,0.8) node[anchor=east,xshift=-55,yshift=-8] {$\sqrt{2}\ket{0}$}; 
    \draw[dashed] (-0.5,-0.3) rectangle (1.5,0.3) node[anchor=east,xshift=-58,yshift=-8] {$X$}; 
    \draw[dashed] (-0.5,-0.3) rectangle (1.5,-0.8) node[anchor=east,xshift=-55,yshift=8] {$\sqrt{2}\bra{0}$}; 
    \begin{scope}[shift={(0,0.5)}]
        \clip (-0.25,0) rectangle (0.25,0.25);
        \draw (0,0) circle(0.25);
    \end{scope}
    \begin{scope}[shift={(0,0.5)}]
        \clip (-0.25+1,0) rectangle (0.25+1,0.25);
        \draw (0+1,0) circle(0.25);
    \end{scope}
    \begin{scope}
        \draw[-] (-0.25,-0.5) -- (-0.25,0.5);
        \draw[-] (0.25,-0.5) -- (0.25,0.5);
        \draw[-] (0.75,-0.5) -- (0.75,0.5);
        \draw[-] (1.25,-0.5) -- (1.25,0.5);
        \fill (-0.25,0) circle (1.5pt);
        \fill (1.25,0) circle (1.5pt);
    \end{scope}
    \begin{scope}[shift={(0,-0.5)}]
        \clip (-0.25,0) rectangle (0.25,-0.25);
        \draw (0,0) circle(0.25);
    \end{scope}
    \begin{scope}[shift={(0,-0.5)}]
        \clip (-0.25+1,0) rectangle (0.25+1,-0.25);
        \draw (0+1,0) circle(0.25);
    \end{scope}
\end{scope}
\end{tikzpicture}
=-i
\begin{tikzpicture}
    \begin{scope}
        \draw (0,0) circle(0.25);
        \fill (-0.25,0) circle (1.5pt);
    \end{scope}
    \begin{scope}[shift={(0.6,0)}]
        \draw (0,0) circle(0.25);
        \fill (-0.25,0) circle (1.5pt);
    \end{scope}
\end{tikzpicture}
=0
\end{equation}

For convenience, we give the graphical representation of common Pauli rotations $R_X$ and $R_ZZ$ gates:
\begin{equation}\label{eq:quon_R}
    \begin{tikzpicture}[scale=0.7]
    \begin{scope}
        \draw[-] (0,0) -- (0,-2);
        \draw[-] (0.5,0) -- (1,-2);
        \draw[-] (1,0) -- (0.5,-2);
        \draw[-] (1.5,0) -- (1.5,-2);
        \node at(0.5,-1) {$\alpha$};
        \fill (0.75,-1) circle (1.5pt);
        \end{scope}
        \node at(3,-1) {$=e^{i\frac{\alpha}{2}}e^{-i\frac{\alpha}{2}X}$};
        \end{tikzpicture}
        ,\qquad
        \begin{tikzpicture}[scale=0.7]
        \begin{scope}
        \draw[-] (-0.5,0) -- (-0.5,-2);
        \draw[-] (0,0) -- (0,-2);
        \draw[-] (0.5,0) -- (2,-2);
        \draw[-] (2,0) -- (0.5,-2);
        \draw[-] (2.5,0) -- (2.5,-2);
        \draw[-] (3,0) -- (3,-2);
        \node at(1,-1) {$\alpha$};
        \fill (1.25,-1) circle (1.5pt);
        \clip (0,0) rectangle (2.5,-2);
        \draw (1.25,0) circle(0.25);
        %\clip (1,-1.75) rectangle (1.5,-2);
        \draw (1.25,-2) circle(0.25);
        \end{scope}
        \node at(5,-1) {$=\frac{1}{\sqrt{2}}e^{i\frac{\alpha}{2}}e^{-i\frac{\alpha}{2}ZZ}$};
    \end{tikzpicture}
\end{equation}
These can be verified by calculating its action on computational basis states.

\section{Quon Graph for Quantum Error Correction Codes}\label{ap:quon_qecc}

In \cite{liu2019quantized}, the author introduces a systematic method for representing stabilizer codes using graphs. This method facilitates the analysis of specific noise effects through graphical representation, thereby enabling a topological understanding of error correction capabilities.

We take the three-qubit repetition code as an example. The codewords map are
\begin{equation}
    \mathcal{E}(\alpha\ket{0}+\beta\ket{1})=\alpha\ket{000}+\beta\ket{111}.
\end{equation}
It encodes one logical qubit by three physical qubits, and corrects one-qubit Pauli $X$ error. In the Quon language, we first claim the three-qubit repetition code can be represented by the graph with three discs:
\begin{equation}\label{ap:eq:quon_3_code}
    \sqrt{2}\ket{+}_L=(\ket{000}+\ket{111})=
    \begin{tikzpicture}
        \begin{scope}
            %\draw (0,0) circle(1);
            \node[circle,draw=blue] (1) at(-1,0) {2};
            \draw[->,thick,blue] (-1.3,0)--(-1.3,0.1);
            \node[circle,draw=blue] (2) at(0.5,0.866) {1};
            \draw[->,thick,blue] (0.8,0.866)--(0.8,0.766);
            \node[circle,draw=blue] (3) at(0.5,-0.866) {3};
            \draw[->,thick,blue] (0.8,-0.866)--(0.8,-0.966);
            \draw [-,black] (1) to [out=90,in=150] (2);
            \draw [-,black] (1) to [out=60,in=180] (2);
            \draw [-,black] (1) to [out=270,in=210] (3);
            \draw [-,black] (1) to [out=300,in=180] (3);
            \draw [-,black] (2) to [out=330,in=30] (3);
            \draw [-,black] (2) to [out=300,in=60] (3);
        \end{scope}
    \end{tikzpicture}, \quad
    \sqrt{2}\ket{-}_R=(\ket{000}-\ket{111})=
    \begin{tikzpicture}
        \begin{scope}
            %\draw (0,0) circle(1);
            \node[circle,draw=blue] (1) at(-1,0) {2};
            \draw[->,thick,blue] (-1.3,0)--(-1.3,0.1);
            \node[circle,draw=blue] (2) at(0.5,0.866) {1};
            \draw[->,thick,blue] (0.8,0.866)--(0.8,0.766);
            \node[circle,draw=blue] (3) at(0.5,-0.866) {3};
            \draw[->,thick,blue] (0.8,-0.866)--(0.8,-0.966);
            \draw [-,black] (1) to [out=90,in=150] (2);
            \draw [-,black] (1) to [out=60,in=180] (2);
            \draw [-,black] (1) to [out=270,in=210] (3);
            \draw [-,black] (1) to [out=300,in=180] (3);
            \draw [-,black] (2) to [out=330,in=30] (3);
            \draw [-,black] (2) to [out=300,in=60] (3);
            \fill (0.825,0) circle (1.6pt);
            \fill (1.125,0) circle (1.6pt);
        \end{scope}
    \end{tikzpicture}.
\end{equation}
Where the three discs represent the three qubits, and the arrows attached to the discs indicate the starting points of strings. The first string pointed to by the arrow is the first one, and so on.

The Eq.~\eqref{ap:eq:quon_3_code} can be verified by calculating the amplitude of computational basis states. For example, the amplitude of $\ket{000}$ can be calculated by put computational basis states Eq.~\eqref{eq:quon_basis} into the graph:
\begin{equation}
    \frac{1}{(\sqrt{2})^3}\begin{tikzpicture}
        \begin{scope}
            %\draw (0,0) circle(1);
            \node[circle,draw=blue,minimum size=18] (1) at(-1,0) {};
            \draw[->,thick,blue] (-1.3,0)--(-1.3,0.1);
            \node[circle,draw=blue,minimum size=18] (2) at(0.5,0.866) {};
            \draw[->,thick,blue] (0.8,0.866)--(0.8,0.766);
            \node[circle,draw=blue,minimum size=18] (3) at(0.5,-0.866) {};
            \draw[->,thick,blue] (0.8,-0.866)--(0.8,-0.966);
            \draw [-,black] (1) to [out=90,in=150] (2);
            \draw [-,black] (1) to [out=60,in=180] (2);
            \draw [-,black] (1) to [out=270,in=210] (3);
            \draw [-,black] (1) to [out=300,in=180] (3);
            \draw [-,black] (2) to [out=330,in=30] (3);
            \draw [-,black] (2) to [out=300,in=60] (3);
            \draw [-,black] (1) to [out=90,in=150] (-1,0);
            \draw [-,black] (1) to [out=60,in=330] (-1,0);
            \draw [-,black] (1) to [out=270,in=210] (-1,0);
            \draw [-,black] (1) to [out=300,in=30] (-1,0);
            \draw [-,black] (2) to [out=330,in=30] (0.5,0.866);
            \draw [-,black] (2) to [out=300,in=210] (0.5,0.866);
            \draw [-,black] (2) to [out=150,in=90] (0.5,0.866);
            \draw [-,black] (2) to [out=180,in=270] (0.5,0.866);
            \draw [-,black] (3) to [out=210,in=330] (0.5,-0.866);
            \draw [-,black] (3) to [out=180,in=150] (0.5,-0.866);
            \draw [-,black] (3) to [out=30,in=270] (0.5,-0.866);
            \draw [-,black] (3) to [out=60,in=90] (0.5,-0.866);
        \end{scope}
    \end{tikzpicture}
    =\frac{1}{(\sqrt{2})^3}
    \begin{tikzpicture}
    \begin{scope}
        \draw (0,0) circle(0.25);
    \end{scope}
    \begin{scope}[shift={(0.6,0)}]
        \draw (0,0) circle(0.25);
    \end{scope}
    \begin{scope}[shift={(1.2,0)}]
        \draw (0,0) circle(0.25);
    \end{scope}
    \end{tikzpicture}
    =1.
\end{equation}
By calculating the amplitude under the basis one by one, the equation can be verified.

The stabilizers of the three-qubit repetition code are $Z_1Z_2$, $Z_2Z_3$, and $Z_1Z_3$. The graph naturally captures these stabilizers as the cycles on the boundary. 
Specifically, let $L$ be the cycle on the boundary between the discs $1$ and $3$. We define the \textit{cycle operator} $O_L$ acting on the logical state $\ket{+}_L$, by adding pairs of charges on the cycle $L$ nearby the discs $1$ and $3$. By diagrammatic operator Eq.~\eqref{eq:quon_pauli}, $O_L=Z_1Z_3$. The cycle operator $O_L$ stabilizes the $\ket{+}_L$, namely $O_L\ket{+}_L = \ket{+}_L$, because each edge in the cycle $L$ contains two changes, which will cancel each other as illustrated in Fig.~\ref{fig:quon_3_stabilizers}.

\begin{figure}[htbp]
    \centering
    \begin{tikzpicture}
    \begin{scope}
        \node[circle,draw=blue] (1) at(-1,0) {2};
        \draw[->,thick,blue] (-1.3,0)--(-1.3,0.1);
        \node[circle,draw=blue] (2) at(0.5,0.866) {1};
        \draw[->,thick,blue] (0.8,0.866)--(0.8,0.766);
        \node[circle,draw=blue] (3) at(0.5,-0.866) {3};
        \draw[->,thick,blue] (0.8,-0.866)--(0.8,-0.966);
        \draw [-,black] (1) to [out=90,in=150] (2);
        \draw [-,black] (1) to [out=60,in=180] (2);
        \draw [-,black] (1) to [out=270,in=210] (3);
        \draw [-,black] (1) to [out=300,in=180] (3);
        \draw [-,orange] (2) to [out=330,in=30] (3);
        \draw [-,orange] (2) to [out=300,in=60] (3);
        \fill (0.75,0.4) circle (1.5pt);
        \fill (0.75,-0.4) circle (1.5pt);
        \fill (0.95,0.55) circle (1.5pt);
        \fill (0.95,-0.55) circle (1.5pt);
    \end{scope}
    \node[] at(1.5,0) {=};
    \begin{scope}[shift={(3.5,0)}]
        \node[circle,draw=blue] (1) at(-1,0) {2};
        \draw[->,thick,blue] (-1.3,0)--(-1.3,0.1);
        \node[circle,draw=blue] (2) at(0.5,0.866) {1};
        \draw[->,thick,blue] (0.8,0.866)--(0.8,0.766);
        \node[circle,draw=blue] (3) at(0.5,-0.866) {3};
        \draw[->,thick,blue] (0.8,-0.866)--(0.8,-0.966);
        \draw [-,black] (1) to [out=90,in=150] (2);
        \draw [-,black] (1) to [out=60,in=180] (2);
        \draw [-,black] (1) to [out=270,in=210] (3);
        \draw [-,black] (1) to [out=300,in=180] (3);
        \draw [-,orange] (2) to [out=330,in=30] (3);
        \draw [-,orange] (2) to [out=300,in=60] (3);
    \end{scope}
    \end{tikzpicture}
        \caption{Cycle operator as stabilizers: $L$ is the cycle on the boundary between the discs $1$ and $3$, corresponding cycle operator $O_L = Z_1Z_3$.}
        \label{fig:quon_3_stabilizers}
    \end{figure}

For the well-known $[[5,1,3]]$ code~\cite{laflamme1996perfect}, it encodes one logical qubit into five physical qubits, and corrects one-qubit arbitrary errors. 
The $[[5,1,3]]$ code is a stabilizer code~\cite{gottesman1997stabilizer} and the code can be fully described by its stabilizer group. 
The stabilizer group of the $[[5,1,3]]$ code has the generators:
\begin{equation}\label{eq:513stabilizer}
    \{X_2Z_3Z_4X_5 , X_1X_3Z_4Z_5 , Z_1X_2X_4Z_5, Z_1Z_2X_3X_5\}.
\end{equation}

The Quon graph of the logical state of $[[5,1,3]]$ code can be represented as
\begin{equation}\label{ap:eq:quon_5_code}
    \begin{tikzpicture}
    \node at(0,0) {$\sqrt{2}\ket{0}_L=$};
    \node[circle,draw=blue] (1) at(1.9,0.94) {1};
    \node[circle,draw=blue] (2) at(0.81,0.56) {2};
    \node[circle,draw=blue] (3) at(0.81,-0.56) {3};
    \node[circle,draw=blue] (4) at(1.9,-0.94) {4};
    \node[circle,draw=blue] (5) at(2.6,0) {5};
    \draw[->,thick,blue] (1.9,1.25)--(1.91,1.25);
    \draw[->,thick,blue] (1.91,-1.25)--(1.9,-1.25);
    \draw[->,thick,blue] (2.9,0)--(2.9,-0.01);
    \draw[->,thick,blue] (0.6,0.8)--(0.61,0.81);
    \draw[->,thick,blue] (0.6,-0.8)--(0.59,-0.79);
    \draw [-,black] (1) to (2);
    \draw [-,black] (2) to (3);
    \draw [-,black] (3) to (4);
    \draw [-,black] (4) to (5);
    \draw [-,black] (5) to (1);
    \draw [-,black] (1) to (1.25,0.05);
    \draw [-,black] (1.18,-0.05) to (3);
    \draw [-,black] (1) to (1.89,0.28);
    \draw [-,black] (1.89,0.15) to (4);
    \draw [-,black] (2) to (1.45,-0.3);
    \draw [-,black] (1.5,-0.4) to (4);
    \draw [-,black] (2) to (1.4,0.36);
    \draw [-,black] (1.53,0.33) to (5);
    \draw [-,black] (3) to (1.83,-0.25);
    \draw [-,black] (1.95,-0.23) to (5);
    \end{tikzpicture}
    \begin{tikzpicture}
    \node at(-0.3,0) {$, \quad \sqrt{2}\ket{1}_L=$};
    \node[circle,draw=blue] (1) at(1.9,0.94) {1};
    \node[circle,draw=blue] (2) at(0.81,0.56) {2};
    \node[circle,draw=blue] (3) at(0.81,-0.56) {3};
    \node[circle,draw=blue] (4) at(1.9,-0.94) {4};
    \node[circle,draw=blue] (5) at(2.6,0) {5};
    \draw[->,thick,blue] (1.9,1.25)--(1.91,1.25);
    \draw[->,thick,blue] (1.91,-1.25)--(1.9,-1.25);
    \draw[->,thick,blue] (2.9,0)--(2.9,-0.01);
    \draw[->,thick,blue] (0.6,0.8)--(0.61,0.81);
    \draw[->,thick,blue] (0.6,-0.8)--(0.59,-0.79);
    \draw [-,black] (1) to (2);
    \draw [-,black] (2) to (3);
    \draw [-,black] (3) to (4);
    \draw [-,black] (4) to (5);
    \draw [-,black] (5) to (1);
    \draw [-,black] (1) to (1.25,0.05);
    \draw [-,black] (1.18,-0.05) to (3);
    \draw [-,black] (1) to (1.89,0.28);
    \draw [-,black] (1.89,0.15) to (4);
    \draw [-,black] (2) to (1.45,-0.3);
    \draw [-,black] (1.5,-0.4) to (4);
    \draw [-,black] (2) to (1.4,0.36);
    \draw [-,black] (1.53,0.33) to (5);
    \draw [-,black] (3) to (1.83,-0.25);
    \draw [-,black] (1.95,-0.23) to (5);
    \fill ($ (1)!.5!(2) $) circle (1.6pt);
    \fill ($ (2)!.5!(3) $) circle (1.6pt);
    \fill ($ (3)!.5!(4) $) circle (1.6pt);
    \fill ($ (4)!.5!(5) $) circle (1.6pt);
    \fill ($ (5)!.5!(1) $) circle (1.6pt);
    \fill ($ (1)!.5!(3) $) circle (1.6pt);
    \fill ($ (1)!.5!(4) $) circle (1.6pt);
    \fill ($ (2)!.5!(4) $) circle (1.6pt);
    \fill ($ (2)!.5!(5) $) circle (1.6pt);
    \fill ($ (3)!.5!(5) $) circle (1.6pt);
    \node at(3.5,-0.2) {.};
    \end{tikzpicture}
\end{equation}
This equation can be checked by the stabilizer group~\eqref{eq:513stabilizer}.
The generators of the stabilizer group~\eqref{eq:513stabilizer} correspond to the four cycles of even length:
\begin{align*}
        L_1:2\longrightarrow 4\longrightarrow 3\longrightarrow 5\longrightarrow 2,\\
        L_2:3\longrightarrow 5\longrightarrow 4\longrightarrow 1\longrightarrow 3,\\
        L_3:4\longrightarrow 1\longrightarrow 5\longrightarrow 2\longrightarrow 4,\\
        L_5:5\longrightarrow 2\longrightarrow 1\longrightarrow 3\longrightarrow 5.
\end{align*}
Specifically, considering the cycle $L_1:2\longrightarrow 4\longrightarrow 3\longrightarrow 5\longrightarrow 2$, we define the \textit{cycle operator} $O_{L_1}$ acting on the Quon graph, and adding pairs of charges on the cycle $L_1$. 
By diagrammatic operator using Eq.~\eqref{eq:quon_pauli}, we have $O_{L_1}=X_2Z_3Z_4X_5$. 
The cycle operator $O_{L_1}$ stabilizes the encoded quantum state $\ket{0}_L, \ket{1}_L$, because each edge in the cycle $L_1$ contains two changes, which will cancel each other:
\begin{equation}
    \begin{tikzpicture}
        \node at(-0.5,0) {$O_{L_1}\sqrt{2}\ket{0}_L=$};
        \node[circle,draw=blue] (1) at(1.9,0.94) {1};
        \node[circle,draw=blue] (2) at(0.81,0.56) {2};
        \node[circle,draw=blue] (3) at(0.81,-0.56) {3};
        \node[circle,draw=blue] (4) at(1.9,-0.94) {4};
        \node[circle,draw=blue] (5) at(2.6,0) {5};
        \draw[->,thick,blue] (1.9,1.25)--(1.91,1.25);
        \draw[->,thick,blue] (1.91,-1.25)--(1.9,-1.25);
        \draw[->,thick,blue] (2.9,0)--(2.9,-0.01);
        \draw[->,thick,blue] (0.6,0.8)--(0.61,0.81);
        \draw[->,thick,blue] (0.6,-0.8)--(0.59,-0.79);
        \draw [-,black] (1) to (2);
        \draw [-,black] (2) to (3);
        \draw [-,orange] (3) to (4);
        \draw [-,black] (4) to (5);
        \draw [-,black] (5) to (1);
        \draw [-,black] (1) to (1.25,0.05);
        \draw [-,black] (1.18,-0.05) to (3);
        \draw [-,black] (1) to (1.89,0.28);
        \draw [-,black] (1.89,0.15) to (4);
        \draw [-,orange] (2) to (1.45,-0.3);
        \draw [-,orange] (1.5,-0.4) to (4);
        \draw [-,orange] (2) to (1.4,0.36);
        \draw [-,orange] (1.53,0.33) to (5);
        \draw [-,orange] (3) to (1.83,-0.25);
        \draw [-,orange] (1.95,-0.23) to (5);
        \fill (1.2,0.43) circle (1.5pt);
        \fill (1.05,0.2) circle (1.5pt);
        \fill (1.2,-0.43) circle (1.5pt);
        \fill (1.2,-0.71) circle (1.5pt);
        \fill (1.5,-0.8) circle (1.5pt);
        \fill (1.65,-0.6) circle (1.5pt);
        \fill (2.2,0.13) circle (1.5pt);
        \fill (2.2,-0.13) circle (1.5pt);
        \end{tikzpicture}
        \begin{tikzpicture}
        \node at(0,0) {$=$};
        \node[circle,draw=blue] (1) at(1.9,0.94) {1};
        \node[circle,draw=blue] (2) at(0.81,0.56) {2};
        \node[circle,draw=blue] (3) at(0.81,-0.56) {3};
        \node[circle,draw=blue] (4) at(1.9,-0.94) {4};
        \node[circle,draw=blue] (5) at(2.6,0) {5};
        \draw[->,thick,blue] (1.9,1.25)--(1.91,1.25);
        \draw[->,thick,blue] (1.91,-1.25)--(1.9,-1.25);
        \draw[->,thick,blue] (2.9,0)--(2.9,-0.01);
        \draw[->,thick,blue] (0.6,0.8)--(0.61,0.81);
        \draw[->,thick,blue] (0.6,-0.8)--(0.59,-0.79);
        \draw [-,black] (1) to (2);
        \draw [-,black] (2) to (3);
        \draw [-,orange] (3) to (4);
        \draw [-,black] (4) to (5);
        \draw [-,black] (5) to (1);
        \draw [-,black] (1) to (1.25,0.05);
        \draw [-,black] (1.18,-0.05) to (3);
        \draw [-,black] (1) to (1.89,0.28);
        \draw [-,black] (1.89,0.15) to (4);
        \draw [-,orange] (2) to (1.45,-0.3);
        \draw [-,orange] (1.5,-0.4) to (4);
        \draw [-,orange] (2) to (1.4,0.36);
        \draw [-,orange] (1.53,0.33) to (5);
        \draw [-,orange] (3) to (1.83,-0.25);
        \draw [-,orange] (1.95,-0.23) to (5);
        \node at(3.0,-0.2) {,};
    \end{tikzpicture}
\end{equation}
where the orange edges represent the cycle ${L_1}$.

It is worth noting that the requirement for an even-length cycle cannot be overlooked in this context. In actuality, for any stabilizer code, a stabilizer corresponds to a cycle in the Quon graph of its encoding map that does not enclose the logical qubit. For simplicity, we did not present the Quon graph of the encoding map for the $[[5,1,3]]$ code here.
The even-length constraint is equivalent to the cycle that does not enclose the logical qubit for the $[[5,1,3]]$ code.

The idea of VGQEC is to introduce parameters to the Quon graph. The definition of parameters in the Quon graph is shown in Eq.~\eqref{eq:quon_crossing_var}.
Parallel strings can be treated as a crossing with parameter $\alpha=0$, and every crossing in the Quon graph contains an implicit parameter $-\frac{\pi}{2}$ or $\frac{\pi}{2}$ corresponding to a positive or negative crossing, respectively.
One potential modification is to replace the implicit parameters of the crossings with explicit variables, thereby enabling variability in braid crossings.
In addition to deforming already existing crossings, we can add variable crossing to graphs, by replacing parallel strings with variable crossings.
These operations make it possible to continuously change the construction of the Quon graph, such as changing the connection between discs, modifying the position of crossings, adding or deleting variable crossings, etc.

\section{Semi-Definite Programming for Optimal Recovery Map}\label{ap:SDP}

For composite channel $\mathcal{M}=\mathcal{R}\circ\mathcal{N}\circ\mathcal{E}$, the channel fidelity $F_C(\mathcal{M})$ quantifies the effectiveness of the code in safeguarding quantum information against noise. 
Assuming the number of logical qubits is $k$, the physical qubits is $n$, then the dimension of the Hilbert space, where $\mathcal{M}$ acts on, is $d=2^k$.
The channel fidelity can be expressed as:
\begin{equation}
    F_C(\mathcal{M})= F_e(\frac{I}{2^k},\mathcal{M}).
\end{equation}

Depending on the fact that $\ket{\psi}=\frac{1}{\sqrt{2^k}} \sum_i \ket{i}\ket{i}$ is a purification of the maximally mixed state $\rho=I/2^k$, the entanglement fidelity $F_e(I/2^k,\mathcal{M})$ can be expressed as:
\begin{equation}
    \begin{aligned}
    F_e(I/2^k,\mathcal{M})&= f(\ketbra{\psi}{\psi},(\mathcal{M}\otimes \mathcal{I})(\ketbra{\psi}{\psi}))=\left(\tr{\sqrt{\sqrt{\ketbra{\psi}{\psi}}(\mathcal{M}\otimes \mathcal{I})(\ketbra{\psi}{\psi})\sqrt{\ketbra{\psi}{\psi}}}} \right)^2\\ 
    &=(\tr{\sqrt{\ketbra{\psi}{\psi}(\mathcal{M}\otimes \mathcal{I})(\ketbra{\psi}{\psi})\ketbra{\psi}{\psi}}})^2=\bra{\psi}(\mathcal{M}\otimes \mathcal{I})(\ketbra{\psi}{\psi})\ket{\psi}\\
    &=\sum_i\bra{\psi} M_i \otimes I \ketbra{\psi}{\psi} M_i^\dagger\otimes I\ket{\psi}= \frac{1}{2^{2k}} \sum_i \abs{\tr{M_i}}^2,
    \end{aligned}
\end{equation}
where $M_i$ are the Kraus operators of the channel $\mathcal{M}$.

Assuming the Kraus operators of the encoding map $\mathcal{E}$, the noise channel $\mathcal{N}$ and the recovery map $\mathcal{R}$ in $\mathcal{M}=\mathcal{R}\circ\mathcal{N}\circ\mathcal{E}$ are $V_i$, $E_j$, and $R_k$, respectively, the channel fidelity $F_C(\mathcal{M})$ can be expressed as:
\begin{equation}\label{ap:eq:Mfidelity}
    F_C(\mathcal{M})=\frac{1}{2^{2k}}\sum_{ijk} \abs{\tr{R_k E_j V_i}}^2.
\end{equation}

Using Choi-Jamiolkowski isomorphism~\cite{choi1975completely}, the channel $\mathcal{R}$ can be represented by the Choi matrix $\chi_{\mathcal{R}}$:
\begin{equation}
    \chi_{\mathcal{R}}=\sum^{2^n-1}_{i,j=0}\sum_{k} \ketbra{i}{j} \otimes R_k\ket{i}\bra{j}R_k^\dagger .
\end{equation}
The Choi matrix $\chi_{\mathcal{R}}$ is a positive semi-definite matrix, and the channel $\mathcal{R}$ is completely positive if and only if the Choi matrix $\chi_{\mathcal{R}}$ is positive semi-definite.

The trace-preserving condition of the channel $\mathcal{R}$ can be expressed as:
\begin{equation}
    \begin{aligned}
        \tr_2{\chi_{\mathcal{R}}}:&= \sum_{w=0}^{2^k-1} \bra{w}\chi_{\mathcal{R}}\ket{w}=  \sum_{w=0}^{2^k-1}\sum^{2^n-1}_{i,j=0}\sum_{k}  \bra{w}R_k\ket{i}\bra{j}R_k^\dagger \ket{w} \ketbra{i}{j}=\sum^{2^n-1}_{i,j=0}\sum_{k} \bra{j} R_k^\dagger R_k\ket{i} \ketbra{i}{j}\\
        &=\sum_{k} (R_k^\dagger R_k)^T=I.
    \end{aligned}
\end{equation}

Using Choi matrix, Eq.~\eqref{ap:eq:Mfidelity} can be rewritten as:
\begin{equation}
    \begin{aligned}
        F_C(\mathcal{M})&=\frac{1}{2^{2k}}\sum_{ijk} \tr(E_j V_i R_k)\tr(R_k^\dagger E_j^\dagger V_i^\dagger)=\frac{1}{2^{2k}} \sum_{ijk} \sum_{p,q=0}^{2^n-1} \sum_{v,w=0}^{2^k-1} \bra{p}E_j V_i \ketbra{v}R_k\ket{p} \bra{q}R_k^\dagger \ketbra{w} E_j^\dagger V_i^\dagger\ket{q}\\
        &=\frac{1}{2^{2k}} \sum_{ij} \sum_{p,q=0}^{2^n-1} \sum_{v,w=0}^{2^k-1} \bra{p}\bra{v}\chi_{\mathcal{R}}\ket{q}\ket{w} \bra{p}E_j V_i \ket{v}\bra{w} E_j^\dagger V_i^\dagger\ket{q}\\
        &=\frac{1}{2^{2k}} \sum_{p,q=0}^{2^n-1} \sum_{v,w=0}^{2^k-1} \bra{p}\bra{v}\chi_{\mathcal{R}}\ket{q}\ket{w} \bra{q}\bra{w} C \ket{p}\ket{v}= \frac{1}{2^{2k}} \tr(\chi_{\mathcal{R}} C),
    \end{aligned}
\end{equation}
where $C=\sum_{ij} \sum_{p,q=0}^{2^n-1} \ketbra{q}{p}\otimes V_i^\dagger E_j^\dagger\ketbra{q}{p} V_i E_j$.

The optimal recovery map $\mathcal{R}_{opt}$ can be obtained by solving the following semi-definite programming (SDP) problem:
\begin{equation}\label{eq:SDP}
    \begin{aligned}
        \text{maximize} \quad &\frac{1}{2^{2k}} \tr(\chi_{\mathcal{R}} C)\\
        \text{subject to} \quad &\chi_{\mathcal{R}}\geq 0,\\
        &\tr_2{\chi_{\mathcal{R}}}=I.
    \end{aligned}
\end{equation}

\section{Three-qubit and Five-qubit Heuristic VGQEC Codes}\label{ap:rep_513}
We modify the three-qubit repetition code and the five-qubit $[[5,1,3]]$ code, getting two VGQEC codes with parameters $\alpha$ and $\{\alpha_i\}_{i=1,\dots,5}$, respectively. In this section, we first introduce the encoding maps of the three-qubit and five-qubit VGQEC codes, in Sec.~\ref{ap:sub:3} and Sec.~\ref{ap:sub:5}, respectively. Then we present the noise model, which is employed to evaluate the VGQEC codes in Sec.~\ref{ap:sub:noise}. The details of the numerical evolution of these VGQEC codes are provided in Sec.~\ref{ap:sub:sdpresults}.
Finally, we present the QVector implementation for comparison in Sec.~\ref{ap:sub:qvector}.

\subsection{Three-qubit VGQEC Code}\label{ap:sub:3}
The three-qubit VGQEC code is constructed by introducing a parameter $\alpha$ to the Quon graph of the three-qubit repetition code:
\begin{equation}\label{ap:eq:3qubit}
    \begin{tikzpicture}
        \begin{scope}
            %\draw (0,0) circle(1);
            \node[circle,draw=blue,minimum size=15] (1) at(-1,0) {2};
            %\draw[->,thick,blue] (-1.3,0)--(-1.3,0.1);
            \node[circle,draw=blue,minimum size=15] (2) at(0.5,0.866) {1};
            %\draw[->,thick,blue] (0.8,0.866)--(0.8,0.766);
            \node[circle,draw=blue,minimum size=15] (3) at(0.5,-0.866) {3};
            %\draw[->,thick,blue] (0.8,-0.866)--(0.8,-0.966);
            \draw [-,black] (1) to [out=90,in=150] (2);    
            \draw [-,black] (1) to [out=300,in=180] (3);
            \draw [-,black] (2) to [out=330,in=30] (3);
            
            \draw [-,black] (1) to [out=270,in=210] (3);
            \draw (1.125,0) circle (1.6pt);
            %\draw (-0.55,0.975) circle (1.6pt);
            %\draw (-0.55,-0.975) circle (1.6pt);
            %\draw (-0.4,-0.725) circle (1.6pt);
            %\draw (0.4,-0.15) circle (1.6pt);
            %\draw (-0.3,0.255) circle (1.6pt);
            \draw (0.55,0.35) circle (1.6pt);
            %\draw [-,black] (1) to [out=60,in=300] (2);
            %\draw [-,black] (2) to [out=180,in=60] (3);
            \draw [-,black] (1) to [out=60,in=195] (0.083,0.216) to [out=15,in=300] (2);
            \draw [-,black] (2) to [out=180,in=105] (0.083,0.216) to [out=285,in=60] (3);
            \node [fill=red,inner sep=1pt,label={[shift={(-0.5,0.15)}]0:$\alpha$}] at (0.083,0.216) {};
            \draw[->,thick,blue] (-1.3,0)--(-1.3,0.1);
            \draw[->,thick,blue] (0.8,0.866)--(0.8,0.766);
            \draw[->,thick,blue] (0.8,-0.866)--(0.8,-0.966);
            %\node [fill=red,inner sep=1pt,label={[shift={(-0.15,0.15)}]0:$\alpha$}] at (0.083,0.216) {};
        \end{scope}
    \end{tikzpicture}.
\end{equation}
When no charges are present, the code graph corresponds to $\sqrt{2}\ket{+}_L$, and when all charges are present, the code graph corresponds to $\sqrt{2}\ket{-}_L$, as shown in Eq.~\eqref{ap:eq:quon_3_code}.

The code graph in Eq.~\eqref{ap:eq:3qubit} is not as easy to transform into a quantum circuit as the five-qubit VGQEC code in Fig.~\ref{fig:513circuit}. To obtain the codewords, we need to use the definition of crossings~\eqref{eq:quon_crossing_var}, and Eq.~\eqref{ap:eq:3qubit} can be represented as:
\begin{equation}
    \begin{tikzpicture}
        \begin{scope}
            %\draw (0,0) circle(1);
            \node[circle,draw=blue,minimum size=15] (1) at(-1,0) {2};
            %\draw[->,thick,blue] (-1.3,0)--(-1.3,0.1);
            \node[circle,draw=blue,minimum size=15] (2) at(0.5,0.866) {1};
            %\draw[->,thick,blue] (0.8,0.866)--(0.8,0.766);
            \node[circle,draw=blue,minimum size=15] (3) at(0.5,-0.866) {3};
            %\draw[->,thick,blue] (0.8,-0.866)--(0.8,-0.966);
            \draw [-,black] (1) to [out=90,in=150] (2);    
            \draw [-,black] (1) to [out=300,in=180] (3);
            \draw [-,black] (2) to [out=330,in=30] (3);
            
            \draw [-,black] (1) to [out=270,in=210] (3);
            \draw (1.125,0) circle (1.6pt);
            %\draw (-0.55,0.975) circle (1.6pt);
            %\draw (-0.55,-0.975) circle (1.6pt);
            %\draw (-0.4,-0.725) circle (1.6pt);
            %\draw (0.4,-0.15) circle (1.6pt);
            %\draw (-0.3,0.255) circle (1.6pt);
            \draw (0.55,0.35) circle (1.6pt);
            %\draw [-,black] (1) to [out=60,in=300] (2);
            %\draw [-,black] (2) to [out=180,in=60] (3);
            \draw [-,black] (1) to [out=60,in=195] (0.083,0.216) to [out=15,in=300] (2);
            \draw [-,black] (2) to [out=180,in=105] (0.083,0.216) to [out=285,in=60] (3);
            \node [fill=red,inner sep=1pt,label={[shift={(-0.5,0.15)}]0:$\alpha$}] at (0.083,0.216) {};
            \draw[->,thick,blue] (-1.3,0)--(-1.3,0.1);
            \draw[->,thick,blue] (0.8,0.866)--(0.8,0.766);
            \draw[->,thick,blue] (0.8,-0.866)--(0.8,-0.966);
            %\node [fill=red,inner sep=1pt,label={[shift={(-0.15,0.15)}]0:$\alpha$}] at (0.083,0.216) {};
        \end{scope}
    \end{tikzpicture}=
    \frac{1+e^{i\alpha}}{2} 
    \begin{tikzpicture}
    \begin{scope}
      %\draw (0,0) circle(1);
      \node[circle,draw=blue,minimum size=15] (1) at(-1,0) {2};
      %\draw[->,thick,blue] (-1.3,0)--(-1.3,0.1);
      \node[circle,draw=blue,minimum size=15] (2) at(0.5,0.866) {1};
      %\draw[->,thick,blue] (0.8,0.866)--(0.8,0.%766);
      \node[circle,draw=blue,minimum size=15] (3) at(0.5,-0.866) {3};
      %\draw[->,thick,blue] (0.8,-0.866)--(0.8,%-0.966);
      \draw [-,black] (1) to [out=90,in=150] (2);
      \draw [-,black] (1) to [out=60,in=180] (2);
      \draw [-,black] (1) to [out=270,in=210] (3);
      \draw [-,black] (1) to [out=300,in=180] (3);
      \draw [-,black] (2) to [out=330,in=30] (3);
      \draw [-,black] (2) to [out=300,in=60] (3);
      %\fill (-0.4,0.725) circle (1.6pt);
      %\draw (-0.4,-0.725) circle (1.6pt);
      %\fill (-0.55,0.975) circle (1.6pt);
      %\draw (-0.55,-0.975) circle (1.6pt);
      \draw (0.825,0) circle (1.6pt);
      \draw (1.125,0) circle (1.6pt);
      \draw[->,thick,blue] (-1.3,0)--(-1.3,0.1);
        \draw[->,thick,blue] (0.8,0.866)--(0.8,0.766);
        \draw[->,thick,blue] (0.8,-0.866)--(0.8,-0.966);
    \end{scope}
    \end{tikzpicture}
+\frac{1-e^{i\alpha}}{2}
\begin{tikzpicture}
    \begin{scope}
      %\draw (0,0) circle(1);
      \node[circle,draw=blue,minimum size=15] (1) at(-1,0) {2};
      %\draw[->,thick,blue] (-1.3,0)--(-1.3,0.1);
      \node[circle,draw=blue,minimum size=15] (2) at(0.5,0.866) {1};
      %\draw[->,thick,blue] (0.8,0.866)--(0.8,0.%766);
      \node[circle,draw=blue,minimum size=15] (3) at(0.5,-0.866) {3};
      %\draw[->,thick,blue] (0.8,-0.866)--(0.8,%-0.966);
      \draw [-,black] (1) to [out=90,in=150] (2);
      \draw [-,black] (1) to [out=60,in=180] (2);
      \draw [-,black] (1) to [out=270,in=210] (3);
      \draw [-,black] (1) to [out=300,in=180] (3);
      \draw [-,black] (2) to [out=330,in=30] (3);
      \draw [-,black] (2) to [out=300,in=60] (3);
      \fill (-0.4,0.725) circle (1.6pt);
      %\draw (-0.4,-0.725) circle (1.6pt);
      %\fill (-0.55,0.975) circle (1.6pt);
      %\draw (-0.55,-0.975) circle (1.6pt);
      \fill (0.825,0) circle (1.6pt);
      \draw (0.825,0.2) circle (1.6pt);
      \draw (1.125,0.2) circle (1.6pt);
      \draw[->,thick,blue] (-1.3,0)--(-1.3,0.1);
    \draw[->,thick,blue] (0.8,0.866)--(0.8,0.766);
    \draw[->,thick,blue] (0.8,-0.866)--(0.8,-0.966);
    \end{scope}
    \end{tikzpicture}.
\end{equation}

Therefore, combining with Eq.~\eqref{ap:eq:quon_3_code} and Eq.~\eqref{eq:quon_pauli}, the logical state $\sqrt{2}\ket{+}_L$ is:
\begin{equation}
    \sqrt{2}\ket{+}_L=\frac{1+e^{i\alpha}}{2}(\ket{000}+\ket{111})+\frac{1-e^{i\alpha}}{2}X_1(\ket{000}+\ket{111})=\frac{1+e^{i\alpha}}{2}(\ket{000}+\ket{111})+\frac{1-e^{i\alpha}}{2}(\ket{100}+\ket{011}).
\end{equation}
Similarly, the logical state $\sqrt{2}\ket{-}_L$ is:
\begin{equation}
    \sqrt{2}\ket{-}_L=\frac{1+e^{i\alpha}}{2}Z_1(\ket{000}+\ket{111})+\frac{1-e^{i\alpha}}{2}Z_1X_1(\ket{000}+\ket{111})=\frac{1+e^{i\alpha}}{2}(\ket{000}-\ket{111})+\frac{1-e^{i\alpha}}{2}(-\ket{100}+\ket{011}).
\end{equation}
And the codewords of the three-qubit VGQEC code are:
\begin{equation}
    \begin{aligned}
        \ket{0}_L&=\frac{1}{\sqrt{2}}(\ket{+}_L+\ket{-}_L)=\frac{1+e^{i\alpha}}{2}\ket{000}+\frac{1-e^{i\alpha}}{2}\ket{011},\\
        \ket{1}_L&=\frac{1}{\sqrt{2}}(\ket{+}_L-\ket{-}_L)=\frac{1+e^{i\alpha}}{2}\ket{111}+\frac{1-e^{i\alpha}}{2}\ket{100}.
    \end{aligned}
\end{equation}

We can define the encoding map $\mathcal{E}$ of the three-qubit VGQEC code as $\mathcal{E}(a\ket{0}+b\ket{1})=a\ket{0}_L+b\ket{1}_L$.

\subsection{Five-qubit VGQEC Code}\label{ap:sub:5}
The five-qubit VGQEC code is constructed by introducing parameters $\{\alpha_i\}_{i=1,\dots,5}$ to the Quon graph of the five-qubit $[[5,1,3]]$ code:
\begin{equation}\label{ap:fig:513}
\begin{tikzpicture}
    \node[circle,draw=blue] (1) at(1.9,0.94) {1};
    \node[circle,draw=blue] (2) at(0.81,0.56) {2};
    \node[circle,draw=blue] (3) at(0.81,-0.56) {3};
    \node[circle,draw=blue] (4) at(1.9,-0.94) {4};
    \node[circle,draw=blue] (5) at(2.6,0) {5};
    \draw[->,thick,blue] (1.9,1.25)--(1.91,1.25);
    \draw[->,thick,blue] (1.91,-1.25)--(1.9,-1.25);
    \draw[->,thick,blue] (2.9,0)--(2.9,-0.01);
    \draw[->,thick,blue] (0.6,0.8)--(0.61,0.81);
    \draw[->,thick,blue] (0.6,-0.8)--(0.59,-0.79);
    \draw [-,black] (1) to (2);
    \draw [-,black] (2) to (3);
    \draw [-,black] (3) to (4);
    \draw [-,black] (4) to (5);
    \draw [-,black] (5) to (1);
    \draw [-,black, name path=13] (1) to (3);
    \draw [-,black, name path=14] (1) to (4);
    \draw [-,black, name path=24] (2) to (4);
    \draw [-,black, name path=25] (2) to (5);
    \draw [-,black, name path=35] (3) to (5);
    \draw ($ (1)!.5!(2) $) circle (1.6pt);
    \draw ($ (2)!.5!(3) $) circle (1.6pt);
    \draw ($ (3)!.5!(4) $) circle (1.6pt);
    \draw ($ (4)!.5!(5) $) circle (1.6pt);
    \draw ($ (5)!.5!(1) $) circle (1.6pt);
    \draw ($ (1)!.5!(3) $) circle (1.6pt);
    \draw ($ (1)!.5!(4) $) circle (1.6pt);
    \draw ($ (2)!.5!(4) $) circle (1.6pt);
    \draw ($ (2)!.5!(5) $) circle (1.6pt);
    \draw ($ (3)!.5!(5) $) circle (1.6pt);
    \path [name intersections={of = 13 and 25,by=1}];
    \node [fill=red,inner sep=1pt,label={[shift={(0.2,0.15)}]190:$\alpha_1$}] at (1) {};
    \path [name intersections={of = 13 and 24,by=2}];
    \node [fill=red,inner sep=1pt,label={[shift={(0,0.1)}]270:$\alpha_2$}] at (2) {};
    \path [name intersections={of = 35 and 24,by=3}];
    \node [fill=red,inner sep=1pt,label={[shift={(-0.15,0.1)}]342:$\alpha_3$}] at (3) {};
    \path [name intersections={of = 35 and 14,by=4}];
    \node [fill=red,inner sep=1pt,label={[shift={(-0.15,-0.15)}]85:$\alpha_4$}] at (4) {};
    \path [name intersections={of = 14 and 25,by=5}];
    \node [fill=red,inner sep=1pt,label={[shift={(0.2,-0.1)}]95:$\alpha_5$}] at (5) {};
    %\draw [name intersections={of=upward line and sloped line, by=x}]
    \end{tikzpicture}
\end{equation}
When no charges are present, the code graph corresponds to $\sqrt{2}\ket{0}_L$, and when all charges are present, the code graph corresponds to $\sqrt{2}\ket{1}_L$, as shown in Eq.~\eqref{ap:eq:quon_5_code}.

\begin{figure}[hbp]
    \includegraphics[width=0.5\textwidth]{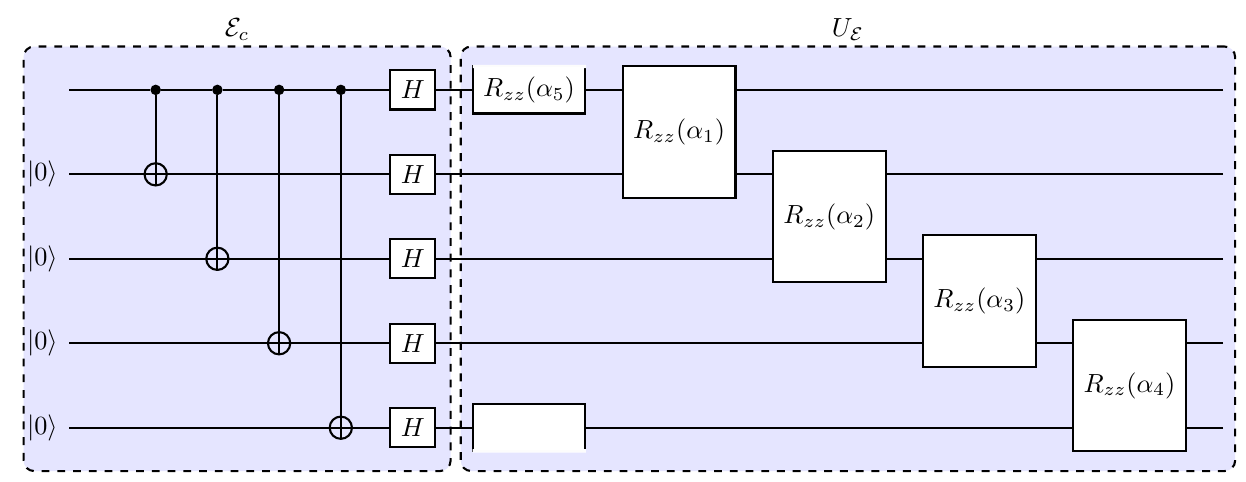}
    \caption{The encoding circuit for an optional five-qubit VGQEC code: The encoding map can be divided into fixed map $\mathcal{E}_c$ and variational quantum circuit $U_\mathcal{E}$. The Quon graph of the VGQEC code is shown in Fig.~\ref{ap:fig:513}.}\label{fig:513circuit}
    \end{figure}

\begin{figure}[hbp]
\includegraphics{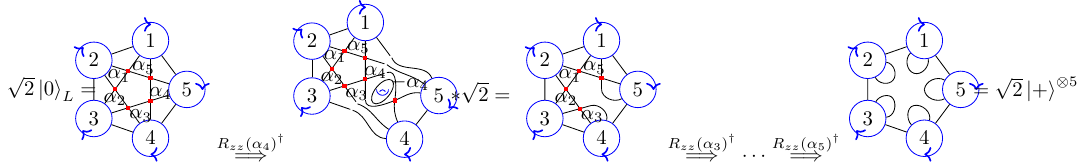}
\caption{The state $ \ket{+}^{\otimes 5} $ is prepared by applying the inverse of the unitary operation $ U_\mathcal{E} $, as depicted in circuit Fig.~\ref{fig:513circuit}, to the logical state $ \ket{0}_L $. In this figure, the global phases are omitted.
The first arrow represents the applying of $ R_{zz}(\alpha_4)^\dagger $ to the logical state $ \ket{0}_L $, which is to glue the graph of $R_{zz}$ to physical qubits 4 and 5.
The gluing operation creates a hole~(shown in the blue graphic), which is contained in a circle. The subgraph of hole in circle is equivalent to $\frac{1}{\sqrt{2}}$, by the property in Eq.~\eqref{eq:quon_genus}. And the factor $\sqrt{2}$ comes from the graph of $R_{zz}$ gate. The first equation is given by replacing the hole with the factor $\frac{1}{\sqrt{2}}$, as well as canceling the two crossings with parameters $\alpha_4$ and $-\alpha_4$.
}
\label{fig:encode_c}
\end{figure}

We claim that the encoding circuit of the VGQEC code is Fig.~\ref{fig:513circuit}. This is because $R_X$ and $R_{ZZ}$ have graphical representations shown in Eq.~\eqref{eq:quon_R}.
Notice that if the inverse of $U_\mathcal{E}$ in Fig.~\ref{fig:513circuit} is applied to the logical state $\ket{0}_L$ of the VGQEC code, it would result in state $\ket{+}^{\otimes 5}$, shown in Fig.~\ref{fig:encode_c}.
Similarly, the result of the inverse of $U_\mathcal{E}$ applying to the logical state $\ket{1}_L$ of the VGQEC code is $\ket{-}^{\otimes 5}$.
And the encoding map of $\ket{\pm}^{\otimes 5}$ is exactly $\mathcal{E}_c$ in Fig.~\ref{fig:513circuit}. Thus we have verified that the encoding circuit of the VGQEC code is Fig.~\ref{fig:513circuit}.

\subsection{Noise Models for Evaluation}\label{ap:sub:noise}
To evaluate the performance of the VGQEC codes, we simulate the performance of the VGQEC code over two specific noise models.

For the three-qubit VGQEC code, we consider that the amplitude damping error occurs on each qubit with the same intensity. The noise channel is expressed as a tensor product of single-qubit amplitude damping channel $\mathcal{N}={ \mathcal{N}^{ad}_s }^ {\otimes 3}$, where $\mathcal{N}^{ad}_s$ is defined as:
\begin{equation}\label{eq:dampingchannel}
\begin{aligned}
    \mathcal{N}^{ad}_s(\rho)&=\sum_{k=0,1}E_k\rho E_k^\dagger,\\
    E_0&=\left[
  \begin{array}{cc}
    1 & 0 \\
    0 & \sqrt{1-\lambda}
  \end{array}
\right]\quad
E_1=\left[
  \begin{array}{cc}
    0 & \sqrt{\lambda} \\
    0 & 0
  \end{array}
\right] ,
\end{aligned}
\end{equation}
with $\lambda\in[0,1]$.

For the case of the five-qubit VGQEC code, the noise is set as a varying noise channel, which is controlled by the parameter $\eta$.
The noise is a composite channel of two parts $\mathcal{N}=\mathcal{N}_2\circ\mathcal{N}^\eta_1$, a variable part $\mathcal{N}^\eta_1$ and fixed part $\mathcal{N}_2$.
The variable noise channel $\mathcal{N}^\eta_1$ is defined as a tensor ${ \mathcal{N}^{\eta}_s }^ {\otimes 5}$ of a single-qubit noise channel:
\begin{equation}
    \mathcal{N}_{s}^\eta(\rho)=0.002 (Z\rho Z+\eta X\rho X +\eta Y\rho Y)+(0.998-0.004\eta)\rho.
\end{equation}
This is a linear interpolation from the dephasing channel to the depolarization channel.
The fixed noise channel $\mathcal{N}_2$ is modeled as a two-qubit correlated Pauli-X error model. Specifically, we apply quantum channel:
\begin{equation}
    \mathcal{N}_{2,i}(\rho)=(1-p_{xx})\rho+p_{xx}X_iX_{i+1}\rho X_iX_{i+1}
\end{equation}
for pairs of nearby qubits, where $i\in\{1,2,3,4,5\}$ and $p_{xx}=0.001$.

\subsection{Performance Evaluation of VGQEC Codes}\label{ap:sub:sdpresults}
For both the three-qubit VGQEC code and the five-qubit VGQEC code, when the parameters $\alpha$ or $\{\alpha_i\}_{i=1,\dots,5}$ are given, the encoding map $\mathcal{E}$ of the VGQEC codes are determined. At each intensity of noise for the three-qubit case and each moment of noise evolution for the five-qubit case, we calculate the optimal channel fidelity of the code:
\begin{equation}\label{ap:eq:optimalfidelity}
    F^{opt}_C\coloneqq\underset{\mathcal{R}}{\max}~F_C(\mathcal{R}\circ\mathcal{N}\circ\mathcal{E})=F_C(\mathcal{R}_{opt}\circ\mathcal{N}\circ\mathcal{E}),
\end{equation}
where $\mathcal{R}_{opt}$ is the optimal recovery map that achieves the maximum channel fidelity $F^{opt}_C$. The optimal recovery map can be efficiently determined by solving a semi-definite programming (SDP) problem~\cite{fletcher2007optimum}.

We take the optimal channel fidelity $F^{opt}_C$ as the object function and input it to the \textit{L-BFGS}~\cite{liu1989limited} optimizer.
The classical optimizer optimizes the parameters and improves the optimal channel fidelity, ultimately yielding the trained VGQEC code.
In our numerical results, for each $\lambda$ in the three-qubit case, the parameters are initialized by $\alpha=0$ with a small random perturbation, starting from the three-qubit repetition code.
The perturbation is applied because $\alpha=0$ (which corresponds to the repetition code) is a saddle point of the optimization problem, and the perturbation helps the optimizer escape from this saddle point.
In the five-qubit case, for each value of $\eta$, the parameters are initialized with random values.
For a fair comparison, we also use the optimal recovery map obtained by SDP to decode the three-qubit repetition code, the five-qubit repetition code, and the $[[5,1,3]]$ code. The optimal channel fidelity for these codes is then calculated.

For the three-qubit case, by numerical results, the optimal parameter is found to be $\alpha=-0.5\pi+k\pi$, where $k\in\mathbb{Z}$. For convenience, we choose $\alpha=-0.5\pi$, and the corresponding codewords are:
\begin{equation}
    \begin{aligned}
        \ket{0}_L&=\frac{1+e^{i\alpha}}{2}\ket{000}+\frac{1-e^{i\alpha}}{2}\ket{011}=\frac{1-i}{2}(\ket{000}+i\ket{011}),\\
        \ket{1}_L&=\frac{1+e^{i\alpha}}{2}\ket{111}+\frac{1-e^{i\alpha}}{2}\ket{100}=\frac{1-i}{2}(i\ket{100}+\ket{111}),
    \end{aligned}
\end{equation}
where the global phase can be omitted.

\subsection{Comparison with other Numerical methods}\label{ap:sub:qvector}
As a comparison, we also use the QVector method~\cite{johnson2017qvector} to protect information against noise.
In original QVector method, both the encoding map and the recovery map are constructed by variational quantum circuits, and the parameters are randomly initialized and then optimized to maximize the channel fidelity. 

For a fair comparison, we discard the original recovery map in the QVector method and instead use the optimal recovery map obtained via SDP for decoding. 
The optimal channel fidelity is taken as the objective function, and the parameters are initialized randomly before being optimized using the \textit{L-BFGS} optimizer.

The encoding circuit is shown in Fig.~\ref{ap:fig:qvector}. For both the three-qubit and five-qubit cases, the repetition number $L$ is set to 2.
In three-qubit case and five-qubit case, there are a total of $12$ and $20$ parameters to be optimized, respectively.

\begin{figure}[hbp]
    \begin{tikzpicture}
        \node[scale=0.7] {
        \begin{quantikz}[row sep=0.3cm,column sep=0.5cm]
        &\targX[draw=red]{0} \gategroup[3,steps=5,style={dashed,rounded corners,fill=blue!5, inner xsep=2pt},background]{$\sc \times L$} &\targX[draw=blue]{0}&\ctrl{1}&\qw&\ctrl{2}&\qw\\
        &\targX[draw=red]{0}&\targX[draw=blue]{0}&\control{0}&\ctrl{1}&\qw&\qw\\
        &\targX[draw=red]{0}&\targX[draw=blue]{0}&\qw&\control{0}&\control{0}&\qw
        \end{quantikz}};
    \end{tikzpicture}
    \caption{The variational quantum circuit used in QVector method: We apply $R_X$-$R_Z$ rotations to each qubit, $CZ$ gates between all pairs of qubits and repeat the previous operation $L=2$ times.
    For convenience, the double qubit gate drawn in black represents $CZ$, the single-qubit gate drawn in the red circle is $R_X$ and the single qubit gate drawn in the blue circle is $R_Z$. As an illustration, we present the three-qubit case.}\label{ap:fig:qvector}
\end{figure}

\section{Construct VGQEC Codes using Variational Quantum Circuits}\label{sec:practical_scheme}

As demonstrated in Sec.~\ref{ap:rep_513}, a code can be represented in a Quon graph, thereby constructing a VGQEC code by introducing parameters to the graph.
For the five-qubit example in Sec.~\ref{ap:rep_513}, the encoding map $\mathcal{E}$ is derived directly from the Quon graph. 
However, transforming a complex Quon graph into a quantum circuit is not always an easy task.

Alternatively, for a given code, besides pre-embedding parameters on the Quon graph of the code to get VGQEC, we can also attach a variable graphical structure to the physical qubit discs in the Quon graph to construct VGQEC code.
Assuming that the encoding map of the original code is $\mathcal{E}_c$, and the attached graphical structure can be compiled into a variational quantum circuit $U_\mathcal{E}$. Then the encoding map of the VGQEC code can be represented as the combination $\mathcal{E}={U}_\mathcal{E}\circ \mathcal{E}_c$.

On the other side, we hope to construct a recovery map of the VGQEC code, which holds the ability to implement the recovery map of the original code and can be further optimized for various noise models.
The recovery map $\mathcal{R}$ of the VGQEC code is constructed by appending $2k$ auxiliary qubits and applying a variational quantum circuit $U_\mathcal{R}$ to the system, followed by a measurement of the auxiliary qubits and the original recovery map $\mathcal{R}_c$. The reason for introducing auxiliary qubits is discussed in Appendix~\ref{ap:aux_qubit}.
The whole error correction scheme is shown in Fig.~\ref{fig:QECC_deformed}.

\begin{figure}[htbp]
\begin{quantikz}
%\lstick{$\rho_{in}$} 
&\gate{\mathcal{E}_c}\qwbundle{k}
\gategroup[1,steps=2,style={dashed,rounded corners,fill=blue!20, inner xsep=2pt},background]{$\sc \mathcal{E}$} 
&\gate{U_\mathcal{E}}\qwbundle{n}
&\gate[1]{\mathcal{N}}\qwbundle{n}
&\qw\gategroup[2,steps=4,style={dashed,rounded corners,fill=blue!20, inner xsep=2pt},background]{$\sc \mathcal{R}$} 
&\qw\qwbundle{n}
&\gate[2]{U_\mathcal{R}}
& \gate{\mathcal{R}_c}\qwbundle{n}
%&\rstick{$\rho_{out}$} 
&\qw\qwbundle{k}\\
 & &  & & & &\lstick{$\ket{0}^{\otimes 2k}$}
  &  \meter [draw=blue]{}\qwbundle{\hspace{-0.4em}2k}
\end{quantikz}
\caption{The encoding map $\mathcal{E}:\mathcal{H}^{\otimes k} \rightarrow \mathcal{H}^{\otimes n}$ is the composition of the original encoding map $\mathcal{E}_c$ and a variational quantum circuit $U_\mathcal{E}$. The recovery map $\mathcal{R}:\mathcal{H}^{\otimes n} \rightarrow \mathcal{H}^{\otimes k}$ is made by introducing $2k$ auxiliary qubits, then acting a variational quantum circuit $U_\mathcal{R}$, after which the auxiliary qubits are traced out and finally the original recovery map $\mathcal{R}_c$ is acted.}\label{fig:QECC_deformed}
\end{figure}

We investigate the performance of three-qubit and five-qubit VGQEC codes modified from the three-qubit repetition code and the $[[5,1,3]]$ code, respectively.
The VGQEC codes in the simulations are constructed by gluing a specially designed structure to the physical qubit discs in the Quon graphs. At this point, the encoding maps $\mathcal{E}$ have a decomposition $\mathcal{E}={U}_\mathcal{E}\circ \mathcal{E}_c$, where $\mathcal{E}_c$ are the fixed encoding maps of the original codes as shown in Fig.~\ref{fig:fixencode} and $U_\mathcal{E}$ are variational quantum circuits.

\begin{figure}[tbp]
\subfigure[]{\begin{tikzpicture}
\node[scale=0.5] {
\begin{quantikz}[row sep=0.5cm,column sep=2.5cm]
     & \ctrl{1} & \ctrl{2} & \qw \\ 
\lstick{$\ket{0}$} & \targ{} & \qw& \qw \\
\lstick{$\ket{0}$}  & \qw & \targ{} &\qw \\
\end{quantikz}};
\end{tikzpicture}}
\subfigure[]{\begin{tikzpicture}
\node[scale=0.5] {
\begin{quantikz}
& \qw & \ctrl{1} & \ctrl{2} & \ctrl{3} & \ctrl{4} & \gate{H} &\gate[]{ R_{zz}(-\frac{\pi}{2})}
&\gate[2]{ R_{zz}(-\frac{\pi}{2})}
&\qw&\qw&\qw&\qw\\ 
\lstick{$\ket{0}$} & \qw & \targ{} & \qw& \qw & \qw & \gate{H} & \qw &\qw&\gate[2]{ R_{zz}(-\frac{\pi}{2})}&\qw&\qw&\qw\\
\lstick{$\ket{0}$} & \qw & \qw & \targ{} &\qw & \qw & \gate{H}&\qw &\qw&\qw&\gate[2]{ R_{zz}(-\frac{\pi}{2})}&\qw&\qw\\
\lstick{$\ket{0}$} & \qw & \qw & \qw & \targ{} & \qw & \gate{H} &\qw&\qw &\qw&\qw&\gate[2]{ R_{zz}(-\frac{\pi}{2})}&\qw\\
\lstick{$\ket{0}$} & \qw & \qw & \qw & \qw & \targ{} & \gate{H} &\gate{ R_{zz}(-\frac{\pi}{2})}&\qw &\qw&\qw&\qw&\qw
\end{quantikz}};
\draw[fill=white, draw=none] (-1.08,1.27) rectangle (-0.299,1.3);
    \draw[fill=white, draw=none] (-1.07,-1) rectangle (-0.305,-1.3);
\end{tikzpicture}}
\caption{The fixed part $\mathcal{E}_c$ in encoding map of the VGQEC codes: (a) For the three-qubit VGQEC code modified from repetition codes, the fixed part is the original encoding maps. The figure shows the circuit for the three-qubit case.
(b) For the five-qubit VGQEC code, it is modified from $[[5,1,3]]$ codes, the fixed part is the encoding map of the $[[5,1,3]]$ code.}\label{fig:fixencode}
\end{figure}

In the above two VGQEC codes, the structure of variational quantum circuits $U_\mathcal{E}$ in the encoding map is inspired by a specially designed Quon graph $\mathcal{G}_\mathcal{E}$. For the case of $n$ qubits, $\mathcal{G}_\mathcal{E}$ has $4n$ strings and $n(2n-1)$ braids crossings with parameters. 
Specifically, the graph features $ 2n $ external strings surrounding it, while the remaining $ 2n $ internal strings intertwine within the graph itself.
If we use $i\rightarrow j$ to indicate that the $i-th$ internal string at the top will reach the $j-th$ position at the bottom in the graph.
Then the linkage relationship of the strings in the graph is $1\rightarrow 2n$, $2\rightarrow 2n-1$, ..., $n \rightarrow n+1$. 
There is an example for $n=3$, shown in Fig.~\ref{fig:matchgate}.

\begin{figure}[htbp]
\begin{tikzpicture}[scale=0.5]
\draw[-] (0,0) -- (0,-1);
\draw[-] (1,0) -- (1,-3);
\draw[-] (2,0) -- (2,-5);
\draw[-] (3,0) -- (3,-4.5);
\draw[-] (4,0) -- (4,-4.5);
\draw[-] (5,0) -- (5,-4.5);
\draw[-] (0,-1) -- (5,-6);
\draw[-] (5,-4.5) -- (0,-10);
\draw[-] (5,-6) -- (5,-11);
\draw[-] (2,-5.5) -- (3,-4.5);
\draw[-] (1,-7.5) -- (4,-4.5);
\draw[-] (0,-11) -- (0,-10);
\draw[-] (4,-6) -- (1,-3);
\draw[-] (4,-6) -- (4,-11);
\draw[-] (3,-6) -- (2,-5);
\draw[-] (3,-6) -- (3,-11);
\draw[-] (1,-7.5) -- (1,-11);
\draw[-] (2,-5.5) -- (2,-11);
%\draw [name intersections={of=upward line and sloped line, by=x}];
\fill (1,-2) circle (3pt);
\fill (2,-3) circle (3pt);
\fill (3,-4) circle (3pt);
\fill (3.75,-4.75) circle (3pt);
\fill (4.3,-5.3) circle (3pt);
\fill (2,-4) circle (3pt);
\fill (2.75,-4.75) circle (3pt);
\fill (2.25,-5.25) circle (3pt);
\fill (3.25,-5.25) circle (3pt);
\fill (3,-6.75) circle (3pt);
\fill (2.75,-5.75) circle (3pt);
\fill (3.8,-5.8) circle (3pt);
\fill (2,-6.5) circle (3pt);
\fill (2,-7.8) circle (3pt);
\fill (1,-8.9) circle (3pt);
%\fill (0.75,0) circle (1.5pt);
\draw[-,blue] (-0.5,0) -- (-0.5,-11);
\draw[-,blue] (5.5,0) -- (5.5,-11);
\clip (-0.5,0) rectangle (5.5,-11);
\draw[blue] (1.5,0) circle(0.25);
\draw[blue] (3.5,0) circle(0.25);
\draw[blue] (3.5,-11) circle(0.25);
\draw[blue] (1.5,-11) circle(0.25);
\end{tikzpicture}
    \caption{``Universal" graph $\mathcal{G}_\mathcal{E}$ for $n=3$: every point in the figure represents a braid crossing with variable. The black and blue strings in the figure represent internal and external strings, respectively, with internal strings intersecting pairwise.}
    \label{fig:matchgate}
\end{figure}

The reason for considering $\mathcal{G}_\mathcal{E}$ is that, the internal $2n$ strings intersect pairwise, offering great structural flexibility when adjusting parameters.
This diagram gives a ``universal" graph for the internal $2n$ strings. 
This means that for any geometry structure of the internal strings, we can use the $\mathcal{G}_\mathcal{E}$ to represent it by setting property parameters.
Firstly, any geometry structure of the internal strings can be decomposed into a series of braid crossings on adjacent strings.
Then the universal property of $\mathcal{G}_\mathcal{E}$ can be illustrated by using \textit{Yang-Baxter} equation \cite{yang1967some,baxter2016exactly} (Fig.~\ref{fig:YBE}) to absorb any braid crossings into $\mathcal{G}_\mathcal{E}$.
Suppose there is a braid crossing on the $2$-nd and $3$-rd positions, then we can use the Yang-Baxter equation to absorb them into $\mathcal{G}_\mathcal{E}$, as shown in Fig.~\ref{fig:universal}.
The additional benefit of $\mathcal{G}_\mathcal{E}$ is its ease of construction through a variational quantum circuit, consisting of a sequence of $R_{ZZ}$ and $R_X$ gates.
This facilitates straightforward implementation on actual quantum hardware.

\begin{figure}[htbp]
\begin{tikzpicture}
\begin{scope}
\draw[-] (0,0) -- (2,-2);
\draw[-] (2,0) -- (0,-2);
\draw[-] (0.5,0) -- (0.5,-2);
%\node at(0.5,-1) {$\alpha$};
\fill (1,-1) circle (1.5pt);
\fill (0.5,-0.5) circle (1.5pt);
\fill (0.5,-1.5) circle (1.5pt);
\end{scope}
\node at(2.5,-1) {$=$};
\begin{scope}[shift={(3,0)}]
\draw[-] (0,0) -- (2,-2);
\draw[-] (2,0) -- (0,-2);
\draw[-] (1.5,0) -- (1.5,-2);
%\node at(0.5,-1) {$\alpha$};
\fill[red] (1,-1) circle (1.5pt);
\fill[red] (1.5,-0.5) circle (1.5pt);
\fill[red] (1.5,-1.5) circle (1.5pt);
\end{scope}
\end{tikzpicture}
    \caption{Graphical interpretation of the Yang-Baxter equation: The black dots in the figure indicate the braid crossings with parameters. We can move the vertical string across the middle crossing to the right side. Simultaneously change the three crossing parameters, with red dots indicating the new parameters.}
    \label{fig:YBE}
\end{figure}

\begin{figure}[htbp]
\begin{tikzpicture}[scale=0.7]
\begin{scope}[scale=0.5]
\draw[-] (0,0) -- (0,-1);
\draw[-] (1,0) -- (1,-3);
\draw[-] (2,0) -- (2,-5);
\draw[-] (3,0) -- (3,-4.5);
\draw[-] (4,0) -- (4,-4.5);
\draw[-] (5,0) -- (5,-4.5);
\draw[-] (0,-1) -- (5,-6);
\draw[-] (5,-4.5) -- (0,-10);
\draw[-] (5,-6) -- (5,-11);
\draw[-] (2,-5.5) -- (3,-4.5);
\draw[-] (1,-7.5) -- (4,-4.5);
\draw[-] (0,-11) -- (0,-10);
\draw[-] (4,-6) -- (1,-3);
\draw[-] (4,-6) -- (4,-11);
\draw[-] (3,-6) -- (2,-5);
\draw[-] (3,-6) -- (3,-11);
\draw[-] (1,-7.5) -- (1,-9);
\draw[-] (2,-5.5) -- (2,-9);
\draw[-] (2,-9) -- (1,-11);
\draw[-] (1,-9) -- (2,-11);
\fill (1.5,-10) circle (3pt);
%\draw [name intersections={of=upward line and sloped line, by=x}];
\fill (1,-2) circle (3pt);
\fill (2,-3) circle (3pt);
\fill (3,-4) circle (3pt);
\fill (3.75,-4.75) circle (3pt);
\fill (4.3,-5.3) circle (3pt);
\fill (2,-4) circle (3pt);
\fill (2.75,-4.75) circle (3pt);
\fill (2.25,-5.25) circle (3pt);
\fill (3.25,-5.25) circle (3pt);
\fill (3,-6.75) circle (3pt);
\fill (2.75,-5.75) circle (3pt);
\fill (3.8,-5.8) circle (3pt);
\fill (2,-6.5) circle (3pt);
\fill (2,-7.8) circle (3pt);
\fill (1,-8.9) circle (3pt);
%\fill (0.75,0) circle (1.5pt);
\draw[-,blue] (-0.5,0) -- (-0.5,-11);
\draw[-,blue] (5.5,0) -- (5.5,-11);
\clip (-0.5,0) rectangle (5.5,-11);
\draw[blue] (1.5,0) circle(0.25);
\draw[blue] (3.5,0) circle(0.25);
\draw[blue] (3.5,-11) circle(0.25);
\draw[blue] (1.5,-11) circle(0.25);
\end{scope}
\node at(3.5,-3) {$\xrightarrow[\text{Equation}]{\text{Yang-Baxter}}$};
\begin{scope}[scale=0.5,shift={(9.5,0)}]
\draw[-] (0,0) -- (0,-1);
\draw[-] (1,0) -- (1,-3);
\draw[-] (2,0) -- (2,-5);
\draw[-] (3,0) -- (3,-4.5);
\draw[-] (4,0) -- (4,-4.5);
\draw[-] (5,0) -- (5,-4.5);
\draw[-] (0,-1) -- (5,-6);
\draw[-] (5,-4.5) -- (0,-10);
\draw[-] (5,-6) -- (5,-11);
\draw[-] (2,-5.5) -- (3,-4.5);
\draw[-] (0,-11) -- (0,-10);
\draw[-] (4,-6) -- (1,-3);
\draw[-] (4,-6) -- (4,-11);
\draw[-] (3,-6) -- (2,-5);
\draw[-] (3,-6) -- (3,-11);
\draw[-] (1.5,-7) -- (4,-4.5);
\draw[-] (1.5,-7) -- (2,-7.8);
\draw[-] (2,-5.5) -- (2,-7);
\draw[-] (2,-7) -- (1,-8.9);
\draw[-] (1,-8.9) -- (1,-11);
\draw[-] (2,-7.8) -- (2,-11);
%\draw [name intersections={of=upward line and sloped line, by=x}];
\fill (1,-2) circle (3pt);
\fill (2,-3) circle (3pt);
\fill (3,-4) circle (3pt);
\fill (3.75,-4.75) circle (3pt);
\fill (4.3,-5.3) circle (3pt);
\fill (2,-4) circle (3pt);
\fill (2.75,-4.75) circle (3pt);
\fill (2.25,-5.25) circle (3pt);
\fill (3.25,-5.25) circle (3pt);
\fill (3,-6.75) circle (3pt);
\fill (2.75,-5.75) circle (3pt);
\fill (3.8,-5.8) circle (3pt);
\fill (2,-6.5) circle (3pt);
\fill[red] (2,-7.8) circle (3pt);
\fill[red] (1,-8.9) circle (3pt);
\fill[red] (1.75,-7.4) circle (3pt);
%\fill (0.75,0) circle (1.5pt);
\draw[-,blue] (-0.5,0) -- (-0.5,-11);
\draw[-,blue] (5.5,0) -- (5.5,-11);
\clip (-0.5,0) rectangle (5.5,-11);
\draw[blue] (1.5,0) circle(0.25);
\draw[blue] (3.5,0) circle(0.25);
\draw[blue] (3.5,-11) circle(0.25);
\draw[blue] (1.5,-11) circle(0.25);
\end{scope}
\node at(8,-3) {$\longrightarrow$};
\begin{scope}[scale=0.25,shift={(31,-9)}]
\draw[-] (0,0) -- (1,-1);
\draw[-] (1,-1) -- (0,-2);
\draw[-] (1,0) -- (0,-1);
\draw[-] (0,-1) -- (1,-2);
\fill (0.5,-0.5) circle (6pt);
\fill (0.5,-1.5) circle (6pt);
\node at(1.5,-1) {$=$};
\draw[-] (2,0) -- (3,-2);
\draw[-] (3,0) -- (2,-2);
\fill[red] (2.5,-1) circle (6pt);
\end{scope}
\begin{scope}[scale=0.5,shift={(18,0)}]
\draw[-] (0,0) -- (0,-1);
\draw[-] (1,0) -- (1,-3);
\draw[-] (2,0) -- (2,-5);
\draw[-] (3,0) -- (3,-4.5);
\draw[-] (4,0) -- (4,-4.5);
\draw[-] (5,0) -- (5,-4.5);
\draw[-] (0,-1) -- (5,-6);
\draw[-] (5,-4.5) -- (0,-10);
\draw[-] (5,-6) -- (5,-11);
\draw[-] (2,-5.5) -- (3,-4.5);
\draw[-] (1,-7.5) -- (4,-4.5);
\draw[-] (0,-11) -- (0,-10);
\draw[-] (4,-6) -- (1,-3);
\draw[-] (4,-6) -- (4,-11);
\draw[-] (3,-6) -- (2,-5);
\draw[-] (3,-6) -- (3,-11);
\draw[-] (1,-7.5) -- (1,-11);
\draw[-] (2,-5.5) -- (2,-11);
%\draw [name intersections={of=upward line and sloped line, by=x}];
\fill (1,-2) circle (3pt);
\fill (2,-3) circle (3pt);
\fill (3,-4) circle (3pt);
\fill (3.75,-4.75) circle (3pt);
\fill (4.3,-5.3) circle (3pt);
\fill (2,-4) circle (3pt);
\fill (2.75,-4.75) circle (3pt);
\fill (2.25,-5.25) circle (3pt);
\fill (3.25,-5.25) circle (3pt);
\fill (3,-6.75) circle (3pt);
\fill (2.75,-5.75) circle (3pt);
\fill (3.8,-5.8) circle (3pt);
\fill[red] (2,-6.5) circle (3pt);
\fill[red] (2,-7.8) circle (3pt);
\fill[red] (1,-8.9) circle (3pt);
%\fill (0.75,0) circle (1.5pt);
\draw[-,blue] (-0.5,0) -- (-0.5,-11);
\draw[-,blue] (5.5,0) -- (5.5,-11);
\clip (-0.5,0) rectangle (5.5,-11);
\draw[blue] (1.5,0) circle(0.25);
\draw[blue] (3.5,0) circle(0.25);
\draw[blue] (3.5,-11) circle(0.25);
\draw[blue] (1.5,-11) circle(0.25);
\end{scope}
\end{tikzpicture}
    \caption{The graph represents the process of  absorbing $R_{Z_1Z_2}$ into $\mathcal{G}_\mathcal{E}$. }
    \label{fig:universal}
\end{figure}

\begin{figure}[htbp]
    \begin{tikzpicture}[scale=0.5]
    \draw[-] (0,0) -- (0,-5);
    \draw[-] (1,0) -- (1,-2);
    \draw[-] (2,0) -- (2,-5);
    \draw[-] (3,0) -- (3,-5);
    \draw[-] (4,0) -- (4,-2);
    \draw[-] (5,0) -- (5,-5);
    \draw[-] (0,-5) -- (5,-10);
    \draw[-] (5,-5) -- (0,-10);
    \draw[-] (5,-10) -- (5,-11);
    \draw[-] (2,-6) -- (3,-5);
    \draw[-] (1,-5) -- (4,-2);
    \draw[-] (0,-11) -- (0,-10);
    \draw[-] (4,-5) -- (1,-2);
    \draw[-] (4,-5) -- (4,-11);
    \draw[-] (3,-6) -- (2,-5);
    \draw[-] (3,-6) -- (3,-11);
    \draw[-] (1,-5) -- (1,-11);
    \draw[-] (2,-6) -- (2,-11);
    %\draw [name intersections={of=upward line and sloped line, by=x}];
    \fill[red] (1,-6) circle (3pt);
    \fill[red] (2,-3) circle (3pt);
    \fill[red] (3,-4) circle (3pt);
    \fill[red] (4,-9) circle (3pt);
    \fill[red] (4,-6) circle (3pt);
    \fill[red] (2,-4) circle (3pt);
    \fill[red] (2.5,-5.5) circle (3pt);
    \fill[red] (2.5,-3.5) circle (3pt);
    \fill[red] (3,-3) circle (3pt);
    \fill[red] (3,-7) circle (3pt);
    \fill[red] (2.5,-7.5) circle (3pt);
    \fill[red] (3,-8) circle (3pt);
    \fill[red] (2,-7) circle (3pt);
    \fill[red] (2,-8) circle (3pt);
    \fill[red] (1,-9) circle (3pt);
    %\fill (0.75,0) circle (1.5pt);
    \draw[-,blue] (-0.5,0) -- (-0.5,-11);
    \draw[-,blue] (5.5,0) -- (5.5,-11);
    \clip (-0.5,0) rectangle (5.5,-11);
    \draw[blue] (1.5,0) circle(0.25);
    \draw[blue] (3.5,0) circle(0.25);
    \draw[blue] (3.5,-11) circle(0.25);
    \draw[blue] (1.5,-11) circle(0.25);
    \end{tikzpicture}
    \caption{Left-right symmetric form of $\mathcal{G}_\mathcal{E}$ for $n=3$: reshaped from Fig.~\ref{fig:matchgate} by using the Yang-Baxter equation Fig.~\ref{fig:YBE}.}
    \label{fig:matchgate2}
    \end{figure}

As discussed above, the graph $\mathcal{G}_\mathcal{E}$ with parameters constructs a ``universal" set. However, the lack of symmetry in this circuit may present challenges when optimizing the parameters of the variational quantum circuit. To address this issue, we utilize the Yang-Baxter equation to transform the graph into a left-right symmetric form, as illustrated in Fig.~\ref{fig:matchgate2}. 
Using Eq.~\ref{eq:quon_R}, we can also transform the left-right symmetric form of $\mathcal{G}_\mathcal{E}$ into a variational quantum circuit consisting of $R_{ZZ}$ and $R_X$ gates.

\begin{figure}[tbp]
\subfigure[variational quantum circuit $U_\mathcal{E}$]{
    \includegraphics{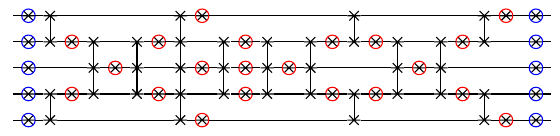}
}
\subfigure[variational quantum circuit $U_\mathcal{R}$]{
    \includegraphics{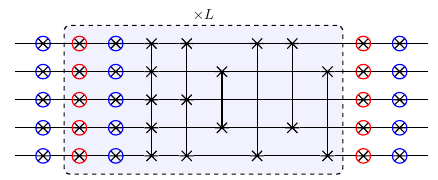}
}
\caption{(a) The variational quantum circuit $U_\mathcal{E}$ contains a layer of $R_z$ rotations to each qubit, a circuit block from the symmetrized $\mathcal{G}_\mathcal{E}$~(Fig.~\ref{fig:matchgate2}) and another layer of $R_z$ rotations at the last.
(b) The variational quantum circuit $U_\mathcal{R}$: We apply $R_X$-$R_Z$ rotations to each qubit, $R_{ZZ}$ gates to all pairs of qubits and repeat the previous operation $L$ times. Finally apply $R_Z$ rotations and $R_X$-$R_Z$ rotations to each qubit at at the beginning and end of the circuit, respectively. 
For convenience, the double qubit gate drawn in black represents $R_{ZZ}$, the single-qubit gate drawn in the red circle is $R_X$ and the single qubit gate drawn in the blue circle is $R_Z$.}\label{fig:encode}
\end{figure}

The variational quantum circuit $U_\mathcal{E}$ in encoding map $\mathcal{E}$ is chosen to be a composition of a layer of $R_z$ rotations to each qubit, a circuit block transformed from the symmetrized $\mathcal{G}_\mathcal{E}$~(Fig.~\ref{fig:matchgate2}) and another layer of $R_z$ rotations at the last. The additional $R_z$ rotations are added to enhance the expressiveness of the variational quantum circuit.
For $n=5$, the variational quantum circuit $U_\mathcal{E}$ is shown in Fig.~\ref{fig:encode}(a).

In our simulations, we tested the VGQEC codes with a single logical qubit. When the code consists of $n$ physical qubits, the variational quantum circuit in the recovery map $U_\mathcal{R}$ has $n+2$ qubits, as discussed in Appendix~\ref{ap:aux_qubit}. The structure of the variational quantum circuit $U_\mathcal{R}$ utilized in the simulation is shown in Fig.~\ref{fig:encode}(b). 
The circuit $U_\mathcal{R}$ can be decomposed into a layer of $R_z$ rotations, $L$ repetitive circuit blocks and alternating layers of single-qubit rotations $R_X$-$R_Z$ acting on all qubits. Each circuit block contains an alternating layer of $R_x$-$R_z$ rotations layer and interactions $R_{ZZ}$ acting on all pairs of qubits.
In principle, any n-qubit unitary evolution can be realized by this ansatz with a sufficiently large $L$, since $\{R_X, R_Z, R_{ZZ}\}$ forms a universal quantum gate set.
For our numerical results, the number of repetitions in $U_\mathcal{R}$ is chosen to be $L=3$.

\section{Details of Numerical Simulation}\label{ap:optimize}

\subsection{Subroutine for Average Fidelity Estimation}\label{ap:subroutine}
To tailor the VGQEC code for a particular noise channel, the \textit{average entanglement fidelity} is employed as the objective function for optimization:
\begin{equation}\label{eq:cost_fun}
    F(\vec{\alpha},\vec{\beta})= {\mathbb{E}}_{\ket{\psi}\sim\mu_H} F_e(\ket{\psi}\bra{\psi},\mathcal{M}),
\end{equation}
where $\mu_H$ is the Haar measure~\cite{mele2024introduction} and $F_e$ is the entanglement fidelity.
The quantum channel $\mathcal{M}$ is chosen to be the noise channel with VGQEC code protection, denoted by $\mathcal{M}=\mathcal{R}\circ\mathcal{N}\circ\mathcal{E}$, where $\mathcal{E}$ and $\mathcal{R}$ represent the encoding and recovery maps of the VGQEC code. 
The parameters vectors $\vec{\alpha}$, $\vec{\beta}$ correspond to the parameters of the variational quantum circuits $U_\mathcal{E}$ and $U_\mathcal{R}$ in $\mathcal{E}$ and $\mathcal{R}$, respectively.

Now, we illustrate a subroutine to estimate the average entanglement fidelity (\ref{eq:cost_fun}). The fidelity estimation algorithm first appeared in~\cite{dankert2009exact}.
An important step in average entanglement fidelity estimation is to generate random quantum states $\ket{\psi}$.
This can be done by acting operator $U$ on the initial quantum state, where $U$ is sampled randomly from the Haar distribution. 
However, since the fidelity depends only on the second-order moments of the distribution, there is no need to sample $U$ from the Haar distribution, but from unitary 2-design distribution~\cite{dankert2009exact}.
A unitary 2-design is a set $X$ on the unitary group $\mathcal{U}(d)$ satisfying
\begin{equation}
    \frac{1}{|X|}\sum_{U\in X} U^{\otimes2}\otimes ({U^\dagger})^{\otimes2}=\int_{\mathcal{U}(d)} U^{\otimes2}\otimes ({U^\dagger})^{\otimes2}d\mu_H(U).
\end{equation}

With a 2-design $X$, the average entanglement fidelity \eqref{eq:cost_fun} of the channel $\mathcal{M}$ is written as
\begin{equation}
     F(\vec{\alpha},\vec{\beta})= \frac{1}{|X|}
    \sum_{U \in X}  \bra{0}U^\dagger\mathcal{M}({U\ket{0}\bra{0}U^\dagger})U\ket{0} .
\end{equation}

The average entanglement fidelity can be estimated by sampling 2-design circuits $U$, getting random states $\ket{\psi}$ by applying $U$ to $\ket{0}^{\otimes k}$, then performing the quantum channel $\mathcal{M}$ on $\ket{\psi}$, applying the inverse of $U$ and measuring all qubits in the computational basis.
The probability of measuring all-0 outcomes is the average entanglement fidelity.
The all-0 outcome and the non-all-0 outcome can be considered as a binary sample, so the estimated probabilities have standard deviation $\mathcal{O}(\frac{1}{\sqrt{N}})$ where $N$ is the number of samples.
The schematic of the algorithm is shown in Fig.~\ref{fig:alg}.

In some cases, it may prove advantageous to utilize an approximate unitary 2-design. A noteworthy example is the $\epsilon$-approximate 2-design~\cite{nakata2017unitary,harrow2023approximate}, and is notably straightforward to execute.

We simulated our scheme by modifying several few-qubit codes to adapt to specific noise models.
In the simulations, the average entanglement fidelity in object function was computed over the projective 2-design quantum states set $\left\{\ket{0},\frac{1}{\sqrt{3}}\ket{0}+\sqrt{\frac{2}{3}}\ket{1},\frac{1}{\sqrt{3}}\ket{0}+\sqrt{\frac{2}{3}}e^{\frac{i2\pi}{3}}\ket{1},\frac{1}{\sqrt{3}}\ket{0}+\sqrt{\frac{2}{3}}e^{\frac{i4\pi}{3}}\ket{1}\right\}$.

\begin{figure}
\subfigure[average entanglement fidelity estimator]{
\begin{tikzpicture}[node distance=15pt]
  \node[draw, rounded corners,style={fill=blue!20}](start)   {Sample $U$ from 2-design};
  \node[draw, rounded corners,below=of start,style={fill=blue!20}](step 1){Put $U$ into Fidelity Sampling Circuit};
  \node[draw, rounded corners, below=of step 1,style={fill=blue!20}](step2)  {Measure the output};
  \node[style={fill=red!10}] at (-1.8,0.8) {Repeat $N$ times};
  \coordinate (choice) at (0,-2.6);
  \node[draw, rounded corners,aspect=2, below=of choice,style={fill=blue!20}]     (choice2)  {Estimate average entanglement fidelity};
  \draw[dashed] ($(step2.south west)-(1.5,0.25)$)rectangle($(start.north east)+(1.25,0.25)$);
  \draw[->] (start)  -- (step 1);
  \draw[->] (step 1) -- (step2);
  \draw[->] (choice) -- (choice2);
\end{tikzpicture}
}
\subfigure[fidelity sampling circuit]{
\begin{quantikz}[row sep=0.3cm,column sep=0.3cm]
 &\lstick{$\ket{0}^{\otimes k}$} &\gate[style={fill=yellow!29}]{U} & \qw & \gate[style={fill=blue!10}]{\mathcal{E}}
 \gategroup[1,steps=3,style={dashed,rounded corners,fill=blue!5, inner xsep=2pt},background]{$\sc \mathcal{M}$} 
&\gate[style={fill=red!10}]{\mathcal{N}} 
&\gate[style={fill=blue!10}]{\mathcal{R}} 
& \qw & \gate[style={fill=yellow!29}]{U^\dagger}  & \meter [draw=blue]{} %\\ 
%&\lstick{$\ket{0}^{\otimes n-k}$} &\qw         &\qw & \qw                      & \qw & \qw                                        & \qw & \qw  & \meter [draw=blue]{} 
\end{quantikz}
}
\caption{Schematic illustration of Fidelity Estimation: (a) average entanglement fidelity Estimator:  The average entanglement fidelity estimator samples a unitary $U$ from the 2-design and then puts $U$ into the fidelity sampling circuit to get the outcomes of the circuit. Repeat this process $N$ times, estimated average entanglement fidelity is the probability of all-0 output.
(b) Fidelity Sampling Circuit: Random quantum states are constructed by $U$ acting on $\ket{0}$, after the random state passes through $\mathcal{M}$ and $U^\dagger$, and then measured on the computational basis.}\label{fig:alg}
\end{figure}

\subsection{Optimization Algorithm}

After estimating the average entanglement fidelity, we take the objective function as the input into the classical optimization algorithm \textit{L-BFGS}~\cite{liu1989limited} to maximize it. 
Following the same strategy as before, we independently initialize the parameters for each noise intensity before optimizing them. 
Our numerical simulations indicate that the zero-point parameter configuration is a saddle point of the objective function. The initial parameters are sampled from a normal distribution with a mean of 0 and a standard deviation of $\pi$.
After optimization, the classical optimizer outputs the optimized parameters $\vec{\alpha}_{opt}$, $\vec{\beta}_{opt}$.
The encoding and recovery maps of the optimized VGQEC code are
\begin{align}
    \mathcal{E}_{out}&=U_{\mathcal{E}}(\vec{\alpha}_{opt})\circ\mathcal{E}_c,\\
    \mathcal{R}_{out}&=\mathcal{R}_c\circ U_\mathcal{R}(\vec{\beta}_{opt}).
\end{align}
The optimized value $F(\vec{\alpha}_{opt},\vec{\beta}_{opt})$ is the average entanglement fidelity of the noise channel with the VGQEC code protection.

\subsection{Thermal relaxation process}\label{app:Thermal_relaxation_process}
A qubit can retain information for only a limited time called \textit{Coherence Time}. There are two metrics to specify the coherence time of a quantum device.
$T_1$ \textit{Coherence Time} is associated with the energy loss that the excited state $\ket{1}$ naturally decays to the ground state $\ket{0}$.
$T_1$ indicates the time for natural relaxation of a qubit. 
Besides this, qubits might interact with the environment and encounter a phase error, and the time constant associated with this error is called the $T_2$ \textit{Coherence Time}.

The thermal relaxation process of a single qubit corresponding to wait time $t$ can be described by the following map :
\begin{equation}
    \rho=\begin{bmatrix}
1-\rho_{11} & \rho_{01} \\
\bar{\rho}_{01} & \rho_{11}
\end{bmatrix} 
\longrightarrow 
\begin{bmatrix}
1-\rho_{11}e^{-\frac{t}{T_1}} & \rho_{01}e^{-\frac{t}{2T_1}-\frac{t}{T_\phi}} \\
\bar{\rho}_{01}e^{-\frac{t}{2T_1}-\frac{t}{T_\phi}} & \rho_{11}e^{-\frac{t}{T_1}}
\end{bmatrix},
\end{equation}
where $\frac{1}{T_\phi}=\frac{1}{T_2}-\frac{1}{2T_1}$ and $T_2\leq 2T_1$.
This process has the following Kraus representation:
\begin{align*}
    \mathcal{N}(\rho)&=\sum_{k=1,2,3}A_k\rho A_k^\dagger,\\
    A_1=&\left[
  \begin{array}{cc}
    1 & 0 \\
    0 & \sqrt{1-\gamma-\lambda}
  \end{array}
\right]
A_2=\left[
  \begin{array}{cc}
    0 & \sqrt{\gamma} \\
    0 & 0
  \end{array}
\right]
A_3=\left[
  \begin{array}{cc}
    0 & 0 \\
    0 & \sqrt{\lambda}
  \end{array}
\right],
\end{align*}
where $\gamma=1-e^{-\frac{t}{T_1}}$ and $\lambda=e^{-\frac{t}{T_1}}-e^{-\frac{t}{2T_2}}$.
Such a CPTP map describes a Phase Amplitude Damping channel.

For comparison purposes, we plot the numerically-optimized results obtained by the iterated convex optimization method~\cite{kosut2009quantum} for the asymmetrical thermal relaxation process.
The main idea is that, for a given noise channel, finding the optimal recovery map to maximum channel fidelity when fixing the encoding map can be formulated as a semi-definite program (SDP), as detailed in Appendix~\ref{ap:SDP}.
Similarly, determining the best encoding map when fixing the recovery map is also an SDP.
The procedure begins with the random selection of an initial encoding map, followed by solving the SDP to obtain the optimal recovery map.
Subsequently, setting the recovery map to this optimized one, the SDP is solved again to determine the corresponding optimal encoding map.
The process is iterative, with the channel fidelity increasing at each step until convergence is achieved.
To make the results sufficiently close to the optimal values, we randomly selected $20$ initial encoding maps, iterated $2000$ times and selected the highest value among them at each wait time $t$.

\section{Photonic Realization of Three-Qubit VGQEC Code: Experimental Details}\label{ap:experimental}
\par
Here, we provide a detailed description of our experimental setup.The three-qubit VGQEC is described by the encoding map $\mathcal{E}(\rho)=E_{\mathcal{E}}\rho E_{\mathcal{E}}^\dagger$ with
\begin{equation}
    E_{\mathcal{E}} = \ket{0}_L\bra{0}+\ket{1}_L\bra{1},
\end{equation}
where $\ket{0}_L$ and $\ket{1}_L$ are defined in:
\begin{equation}\label{eq:threequbitad}
    \begin{aligned}
        &\ket{0}_L=\frac{1}{\sqrt{2}}(\ket{000}+i\ket{011}), \\
        &\ket{1}_L=\frac{1}{\sqrt{2}}(i\ket{100}+\ket{111}).
    \end{aligned}
\end{equation}
Following the encoding map, the logical qubit space, a four-dimensional subspace, is referred to as the encoding subspace. In the absence of external disturbances and noise, the logical qubit remains within this encoding subspace. The projection onto the encoding subspace is given by
\begin{equation}
\begin{aligned}
\Pi'_{\mathcal{E}}=&\ket{000}\!\!\bra{000}+\ket{001}\!\!\bra{001}+\ket{110}\!\!\bra{110}+\ket{111}\!\!\bra{111}.
\end{aligned}
\end{equation}
The encoding map combined with the projection $\Pi'_\mathcal{E}$, is consistent with $\mathcal{E}$, as $\Pi'_\mathcal{E}E_{\mathcal{E}} = E_{\mathcal{E}}$.
\par
The average fidelity is 2-design and applying the elements of a 2-design to a fixed pure state produces a set of vectors, which can be expressed by the mutually unbiased bases (MUBs). MUBs for a single qubit are given by the eigenvectors of the Pauli operators $\sigma_x$, $\sigma_y$ and $\sigma_z$ \cite{10.1117/12.615759}. The corresponding orthogonal basis is $\ket{H}$, $\ket{V}$, $\ket{D}$, $\ket{A}$, $\ket{L}$ and $\ket{R}$. We define $\ket{H}$ as $\ket{0}$ and $\ket{V}$ as $\ket{1}$. Then the other MUBs are $\ket{D},\ket{A} = (\ket{0}\pm \ket{1})/\sqrt{2}$ and $\ket{L},\ket{R} = (\ket{0}\pm i\ket{1})/\sqrt{2}$.
\par
The three-qubit VGQEC experimentally demonstrated, exhibits outstanding performance against amplitude damping noise. To show its advantage, we design an optical circuit to simulate three-qubit amplitude damping noise. The Kraus operators $E_0$ and $E_1$ describing the single-qubit noise are given in Eq.~\eqref{eq:dampingchannel}. The three-qubit noise channel $\mathcal{N}$ is then defined as
\begin{equation}
\mathcal{N}(\rho)=\sum_{a,b,c} (E_a \otimes E_b \otimes E_c) \rho (E_a^\dagger \otimes E_b^\dagger \otimes E_c^\dagger),
\end{equation}
where $a,b,c \in \{0, 1\}$.  For compactness, we represent the tensor products of $E_0$ and $E_1$ using the Kraus operators $N_i$, indexed by $i \in \{1,2, \dots, 8\}$. The noise channel can then be expressed as: $\mathcal{N}(\rho) = \sum_{i=1}^8 N_i \rho N_i^{\dagger}$. We also note that the $E_0$ and $E_1$ can be expressed as follows:
\begin{equation}
\begin{split}
E_0  = \begin{bmatrix}
    1 & 0 \\
    0 & 1
\end{bmatrix}
\begin{bmatrix}
1 & 0 \\
0 & \sqrt{1-\lambda}
\end{bmatrix},\ 
E_1  = \begin{bmatrix}
    0 & 1 \\
    1 & 0
\end{bmatrix}
\begin{bmatrix}
0 & 0 \\
0 & \sqrt{\lambda}
\end{bmatrix},
\end{split}
\end{equation}
where each Kraus operator is a diagonal damping operator follows by a permutation. Therefore, each three-qubit Kraus operator $N_i$ can be factored into a diagonal damping operator $A_i$ followed by a permutation unitary $B_i$. The noise channel can then be expressed as:
\begin{equation}
    \mathcal{N}(\rho) = \sum_{i=1}^8 B_i A_i\rho B_i^{\dagger}A_i^{\dagger}.
\end{equation}
\par
The optimal recovery map $\mathcal{R}_{opt}$, obtained by solving a SDP problem, is represented by $16 \times 16$ Choi matrix with rank 5. This implies that the recovery map can be implemented using five Kraus operators, each represented by a $2 \times 8$ matrix, denoted as $R_i$, corresponding to the eigenvectors associated with the five non-zero eigenvalues of the Choi matrix. $R_i$ can be denoted by inner attribute of the Choi matrix. Three of the five recovery map Kraus operator are parameterized:
\begin{equation}
%\begin{array}{c}
%\begin{split}
R_0  = \begin{bmatrix}
    0 & 0 & 0 & 0 & 0 & 0 & 0 & 0\\
    c_1 & 0 & 0 & 0 & 0 & 0 & i c_2 & 0
\end{bmatrix},\ 
R_1  = \begin{bmatrix}
    \epsilon_1 & 0 & 0 & 0 & 0 & 0 & i\epsilon_2 & 0\\
    0 & i c_3 & 0 & 0 & 0 & 0 & 0 & c_4
\end{bmatrix},\
R_2  = \begin{bmatrix}
    c_2 & 0 & 0 & 0 & 0 & 0 & -i c_1 & 0\\
    0 & -i c_4 & 0 & 0 & 0 & 0 & 0 & c_3
\end{bmatrix},\ 
%\end{array}
\end{equation}
Two of them are parameter-invariant:
\begin{equation}
R_3  = \begin{bmatrix}
    0 & 0 & 1 & 0 & 0 & 0 & 0 & 0\\
    0 & 0 & 0 & i & 0 & 0 & 0 & 0
\end{bmatrix},\
R_4 = \begin{bmatrix}
    0 & 0 & 0 & 0 & 1 & 0 & 0 & 0\\
    0 & 0 & 0 & 0 & 0 & i & 0 & 0
\end{bmatrix},\
\end{equation}
where the six variable parameters satisfy $\{c_1,c_2,c_3,c_4,\epsilon_1,\epsilon_1\} \in [0,1]$ and $c^2_1+c_2^2=c_3^2+c_4^2=1$.  $c_1=0.0663\lambda^3+0.0952\lambda^2-0.3517\lambda+0.7067$ and $c_3=0.0947\lambda^3-0.1014\lambda^2-0.3482\lambda+0.7067$ are obtained by a third-order fitting. $\epsilon_1$ and $\epsilon_2$ are one order of magnitude smaller than $\{c_1,c_2,c_3,c_4\}$.
The optimal recovery map is then given by:
\begin{equation}
    \mathcal{R}_{opt}(\rho)=\sum_{i=1}^5 R_i \rho R_i^\dagger.
\end{equation}
\par
The optimal composite channel, encompassing the encoding, noise channel, and optimal recovery, is given by:
\begin{equation}
    \mathcal{M}_{opt}(\rho)=\mathcal{R}_{opt} \circ \mathcal{N} \circ \mathcal{E} =\sum_{i,j} R_j N_i E_{\mathcal{E}} \rho E_{\mathcal{E}}^\dagger N_i^\dagger R_j^\dagger.
\end{equation}
Considering the encoding subspace and the decomposition of the noise channel, we can further rewrite the composite channel as
\begin{equation}
\mathcal{M}_{opt}(\rho)=\sum_{i,j} R_j B_i A_i \Pi'_{\mathcal{E}} E_{\mathcal{E}} \rho E_{\mathcal{E}}^\dagger \Pi'_{\mathcal{E}} A_i^\dagger B_i^{\dagger}  R_j^\dagger,\\
\end{equation}
with $\Pi'_{\mathcal{E}} E_{\mathcal{E}}=E_{\mathcal{E}}$ and $N_i = B_i A_i$. While the composite of the noise channel and the recovery map has 40 different Kraus operators (8 $N_i$ and 5 $R_j$), the projector $\Pi'_{\mathcal{E}}$ rules out lots of them, and leaves only 14 non-zero Kraus operators.
We define a new group of Kraus operators $K_l = C_{j_l}A_{i_l} E_{\mathcal{E}}$ with $C_{j_l} = R_{j_l}B_{i_l}$, where $l$ denotes the index of non-zero Kraus operators for $\mathcal{M}_{opt}$. The optimal composite channel naturally becomes
\begin{equation}
\begin{aligned}
\mathcal{M}_{opt} (\rho)
=\sum_{l=1}^{14} K_l \mathcal{E}(\rho) K_l^\dagger =\sum_{l=1}^{14} C_{j_l}A_{i_l} E_{\mathcal{E}} \rho E_{\mathcal{E}}^\dagger C_{j_l}^\dagger A_{i_l}^\dagger, \\
\end{aligned}
\end{equation}
where $E_{\mathcal{E}}$, $A_{i_l}$ and $C_{j_l}$ are realized in Module-1, Module-2 and Module-3 of our experimental setup, respectively.

\section{Auxiliary Qubits in Variational Circuit}\label{ap:aux_qubit}
In this section, we discuss the demand of auxiliary qubits when implementing arbitary CPTP maps $\mathcal{T}:\mathcal{L}(\mathcal{H}^{\otimes n})\rightarrow \mathcal{L}(\mathcal{H}^{\otimes k})$ using a unitary and paratial trace operation. Specifically, we will show that there existing a unitary $U_\mathcal{T}$ acting on $\mathcal{H}^{\otimes n+2k}$ such that the CPTP map $\mathcal{T}$ can be implemented by the unitary $U_T$ and a partial trace operation:
\begin{equation}
    \mathcal{T}(\rho)=\tr_{\mathcal{H}^{\otimes n+k}}[U_\mathcal{T}(\rho\otimes \ketbra{0}^{\otimes 2k})U_\mathcal{T}^\dagger],
\end{equation}
for any $\rho\in\mathcal{L}(\mathcal{H}^{\otimes n})$. Thus inserting $2k$ auxiliary qubits is sufficient to implement the recovery map.

For convenience, we use a general method of converting operators to ket notation. Let $C$ be a bounded linear operator from $\mathcal{H}_2$ to $\mathcal{H}_1$,  $C\in\mathcal{L}(\mathcal{H}_2,\mathcal{H}_1)$.
The ket notation in the Hilbert space $\mathcal{H}_1\otimes\mathcal{H}_2$ associated with $C$ is defined as
\begin{equation}\label{eq:notation}
    | C \rangle\rangle=\sum_{ij}c_{ij}\ket{i}_1\ket{j}_2,
\end{equation}
where $\{\ket{i}_1\}$ and $\{\ket{j}_2\}$ are orthonormal bases for $\mathcal{H}_1$ and $\mathcal{H}_2$, and $c_{ij}= { }_1\langle i|C| j\rangle_2$ is the matrix element of $C$ on these bases.

In general, a completely positive~(CP) map $\mathcal{T}:\mathcal{L}(\mathcal{H}^{\otimes n})\rightarrow \mathcal{L}(\mathcal{H}^{\otimes k})$ can be represented by Kraus decomposition, where the map is specified by a set of operators $\{T_s\}$. 
But the representation is not one-to-one between sets of operator elements and channel maps, even the sizes of operator sets aren't fixed.
Another alternative representation is the \textit{Choi matrix}~\cite{de1967linear}, which represents completely positive map $\mathcal{T}: \mathcal{L}(\mathcal{H}^{\otimes n})\rightarrow \mathcal{L}(\mathcal{H}^{\otimes k})$ as a positive operator $R_\mathcal{T}\in \mathcal{L}(\mathcal{H}^{\otimes n+k})$ as follows
\begin{equation}
    R_\mathcal{T}\equiv \mathcal{T}\otimes\mathcal{I}(|I \rangle\rangle \langle\langle I |),
\end{equation}
where $\mathcal{I}$ denotes the identical map over the extension space $\mathcal{H}^{\otimes n}$, and the vector $|I \rangle\rangle \in \mathcal{H}^{\otimes 2n}$ is the ket notation (\ref{eq:notation}) for identity matrix in $\mathcal{H}^{\otimes n}$.
As described in Ref.~\cite{d2001optimal}, the correspondence between CP maps from $\mathcal{L}(\mathcal{H}^{\otimes n})$ to $\mathcal{L}(\mathcal{H}^{\otimes k})$ and positive operators on $\mathcal{H}^{\otimes n+k}$ is one-to-one, and the Kraus decomposition $\{T_s\}$ of the CP map is associated with the positive operators $R_\mathcal{T}$ corresponding to the CP map as follows:
%\wfc{wfc: The correspondence between CP maps from $\mathcal{L}(\mathcal{H}^{\otimes n})$ to $\mathcal{L}(\mathcal{H}^{\otimes k})$ and positive operators on $\mathcal{H}^{\otimes n+k}$ is one-to-one \cite{d2001optimal}. The Kraus decomposition $\{T_s\}$ of the CP map is associated with the positive operators $R_\mathcal{T}$ corresponding to the CP map as follows \cite{d2001optimal}:}
\begin{equation}
    R_\mathcal{T}=\sum_s |T_s \rangle\rangle \langle\langle T_s |,
\end{equation}
and 
\begin{equation}
    \mathcal{T}(\rho)=\sum_s T_s\rho T_s^\dagger=\tr_{\mathcal{H}^{\otimes n}}[I\otimes \rho^T R_\mathcal{T}].
\end{equation}

The trace-preserving condition for $\mathcal{T}$ 
\begin{equation}
    \tr_{\mathcal{H}^{\otimes k}}{\mathcal{T}(\rho)}=1=\tr_{\mathcal{H}^{\otimes n}}[\rho^T \tr_{\mathcal{H}^{\otimes k}}[R_\mathcal{T}]]
\end{equation}
equivalently requires that the Choi state $R_\mathcal{T}$ satisfies
\begin{equation}\label{eq:CPTP}
    \tr_{\mathcal{H}^{\otimes k}}[R_\mathcal{T}]=I\in \mathcal{H}^{\otimes n}.
\end{equation}

By doing the spectral decomposition of $R_\mathcal{T}\in \mathcal{L}(\mathcal{H}^{\otimes n+k})$, we can obtain a special Kraus decomposition:
\begin{equation}
    R_\mathcal{T}=\sum_s |\hat{T}_s \rangle\rangle \langle\langle \hat{T}_s |,
\end{equation}
where $\hat{T}_s \in \mathcal{L}(\mathcal{H}^{\otimes n},\mathcal{H}^{\otimes k})$, and the size of operator set $\{\hat{T}_s\}$ is the same as the dimension of Hilbert space $\mathcal{H}^{\otimes n+k}$.
Specifically, when $\mathcal{T}$ is a CPTP map by (\ref{eq:CPTP}), we have:
\begin{equation}
    \sum_s \hat{T}^\dagger_s\hat{T}_s=\sum_s\tr_{\mathcal{H}^{\otimes k}}{|\hat{T}_s \rangle\rangle \langle\langle \hat{T}_s |}=I \in \mathcal{L}(\mathcal{H}^{\otimes n}).
\end{equation}
We can define an operator $U_\mathcal{T}\in\mathcal{L}(\mathcal{H}^{\otimes n+2k})$, which satisfying:
\begin{equation}\label{eq:ut}
    U_\mathcal{T}\ket{\psi}\ket{0}^{\otimes 2k}=\sum_s \hat{T}_s\ket{\psi} \ket{e_s},
\end{equation}
where $\ket{\psi}\in\mathcal{H}^{\otimes n}$  and $\{\ket{e_s}\}$ are orthonormal basis for $\mathcal{H}^{\otimes n+k}$. For arbitary states $\ket{\psi},\ket{\phi}\in\mathcal{H}^{\otimes n}$, there is the following relationship:
\begin{equation}
    \bra{\psi}\bra{0}^{\otimes{k}} U_\mathcal{T}^\dagger U_\mathcal{T}\ket{\phi}\ket{0}^{\otimes{k}}=\sum_s\bra{\psi}\hat{T}_s^\dagger\hat{T}_s\ket{\phi}=\bra{\psi}\ket{\phi}.
\end{equation}
Thus $U_\mathcal{T}$ can be extended to a unitary operator acting on the space $\mathcal{H}^{\otimes n+2k}$. It's easy to verify that
\begin{equation}
    \tr_{\mathcal{H}^{\otimes n+k}}{[U_\mathcal{T}(\rho\otimes\ketbra{0}^{\otimes 2k})U_\mathcal{T}^\dagger]}=\sum_s \hat{T}_s\rho \hat{T}_s^\dagger.
\end{equation}

In summary, the above derivation yields the following results:
Let $\mathcal{T}: \mathcal{L}(\mathcal{H}^{\otimes n})\rightarrow \mathcal{L}(\mathcal{H}^{\otimes k})$ be a completely positive and trace-preserving linear map. Then there is a unitary $U_\mathcal{T}\in \mathcal{L}(\mathcal{H}^{\otimes n+2k})$ such that 
\begin{equation}
    \mathcal{T}(\rho)=\tr_{\mathcal{H}^{\otimes n+k}}{[U_\mathcal{T}(\rho\otimes\ketbra{0}^{\otimes 2k})U_\mathcal{T}^\dagger]}.
\end{equation}

In particular, for $[n,k]$ stabilizer codes, the size of Kraus operators set corresponding to the recovery map of the stabilizer code is $2^n/2^k$. Mirroring the construction of $U_\mathcal{T}$ in Eq.~\eqref{eq:ut}, the recovery map for stabilizer codes can be achieved by applying a unitary $V\in \mathcal{L}(\mathcal{H}^{\otimes n})$ and subsequently measuring the last $n-k$ qubits.
%As described in Ref.~\cite{fletcher2007optimum}, when given input ensemble, encoding map and noise channel, 
We assume the optimal recovery map $\mathcal{R}_{opt}$ which maximize channel fidelity, corresponds to unitary $U_{opt}\in\mathcal{L}(\mathcal{H}^{\otimes{n+2k}})$.
When considering the recovery map of VGQEC codes, the optimal recovery map can be realized by introducing $2k$ auxiliary qubits, applying a unitary operator $U_\mathcal{R}=(V^\dagger \otimes I)U_{opt} \in\mathcal{L}(\mathcal{H}^{\otimes n+2k})$, then measuring the auxiliary qubits, and subsequently employing the stabilizer code's original recovery map afterward. 
The process can be formulated as follows:
\begin{equation}
    \mathcal{R}(\rho)=\tr_{\mathcal{H}^{\otimes n+k}}{[(V\otimes I) U_\mathcal{R}(\rho\otimes\ketbra{0}^{\otimes 2k})U_\mathcal{R}^\dagger(V^\dagger \otimes I)]}=\tr_{\mathcal{H}^{\otimes n+k}}{[U_{opt}(\rho\otimes\ketbra{0}^{\otimes 2k})U_{opt}^\dagger]},
\end{equation}
where the second equation is due to $(V \otimes I)(V^\dagger \otimes I)U_{opt}=U_{opt}$, the desired recovery map $\mathcal{R}_{opt}$ is applied.

As a result, introducing $2k$ auxiliary qubits is sufficient.
% \wfc{The term $V^\dagger \otimes I$ cancels the original recovery map, }
% \wfc{$V^\dagger \otimes I$ looks a little weird}
% \wfc{dao li wo dou dong, but the recovery process still looks a little luan.}

\section{Another three-qubit code for the amplitude damping noise}\label{ap:three_qubit}

Ref.~\cite{dutta2024smallest} presents another three-qubit code designed to address amplitude damping noise. The logical states are defined as follows:
\begin{equation}
    \ket{0}_L=\frac{1}{\sqrt{3}}(\ket{100}+\ket{010}+\ket{001}),\quad \ket{1}_L=\ket{111}.
\end{equation}

In the reference, the authors explicitly construct a recovery map for amplitude damping noise, which is determined through syndrome measurement:
\begin{align}
    P_0&=\ketbra{0_L}+\ketbra{1_L}, \\
    P_1&=\ketbra{000}+\ketbra{110}+\ketbra{101}+\ketbra{011}, \\
    P_f&=I-P_0-P_1,
\end{align}
where $P_0$ and $P_1$ correspond to scenarios with no damping error and a single damping error, respectively. The recovery operator associated with the syndrome measurement is given by:
\begin{align}
    R_0&=\sqrt{1-\lambda}\ketbra{0_L}+\ketbra{1_L}, \\
    R_1&=\ketbra{0_L}{000}+\frac{1}{(1-\lambda)\sqrt{3}}\ket{1_L}(\bra{110}+\bra{101}+\bra{011}),
\end{align}
where $\lambda$ is the damping parameter. Here, $R_0$ corrects for the distortion due to the no-damping error when the syndrome measurement yields $P_0$, while $R_1$ addresses distortions from a single damping error when the measurement output $P_1$.
Based on the above explicit recovery map, Ref.~\cite{dutta2024smallest} demonstrates that the worst-case fidelity of the code is $1-\lambda^2+\Omega(\lambda^3)$, which contains no linear term in $\lambda$.

\begin{figure}[htbp]
\centering
\includegraphics[width=0.5\textwidth]{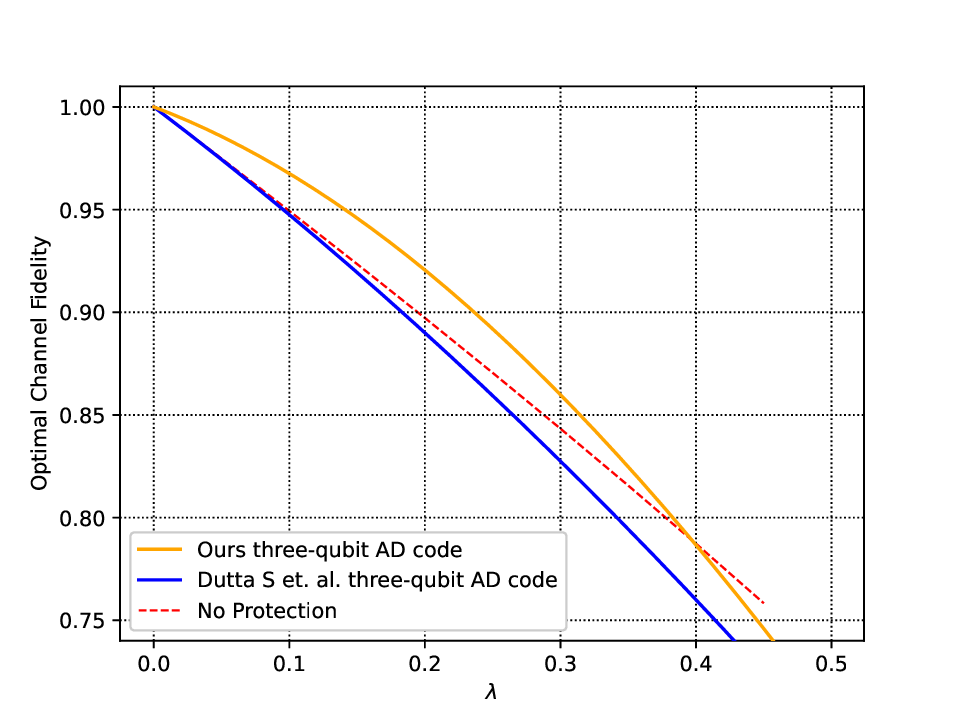}
\caption{The optimal channel fidelity of the three-qubit code for amplitude damping noise. The blue curve represents the three-qubit code introduced in Ref.~\cite{dutta2024smallest}, while the orange curve represents the three-qubit code obtained from the VGQEC scheme.}\label{fig:three_qubit}
\end{figure}

In contrast, since our three-qubit code is derived differently and the recovery map is not explicitly provided, we are unable to directly formulate an analytic expression for fidelity. For the convenience of the readers, we numerically calculate the optimal channel fidelity for both three-qubit codes under amplitude damping noise. The results are shown in Fig.~\ref{fig:three_qubit}.

\end{document}